\documentclass[manuscript,screen]{acmart}
\usepackage{booktabs} % For formal tables

\usepackage[ruled]{algorithm2e} % For algorithms
\usepackage{amsmath,amssymb,amsfonts}
\usepackage{graphicx}
\usepackage{multirow}
\usepackage{subfig}

\begin{document}
% Title portion
\title{Malicious Overtones: hunting data theft in the frequency domain with one-class learning}

\author{Brian~A.~Powell}
\affiliation{%
  \institution{Johns Hopkins University Applied Physics Laboratory}
  \streetaddress{11100 Johns Hopkins Rd.}
  \city{Laurel}
  \state{MD}
  \postcode{20723}
  \country{USA}}
\email{brian.a.powell@jhuapl.edu}

\begin{abstract}
A method for detecting electronic data theft from computer networks is described, capable of recognizing patterns of remote exfiltration occurring over days to weeks.  Normal traffic flow data, in the form of a host's ingress and egress bytes over time, is used to train an ensemble of one-class learners.  The detection ensemble is modular, with individual classifiers trained on different traffic features thought to characterize malicious data transfers.  We select features that model the egress to ingress byte balance over time, periodicity, short time-scale irregularity, and density of the traffic. The features are most efficiently modeled in the frequency domain, which has the added benefit that variable duration flows are transformed to a fixed-size feature vector, and by sampling the frequency space appropriately, long-duration flows can be tested.  When trained on days- or weeks-worth of traffic from individual hosts, our ensemble achieves a low false positive rate ($<2\%)$ on a range of different system types.  Simulated exfiltration samples with a variety of different timing and data characteristics were generated and used to test ensemble performance on different kinds of systems:  when trained on a client workstation's external traffic, the ensemble was generally successful at detecting exfiltration that is not simultaneously ingress-heavy, connection-sparse, and of short duration---a combination that is not optimal for attackers seeking to transfer large amounts of data.   Remote exfiltration is more difficult to detect from egress-heavy systems, like web servers, with normal traffic exhibiting timing characteristics similar to a wide range of exfiltration types.   
%The modular ensemble can be customized to target exfiltration of different types or sophistication, or even different kinds of anomalous traffic.
\end{abstract}

%
% The code below should be generated by the tool at
% http://dl.acm.org/ccs.cfm
% Please copy and paste the code instead of the example below.
%

%
% End generated code
%


\maketitle

\section{Introduction}
\label{sec:introduction}
In 2013, Yahoo! was compromised by a state-sponsored actor using forged web cookies.  The ensuing data breach was the largest in history, including the names, email addresses, birth dates, telephone numbers, and password hashes of 3 billion users.  The breach devalued the company an estimated \$350 million \cite{Yahoo}. In 2012, attackers used valid credentials to access databases at the United States Office of Personnel Management, exfiltrating finger prints, social security numbers, and personal details of 21.5 million cleared defense contractors.  Credit monitoring costs alone for the affected individuals are upwards of \$500 million \cite{OPM}.  More recently, Equifax fell victim to a vulnerable web application that resulted in the theft of sensitive data like social security numbers and birth dates of some 147 million customers.  When lawsuits, investigations, and credit monitoring services are payed out, costs could top \$600 million \cite{Equifax}.  Many more breaches go unreported in the media. A global 2017 study of 419 data breaches of varying size reports an average cost to companies of \$3.62 million, with an average cost of \$141 per stolen record \cite{pon}. In short, date theft is {\it expensive}.

Risk is traditionally considered the product of likelihood of occurrence and severity of outcome.  The effects of data theft are undoubtedly severe, and it is becoming increasingly common: there were 1579 U.S. data breaches reported in 2017, up 44.7\% from 2016 \cite{ITRC}.  It is further estimated that victims of a data breach have a 27\% likelihood of suffering another one within 24 months \cite{pon}.  Virtually every sector has experienced some kind of data breach: defense, financial, manufacturing, health, and notably retail sectors have all been targeted in the past several years \cite{verizon}.  Motivations span corporate espionage, to organized financial crime, to state-sponsored information warfare.  In short, data theft is increasingly a {\it high-risk eventuality}.

Data theft comes in many forms, from the physical removal of a disk drive from a secure area to the remote exfiltration of electronic files over an internet connection.  It is this latter method that concerns us here.  It is technically very easy to do: tools include native command line utilities and other freely available open source capabilities. It is also notoriously difficult to detect: data compression and encryption are easily implemented, rendering exfiltration traffic especially hard to find among the crowded network pipes of typical organizations.  The significance of the challenge is born out in the numbers: in 2017, the average time to detect a breach was 191 days, and another 66 days to contain it \cite{pon}.  Meanwhile, the actual exfiltration activity in the majority of cases lasted on the order of minutes-to-hours \cite{verizon}.  

That's not to say there aren't defensive options \cite{MWR,Ullah}. Deep packet inspection of traffic over common channels like web or email can identify signatures, keywords, or other indicators of exfiltrated data in unencrypted traffic \cite{Liu}, but the wide-spread use of encryption poses a challenge to these methods.  Domain Name Service reputation scoring and other filters might reveal remote sites as plausibly malicious, though domain fronting and the proliferation of cloud services has made this technique less effective in recent times.  Approaches based on static rules aren't generally flexible enough to recognize normal traffic that might resemble exfiltration ({\it e.g.} file uploads to Google Drive or remote data backup), generating undesirable rates of false positives \cite{Mazzola}.  

Academic research has provided a wealth of approaches to the problem of data exfiltration, though most proposals tend to focus on a specific application, threat model, or exfiltration scenario.  Methods of host-based data theft detection seek to identify malicious users ({\it e.g.} by looking for anomalous patterns in system call data \cite{Jewell}) or unusual log files ({\it e.g.} to find unauthorized database transactions \cite{Kim3}).  The complementary approach of network-based data theft detection has also been explored in the literature. Though more general, approaches tend to focus on specific targets, like malicious insiders \cite{Koch}; use-cases, like encrypted communications \cite{Koch2} or entire-network detection \cite{Marchetti}; or environments, like tactical wireless networks \cite{Sigholm}.  Those focused on network traffic prove effective against only specific kinds of data exfiltration, like large volume transfers \cite{Marchetti} or long-duration flows \cite{Wang}.   

Meanwhile, there has been substantial research on more generic traffic anomaly detection, though many of these methods are designed to detect volume anomalies (like denial of service or worm propagation) or local traffic irregularities (like performance spikes or outages) that appear as extreme values about a normal traffic background.  Data exfiltration need not be a localized, aberrant event like these, but can instead blend into the normal ebb and flow of a network's egress traffic as a complete communication on its own. As a post-exploitation activity, data exfiltration is not included in the network intrusion benchmarks like the DARPA Intrusion Detection Evaluation-derived KDD Cup 1999 data set \cite{KDDCUP} that spurred the development of much published work on network anomaly detection.  There is thus a relative dearth of methods tailored to the problem of detecting general data exfiltration at the network-level, and a pressing need to develop them given today's threat landscape. The absence of clear indicators or triggers for data exfiltration, which can be carried out quietly, at low data rates, over long periods of time to otherwise legitimate remote sites, argues for approaches based on adaptive pattern recognition of normal network traffic so that data exfiltration, as an atypical activity, will be recognized as such.  

In this paper we describe a {\it flow-based, transform-domain one-class learning} method that learns the patterns of behavior of a given host's normal traffic over potentially long time periods.  It seeks to detect overt data exfiltration\footnote{In contrast to {\it covert} exfiltration, in which data is indirectly communicated, via, for example, timing channels or network steganography.} of a general nature, in which the data itself is communicated out of the network using protocols like web/secure web (HTTP/HTTPS), file transfer protocol (FTP), or internet relay chat (IRC) over time periods spanning minutes, days, or weeks.  The method considers {\it only} the ingress and egress bytes of the device over time, data easily obtained from network flow records or packet captures.  It is protocol agnostic and does not consider the contents of the traffic. It is applicable to encrypted, compressed, encoded, or otherwise obfuscated data. Because data exfiltration need not be a localized anomaly about normal behavior, this method examines traffic between a single host of interest and each remote system to which it connects separately, labeling each entire flow as either normal or exfiltration.  This approach is therefore based on classification rather than traditional statistical outlier detection.

Starting with ingress and egress bytes over time, we derive a set of features characterizing each time series in the frequency domain.  The adoption of transform-domain features is motivated by two key considerations: frequency-space representations like the Fourier transform efficiently encode temporal correlations in time series data, enabling learning algorithms to learn more complex correlations with fewer features than those trained on temporal domain data.  The second consideration is our desire to test traffic flows for long-duration structure, indicative of data exfiltration occurring over days or weeks.  A several-weeks long flow at seconds-resolution will have millions of time steps, and so any classifier trained in the time-domain will have millions of features to contend with.  Downsampling can help, but only at the cost of stamping out local variations in the time series.  In the frequency domain, based on the desired resolution at each time scale ({\it e.g.} seconds-resolution on minute time scales, hourly resolution on daily time scales, etc) a frequency binning is chosen and fixed and each time series is then mapped to the same dimension in frequency space.  This allows us to test flows of varying length, including those of long duration. 

This method establishes a baseline of putative normal network traffic by training a set of one-class learners on a historical record of a host's flow data.  Normal traffic is so-defined because it is not malicious, not because it fits a particular pattern or all looks the same.  In contrast to unsupervised anomaly detection methods, which work by finding outliers relative to bulk clusters or distributions of similar points, one-class learning allows the user to {\it define} what normal is, with the allowed variability of the class only limited by the capacity of the learning algorithm.  In contrast to supervised learning, the advantage of one-class models is that one does not need to define up-front what an anomaly looks like.

In this work, we derive four sets of features from the regular and short-time discrete Fourier transform.  We employ an ensemble of one-class learners, each trained on a different feature representation, such that errors across ensemble components are approximately decorrelated. The approach is {\it modular}, in that classifiers can be added or removed from the ensemble to test for different combinations of features.  The chosen feature sets encode different traffic characteristics, including the balance of ingress to egress bytes; harmonic structure, like periodicity; short time-scale byte irregularities; and byte density of the traffic.   

To test the ensemble, we generate a sample set of simulated exfiltration traffic, covering a variety of different egress patterns organized into four classes based on timing and data characteristics: periodic/constant egress data, aperiodic/constant egress data, periodic/variable egress data, and aperiodic/variable egress data.  We test the ensemble on a variety of systems: a client workstation, an email gateway with mixed client and server behavior, and an outward-facing web server.  We find that the ensemble works best on client systems, and is generally effective at detecting exfiltration that is not simultaneously {\it ingress-heavy, connection-sparse, and of relatively short duration}---a combination that conspires to make it difficult for the adversary to get appreciable amounts of data out of the network.  The ensemble performs moderately-well for certain kinds of exfiltration on the email gateway,  but struggles to detect exfiltration from egress-heavy systems, like web servers, whose normal traffic exhibits timing characteristics similar to exfiltration.  

%To the author's knowledge, this is the first time these elements---flow-based, transform-domain, one-class learning---have been applied together to the problem of data exfiltration detection, though there is a vast body of related prior work on the subject, which we now discuss.
\section{Related Work}
\subsection{Data exfiltration proper}
There has been much recent work on the problem of detecting data exfiltration, including host- and network-based approaches targeting both insider and remote threats (see \cite{Ullah} for a modern and comprehensive review).   Host-based approaches typically focus on identifying abnormal user or system behavior suggestive of data theft.  In \cite{Jewell}, user system call behavior is analyzed for anomalous activity and the authors demonstrate that certain actions relevant to data exfiltration, like the unauthorized copying of files to removable drives, can be detected.  \cite{Kim3} employ density-based outlier detection to the problem of database access monitoring, with the hopes of identifying malicious database interactions indicative of data theft.  This work is focused primarily on developing a computationally efficient means of performing this analysis, though the method is currently unable to alert in real-time as transactions occur.  Databases are also the focus of \cite{Shan}, who propose an adaptive SQL injection protection framework.  

Complementary network-based approaches have also been developed.  Koch and Rodosek \cite{Koch} propose an insider detection scheme that examines packet sizes and data rates to ascertain user typing characteristics in order to create user profiles.  The method is applicable to encrypted communications because it does not leverage data contents.  Anomalous activity can be identified as that not associated with any known user profile.  \cite{Marchetti} seek to identify anomalous systems across the network with respect to a set of traffic characteristics indicative of some types of data exfiltration: the number of egress bytes, number of flows initiated by internal hosts, and the number of unique external address contacted by internal hosts within a given time period.  Hosts are then ranked according to a ``suspiciousness'' score which is shown to be a reliable metric for detecting data exfiltration in excess of 500 Mb per day.  \cite{Sigholm} attempt to detect data exfiltration occurring in tactical mobile ad-hoc networs.  They apply clustering to a set of flow and packet header data and identify outlier flows as those lying beyond some threshold distance to a nearest cluster.   In \cite{Pu}, a behavior-based exfiltration detection system is devised that considers egress/ingress ratio, long duration connections, and client/server roles in establishing connections as important elements for identifying data theft.  The method seems promising but the system's performance is not well-established in \cite{Pu}.  In \cite{Wang}, a visual analytic illustrating the payload size and duration of connections is proposed as a means of manually ascertaining exfiltration, where it is argued that exfiltration should occur in the high payload size/long duration sector of the graph.  Of course, this is true of only a subset of possible exfiltration scenarios, and the thresholds are not informed by any behavior-based profiling of normal data.  Finally, in \cite{Ramachandran} a hybrid approach incorporating both network and host profiling is introduced.  Kernel density estimation of incoming and outgoing bytes and the nature of packets over time is used to establish normal behavior of individual hosts and tested against densities estimated from current data.  

Our proposed method extends these approaches, which are primarily based on ingress and egress volume statistics, by considering the temporal characteristics of the traffic flows.  It builds on ideas from the general problem of network anomaly detection, such as flow-based assessment, transform-domain features, and classification as a means of detection.  We now summarize some of this work. 
\subsection{Network intrusion detection, more generally}
Network anomaly detection has traditionally been applied to the problems of network intrusion detection and network management.  Though the nature of the anomalies might differ between these two applications, each is concerned with identifying aberrant or unusual patterns in network traffic data.  Methods differ in the datasets used ({\it e.g.} packet header settings like protocol flags, connection details like IP addresses and ports, or traffic statistics like average bytes per flow), scope of application ({\it e.g.} network-wide or single-link), type of anomalies sought ({\it e.g.} volume, change points, or additive outliers), features ({\it e.g.} time, frequency, or time-frequency domains), and detection methodologies ({\it e.g.} statistical, clustering, or classification). 

While deep packet inspection and packet header data have been used to discover cyber attacks, traffic anomaly detection tends to leverage the link and connection statistics provided by flow-level data (see \cite{Drasar,Sperotto,Li} for recent reviews of flow-based anomaly detection methods).  

The work of \cite{Dubendorfer} uses the ratio of incoming to outgoing bytes, and the numbers of bidirectional and outgoing connections to organize systems into three sets according to thresholds placed on each feature.  Anomalies are indicated by the sudden increase in cardinality of one or more of these sets; this detection scheme is most sensitive to rapid, global network changes, like worm propagation.  \cite{Kim} considers a range of flow-level statistics to characterize traffic patterns, and defines a set of ``detecting functions'' that are weighted combinations of these features.  The weights are tuned empirically to detect attacks indicated by extreme values of one or more of these features, like scanning and flooding. The work of \cite{Shyu} takes an unsupervised approach to detecting the KDD Cup intrusion types by examining the principal component analysis (PCA) reconstruction error of samples.   Another unsupervised approach to KDD Cup anomaly detection includes $k$-means clustering with an evolutionary approach to determining the number of clusters, which is key for not over-clustering normal samples \cite{Lu2}.  Other clustering-based method include \cite{Eskin,Leung}.  Supervised approaches to the KDD Cup benchmark include the application of support vector machines to the entropy of connection features \cite{Yan} and to features extracted from Kalman filter-optimized cluster-labeled samples \cite{Garg}. In \cite{Lakhina2} and \cite{Qian}, the distributions of four features: source address, destination address, source port, and destination port are examined for outliers using sample entropy.  Detectable anomalies include large data transfers, denial of service (DoS), traffic bursts, scanning, outages, and worms.   

Traffic anomalies have also been assessed by concentrating on the patterns found in various time series constructed from flow-level data.  The seminal work of \cite{Brutag} employed Holt-Winters forecasting error of byte time series to predict network outages. This kind of forecasting is sensitive to extreme values in the time series, and so is appropriate to localized anomalies. More recently, \cite{Nguyen} expanded this methodology to multivariate time series by including metrics sensitive to socket and port variability, enabling them to detect port scans and TCP SYN floods.  \cite{Lakhina} use PCA to reduce raw traffic time series and then seek local anomalies by computing the prediction error of the residuals, which encoded in the higher-components of the PCA decomposition.  This method allows for the detection of volume anomalies.  The work of \cite{Soule} studies the prediction error of a Kalman filter applied to link traffic time series, and is useful for finding volume anomalies in large networks.  Network anomalies have been investigated using recurrent neural networks with long short-term memory (LSTM) units in \cite{Kim2} and \cite{Cheng}. Both studies are supervised, with the models trained on labeled anomalous data, the former work focused on web traffic and the latter on border gateway protocol. These models can identify anomalies localized to the duration of the window used to define samples, which in these studies included fewer than 100 time steps.  LSTMs were also used in \cite{Radford} to learn traffic patterns between distinct IP pairs over a 10 time step sliding window, and anomalies were flagged based on prediction error of the subsequent time step.   In \cite{Casas2}, traffic time series are examined at various levels of resolution (all traffic versus traffic from different subsets of source and destination address) for change points, and then all flows at a change point are analyzed using sub-space clustering and evidence accumulation.  This method is successful at detecting DoS, scanning, and worms.

Traffic anomaly detection has also been treated in the transform domain.  In applications seeking to detect local aberrations in traffic behavior, wavelets have been employed as an efficient means of separating out normal (low-frequency) baseline traffic in order to isolate the (high-frequency) anomalies.  In \cite{Barford}, traffic time series are decomposed into low- and high-frequency components and local variability of the high-frequency component is examined for anomalies.  The types of anomalies detectable with this method are limited in duration by the desired temporal resolution, here between 3-4 hours.  \cite{Huang} applied the wavelet transform to the time series of packets over time to identify DoS and scanning activity ocurring over relatively short time scales (no longer than 5 minutes).  In \cite{Lu}, an autoregressive model is used to learn the wavelet representation of normal traffic, and anomalous flows are associated with wavelet coefficients that are outliers with respect to a Gaussian mixture model.  This method is applied successfully to the KDD Cup attack types. \cite{Jiang} implement the S-transform of the bytes versus time signal and correlations are computed among the different flows to find those that are anomalous.  This approach applies to network-wide traffic analysis, and can identify distributed DoS, traffic surges and outages, and worm propagation.  Due to their success in discovering volume anomalies in traffic data, wavelets have been widely applied to the problem of DoS discovery, \cite{Ramanathan,Li2,Dainotti}.  \cite{Chimetseren} employs the discrete Fourier transform to differentiate network intrusions from normal traffic, which is treated as noise.  A kind of difference between spectra is used to identify suspected intrusions, resulting in a 5\% false positive and 100\% true positive rate.  In \cite{Meng}, the discrete Fourier transform was used to detect the periodic heartbeat communications of certain trojans.  

As noted throughout, the above research is focused on either volume-type or localized anomalies indicative of DoS, scanning, and other remote unauthorized activity.  Meanwhile, data exfiltration can be a quiet, long-duration activity that requires a different set of detection techniques.   
\section{Data Exfiltration Tradecraft}
Data exfiltration is traditionally a post-exploitation maneuver: an adversary infiltrates a network and subsequently extracts target information.  The initial breach can be accomplished any number of ways, and we aren't interested in these details here.  We assume that an attacker has gained a foothold on the network and has established command and control (C2) with some remote infrastructure.  The C2 channel is facilitated by software on the compromised host and is used by the attacker to issue commands and other instructions to the host: a variety of C2 architectures are available, including IRC, DNS, and HTTP/HTTPS \cite{Giani,Antwerp}.  Often, the compromised host sits behind a firewall or proxy and is not directly accessible from outside the network; C2 agents then must occasionally originate connections outbound from the compromised network back to the attacker's infrastructure.  This is called {\it beaconing}.  Not all attacks will involve C2 infrastructure: autonomous malware can be written to run through a prescribed list of instructions and can establish data channels for the one-way transfer of data out of the network.  Separate data channels for exfiltration might also be established in addition to C2 channels.  

A variety of tools are available for establishing remote access and C2. If secure shell (SSH) or FTP are available on the victim system and allowed out of the network, perhaps via protocol tunneling, these are generally simple and effective options.  If native utilities are not available, lightweight tools like \texttt{netcat} or \texttt{socat} can be installed and used to pull data off the system over arbitrary ports.  Post-exploitation payloads like Meterpreter \cite{meterpreter} or Intersect \cite{intersect} have simple download options.  In addition to its exfiltration capability, Cobalt Strike's Beacon \cite{cobalt} payload supports a variety of C2 functions.  More advanced threats write their own agents that integrate persistent access with a range of post-exploitation tools, including callback and lateral movement capabilities.  

Once a foothold and C2 channel have been established, the adversary will move laterally and escalate privileges in search of target information. If it exists across several systems, the data will be aggregated on one or relatively few systems on the network in preparation for exfiltration.  This is called {\it staging}, and it is done in the interest of stealth to minimize the number of systems from which data will be transferred out of the network. A variety of protocols might be used to stage data, though in enterprise networks where file sharing is abundant, shares can be easily mounted and accessed using native utilities and protocols, like server message block (SMB), offering an efficient and stealthy means of moving data from system to system. Before egress the data might be encrypted (so it cannot be inspected) or obfuscated (by encoding it to prevent detection based on string matching), compressed (to reduce its size), or chunked (to break it into pieces rather than a single, large extraction). These techniques all work to make data exfiltration look inconspicuous, allowing it to ``hide in the noise'' of normal traffic. 

Data will then be exfiltrated over a particular protocol: ultimately, the attacker is constrained by the kinds of traffic that are allowed to egress the network.  Common protocols like web, FTP, email, and chat are popular candidates (though see \cite{Antwerp} for others), making up over half of the protocols used in incidents investigated in 2014 \cite{MWR}.  Data will be exfiltrated to a system controlled by the attacker, typically a device hosted by a common cloud or file sharing service, the rationale being that these are trusted domains and won't raise any red flags.  Depending on the size of the heist, the exfiltration might occur over the span of only a few minutes, or could take several days with data getting pulled out of the network piecemeal.  A variety of exfiltration patterns are achievable by varying the timing of data pulls and the amounts egressed over time.  The objective of this work is to identify exfiltration by learning these patterns of behavior. 
\begin{figure*}[htp]
\begin{tabular}{cc}
\subfloat[]{\includegraphics[width = 3in]{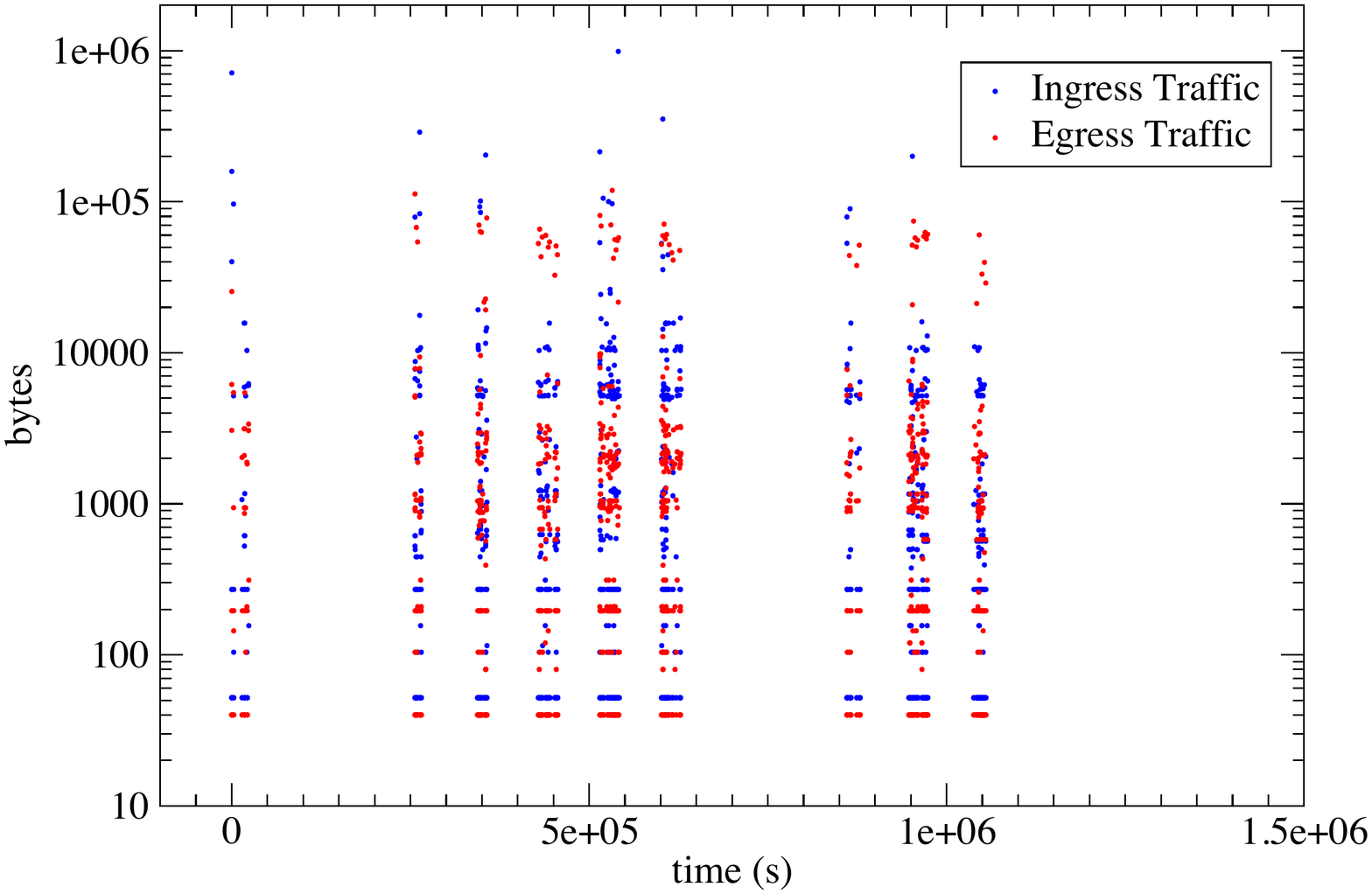}} &
\subfloat[]{\includegraphics[width = 3in]{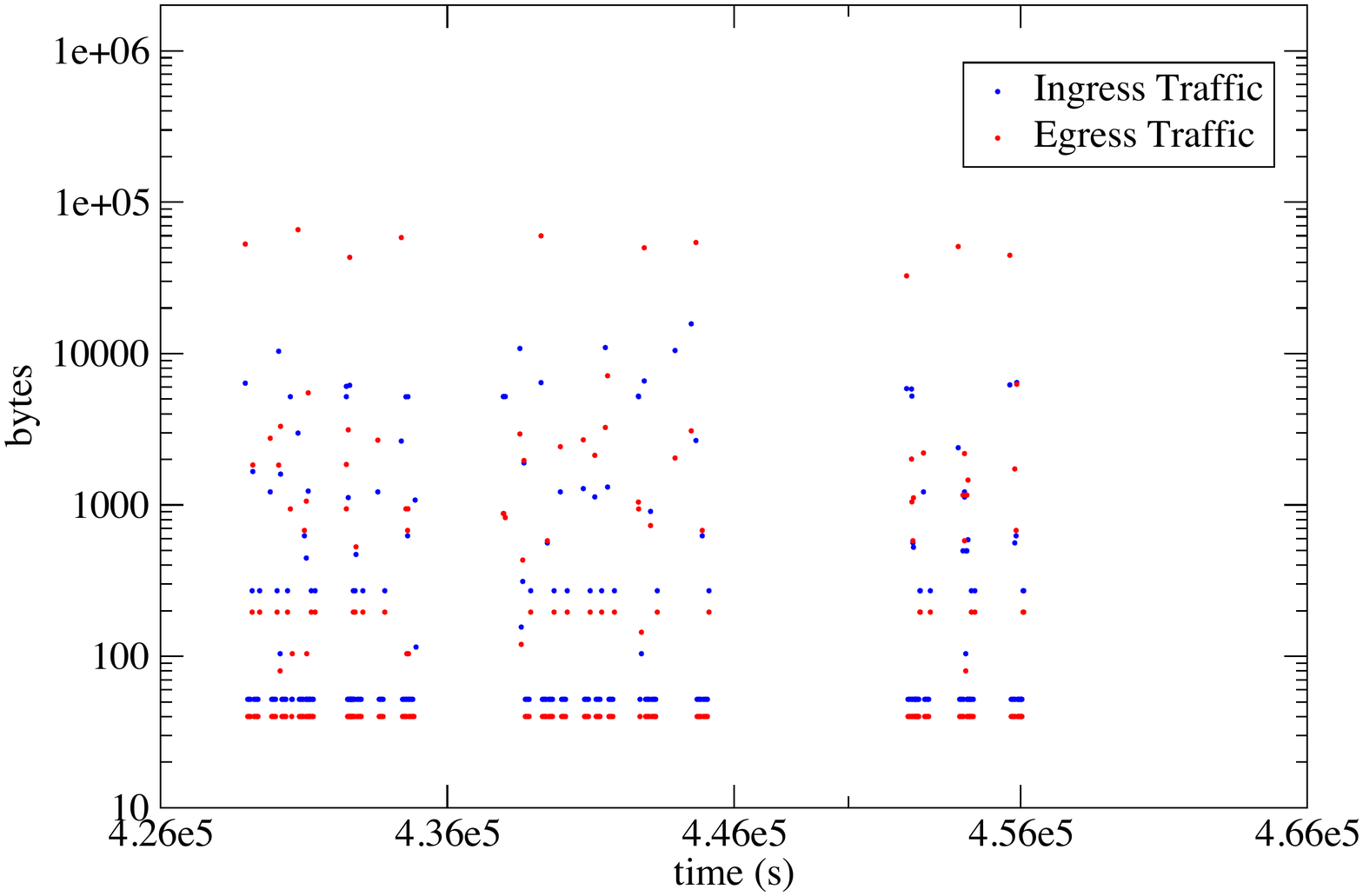}}\\
\end{tabular}
\caption{Sample communication. (a) Traffic spanning a 2-week period. Note the clear diurnal activity. (b) Detail of one 12-hour period.}
\label{1}
\end{figure*}
\section{Data Source and Feature Representations}
The detection capability described here is designed to monitor individual hosts.  We refer to these hosts throughout this work as {\it internal systems}, to distinguish them from the many {\it external}, or remote, systems with which they exchange data. Data will flow out from internal systems: we call this {\it egress}, and it will flow in from external systems: we call this {\it ingress}. All systems are identified by their internet protocol (IP) address.  We are interested in collecting and characterizing ingress and egress network traffic between a given internal system and all remote systems to which it connects over a prescribed time period.  The data object of interest is the pair of functions $(b_I(t),b_E(t))$, describing the ingress and egress bytes transferred over time, between the internal host and a single external host on a single service port over the time period of interest.  If the internal system is a typical workstation on an enterprise network, connections that originate from outside the network typically cannot reach it (because of network address translation or firewall rules), and so the external systems in these connections are assumed to be destination systems (\texttt{dst\_ip}) and the internal system is then the source (\texttt{src\_ip}).  The external port is then the service, or destination, port (\texttt{dst\_port}).  In contrast, for internal servers in the DMZ or email gateways that are outward-facing, then we expect remote systems to sometimes be sources and the internal system to be the destination.  Because traffic data has finite time resolution, the functions $(b_I(t),b_E(t))$ are discrete.  We refer to the data exchanged at a single instant in time  as a {\it connection}.   We call the functions $(b_I(t),b_E(t))$ for a triplet \texttt{(src\_ip,dst\_ip,dst\_port)} a {\it communication}, and so a communication is generally comprised of many individual connections, each typically including both ingress and egress data.

The traffic data required to derive the pair $(b_I(t),b_E(t))$ for the triplet \texttt{(src\_ip,dst\_ip,dst\_port)} can be collected from a variety of flow-oriented sources, including Cisco NetFlow, sFlow, IP Flow Information Export (IPFIX), or Simple Network Management Protocol (SNMP).  For this study we use unsampled\footnote{Since sampled NetFlow results in an incomplete picture of network activity that has not been tested in this analysis, we recommend caution when adapting this methodology to sampled netflow.} Cisco NetFlow, which provides summary data on all connections handled by NetFlow-enabled network devices.  A single NetFlow record is unidirectional and describes one-half of either a complete session, or a segment of a long session, between two endpoints. Each record contains a range of information including the source and destination IP addresses, source and destination ports, start time and duration of the connection, bytes exchanged, and so on\footnote{See \cite{nfv9} for details on reported fields.}.  An alternative, host-based option would be to derive summary data from packet captures\footnote{For example, using the tool \texttt{nfdump} \cite{nfdump}.}.   All flow data used in this study was obtained from APL's real computer network.

A sample communication is shown in Figure \ref{1}.  Two important things to note: 1) the traffic data can be sparse. NetFlow records are discrete events, and so there is no data when there are no connections, and 2) while the NetFlow records are unidirectional, the egress and ingress components  comprising the bidirectional flow have the same timestamp, and so the communication $(b_I(t),b_E(t))$ is generally double-valued.

The objective of this study is to use one-class learning to identify data exfiltration.  A critical step in any classification problem, before even selecting learning algorithms, is deciding on a feature representation of the data.  Communication data in the time domain, as shown in Figure \ref{1}, is easy to visualize and understand but poses some immediate challenges to automated classification.  First and primarily, longer communications must have more features than shorter communications at a given time resolution.  This has implications for the maximum duration communications that we can test: for example, a two-week communication with 1-second resolution contains $3600\times24\times14 \approx$ 1 million features.  Second, certain patterns, like periodicity, exhibited by communications over time are not efficiently represented in the time domain: for example, a two-week communication with an hourly spike in egress traffic would have over 300 involved features, and a classifier would need to learn this correlation.  For these reasons, we choose to consider the problem of exfiltration detection in the frequency domain.

Since the functions $(b_I(t),b_E(t))$ are non-uniform in time we employ the {\it non-uniform discrete Fourier transform}, which takes time {\it and/or} frequency vectors with arbitrary binnings,  
\begin{equation}
\label{ndft}
Y_{E,I}(f_m) = \sum_{n=0}^{N-1}b_{E,I}(t_n)\exp\left(-2\pi i f_m t_n\right),
\end{equation}
where the $n \in [0,N-1]$ are time and $m \in [0,\text{dim} \,{\bf \text{f}}]$ are frequency indices.  An immediate advantage of working in the frequency domain is that communications with arbitrary durations all map to the same, fixed dimension spectra in frequency space.  Of course, the dimension of the frequency feature vector depends on this duration in order that the frequencies of interest (both low and high bands) are covered; however, we have considerable freedom in how we sample this frequency vector.  The prescription adopted here is motivated by the desire to have seconds-resolution for components with periods on the order of seconds, but with considerably less resolution for lower-frequency components, with periods on the order of days.  A method\footnote{Proposed by colleague Max Kresch.} that achieves this goes as follows: generate the sequence $a_i = (2^i)$ with $i \in [0,\log_2 T]$, where $T$ is the total duration (in seconds) of the time window of interest ({\it e.g.} $[1,2,4,8,16,32,...]$).  Then, successively bisect consecutive elements of the sequence ({\it e.g.} $[1,2,3,4,6,8,12,16,24,32,...]$ after one bisection round).  Repeat until you have the desired number of frequency bins, where each frequency is given by $f_i = a^{-1}_i$.  The resultant binning is dense for high frequencies, falling off gently in the low-frequency regime.  This freedom in how we sample our feature vector is a perk of working in the frequency domain; it's not clear how a similar compression would be achieved in the time domain. 

The first set of features that we consider characterizes the relationship between the ingress and egress traffic rates of each flow.  These are efficiently summarized by the average amplitudes of the respective Fourier transforms, $(\overline{|Y_{I}(f)|},\overline{|Y_E(f)|})$, forming a two-dimensional representation for each flow.  While the average ingress and egress bytes taken over the duration of the flow, $(\overline{b_I(t)},\overline{b_E(t)})$, might seem a simpler representation, we will show later in Section 9 that the spectral averages lead to better performance.  Since the average amplitudes can range over many orders of magnitude in either dimension we take the logarithm, $(\log\,\overline{|Y_{I}(f)|},\log\,\overline{|Y_E(f)|})$.
\begin{figure*}[htp]
\subfloat[]{\includegraphics[width = 3in]{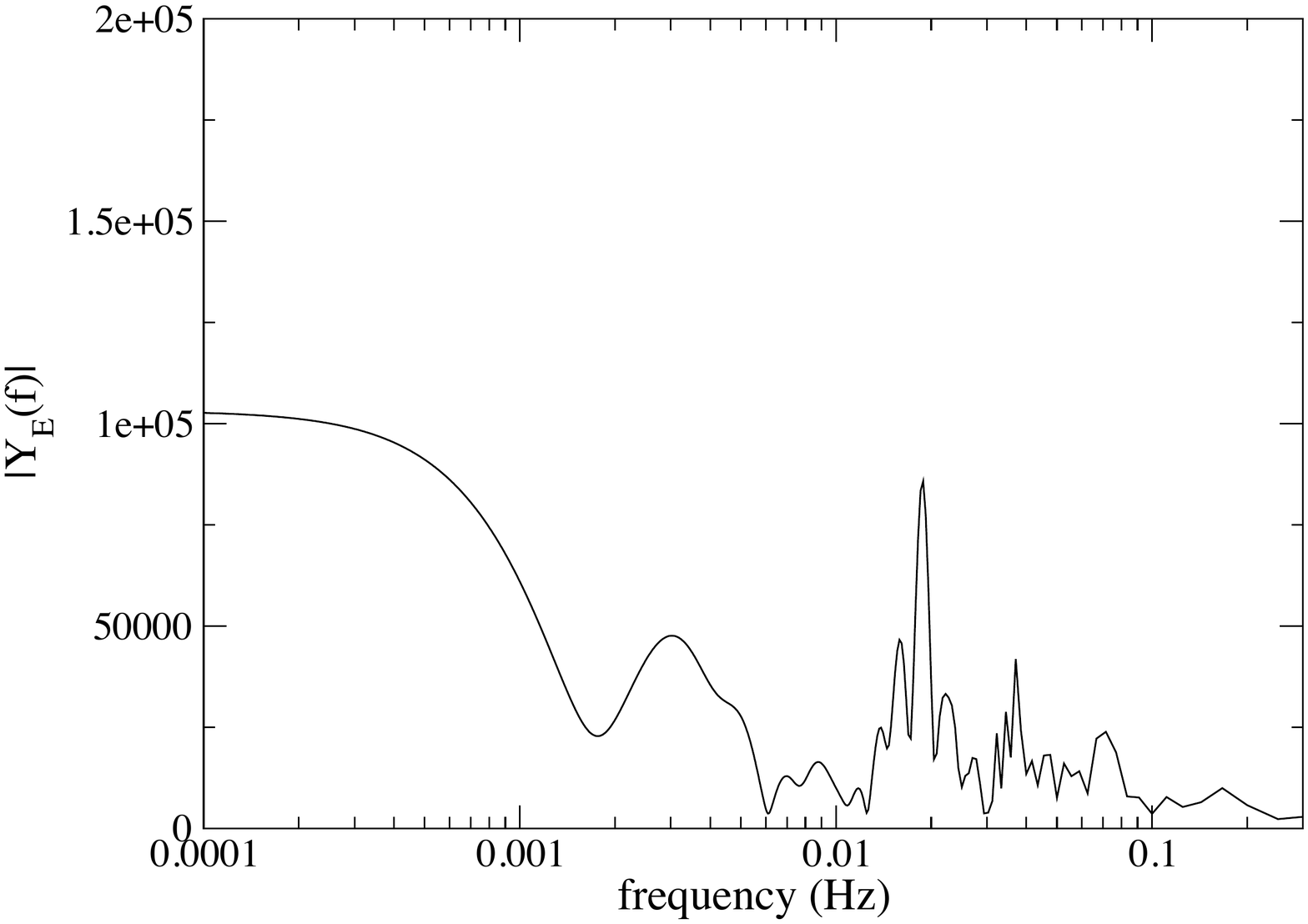}}
\subfloat[]{\includegraphics[width = 3in]{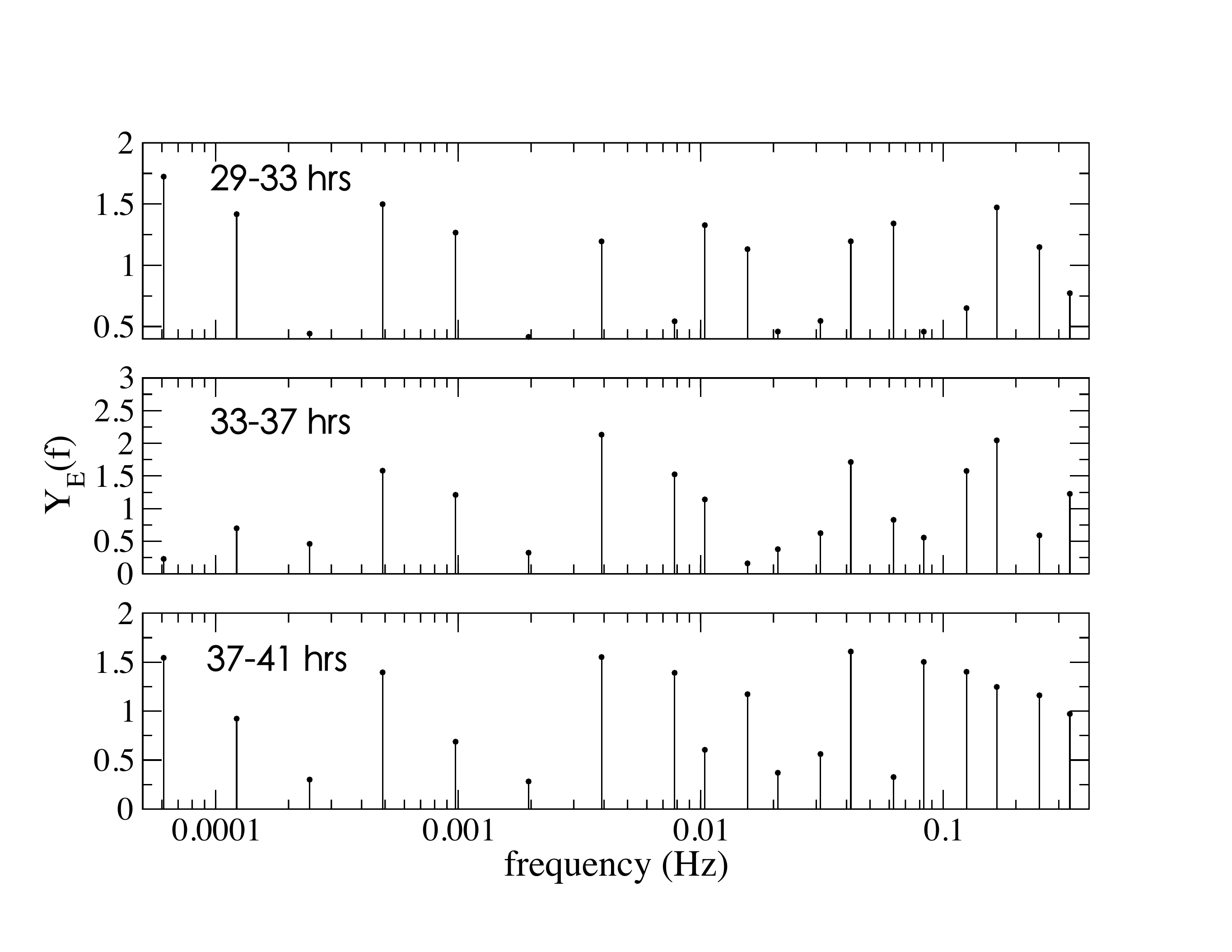}} 
\caption{Transform domain feature representations of the sample communication in Figure \ref{1}. (a) Egress Fourier amplitude, $|Y_E(f)|$. (b) Short-time Fourier transform, $F(\tau,f_m)$ \newline (3 time windows shown).}
\label{2}
\end{figure*}

To characterize the temporal structure, including periodicity and other correlations, we next consider the normalized amplitudes of the individual egress and ingress Fourier transforms, 
\begin{equation}
|Y_{E,I}(f)|= \sqrt{\text{Re}(Y_{E,I}(f))^2 + \text{Im}(Y_{E,I}(f))^2}\Big/\overline{|Y_{E,I}(f)|},
\label{NORMamp}
\end{equation}
The normalization is desirable because we are particularly interested in the spectral content of the communication, not the absolute number of bytes transferred which is already summarized in the representation\footnote{These are {\it unnormalized} quantities sharing the same notation as Eq. (\ref{NORMamp}); this distinction will be clearly stated whenever there is a chance of confusion.} when  $(\log\,\overline{|Y_{I}(f)|},\log\,\overline{|Y_E(f)|})$.  The Fourier amplitudes $|Y_{E,I}(f)|$ reveal the full harmonic structure of the communication over the chosen frequency range, and so are good for more than just testing periodicity: any harmonic idiosyncrasy exhibited by network traffic will be conveyed to the classifier via this feature set.  This feature representation has dimensions equal to the number of frequency bins selected.   

The Fourier amplitude encodes the harmonic structure of the communication across its full duration, but it struggles to find periodic components or substructures that are time variable: for example, a communication with hourly connections over a certain eight-hour period for, say, three consecutive days and is otherwise idle. This is certainly a pattern worth recognizing, but the Fourier amplitude will have a difficult time resolving it. We therefor consider a third feature representation, one that finds its place in the time-frequency domain: the {\it short-time Fourier transform} (STFT).  The idea is to break a long signal into smaller chunks and compute the Fourier transform of each chunk individually.  The collection of Fourier transforms then gives a measurement of how the component frequencies are changing with time: if there is a localized burst of traffic, or transient periodicity, it will show up in the STFT.  Formally, the STFT is defined 
\begin{equation}
\label{stft}
F(\tau,f_m) =
 \sum_{n=0}^{N-1}b_{E,I}(t_n) w_a(t_n-\tau)\exp\left(-2\pi i f_m t_n\right), 
\end{equation}
where $w_a(t_n-\tau)$ is a window function that is nonzero only over a portion of the function $b_{E,I}$.  In what follows, we consider a rectangular window 
\begin{equation}
\label{window}
w_a(t_n-\tau) = \begin{cases}1,& \text{if } \,\tau-\frac{a}{2} < t_n < \tau + \frac{a}{2}\\
0,& \text{otherwise},
\end{cases}
\end{equation}
that is zero when $t_n$ is outside the window of width $a$ centered at time $\tau$.   The width, $a$, of the window sets the time resolution of the component frequencies.  We desire good time resolution to complement the good frequency resolution already captured in $|Y_{E,I}(f)|$, and so we want to take $a$ small compared to the full duration of the communication (this is a so-called {\it wideband} STFT). While the windows should generally be taken to overlap to minimize edge-effects, we find that this isn't necessary given the sparseness of the communications and the desired resolution of each component Fourier transform.  We therefore start at $t_0$ with $\tau = a/2$, and move the window along in increments of $a$ units so that the $F(\tau,f_m)$ of a communication with $N$ time steps will have $\lceil{N/a}\rceil$ component Fourier transforms.  The STFT is typically given in terms of the {\it spectrogram},
\begin{equation}
S(\tau,f) = |F(\tau,f)|^2,
\end{equation}
which is a discrete function in two-dimensions (time and frequency); it can be thought of as a matrix with rows given by the $f_m$ and columns by time slices $t_n$. The spectrogram feature vector is formed by ``flattening'' this matrix, concatenating the rows to form a single $\text{dim}\,\text{{\bf f}} \times \lceil{N/a}\rceil$-dimensional array.  It is important to point out that the STFT is tunable to the data set---both the length of the time window and the frequency resolution must be chosen; we will discuss these optimizations when we perform our experiments in Section 7.1.
%however, the STFT must be run on all samples for each setting and so only a fairly coarse grid search is practical during training.  

Examples of the DFT and STFT feature representations are shown in Fig \ref{2} for the communication of Fig \ref{1}.  

\begin{figure*}[htp]
\begin{tabular}{cc}
\subfloat[]{\includegraphics[width = 3in]{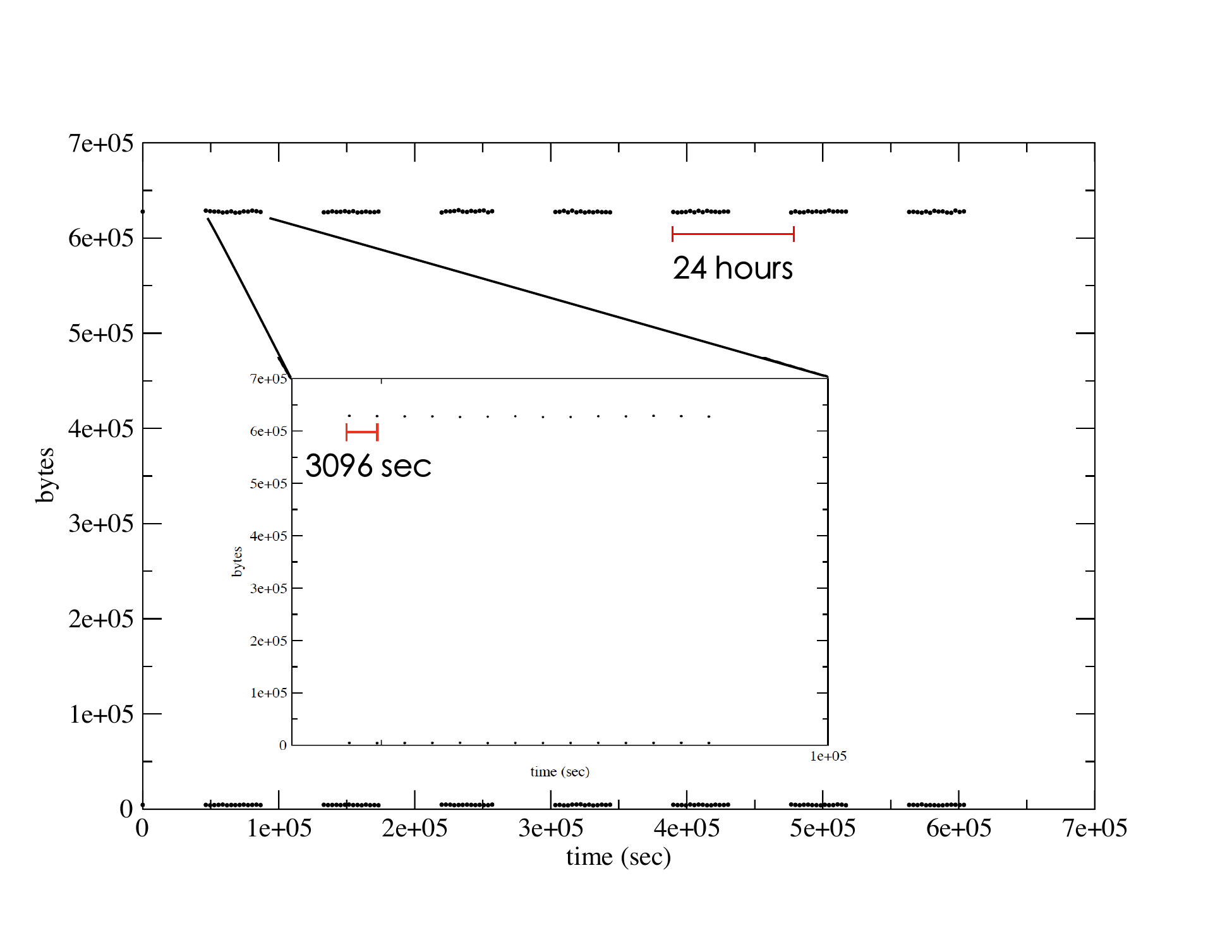}} &
\subfloat[]{\includegraphics[width = 3in]{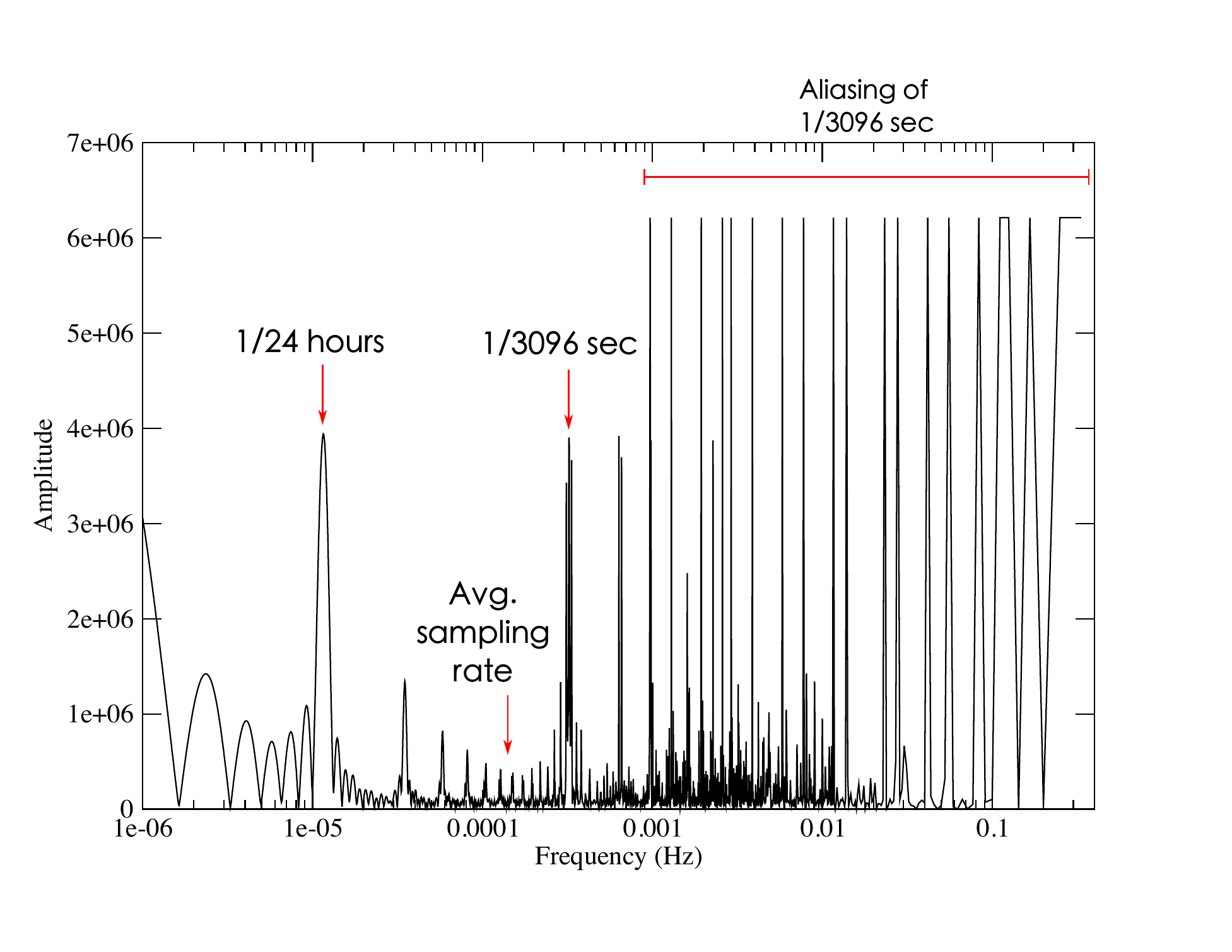}}\\
\end{tabular}
\caption{Aliasing in a communication. (a) Simulated periodic communication, focusing on egress traffic on top. (b) Fourier amplitude of egress traffic in (a), with aliasing clearly evident.}
\label{3}
\end{figure*}
Before moving on, an important caveat is in order.  In signal processing, whenever the Fourier analysis includes component frequencies greater than the sampling rate of the discrete signal, a phenomenon called {\it aliasing} occurs. It is so-named for the presence of additional, spurious frequencies in the spectrum at integer multiples of the sampling rate that all refer to the same harmonic\footnote{See \cite{sp} for a gentle and clear discussion of this and related principles of signal processing.}.  Aliasing can be generally avoided by restricting the analysis frequency range to be less than half the sampling frequency, $f_i < f_s/2$, a condition called the {\it Nyquist criterion.} With non-uniform time sampling of type used here, though there is no constant sampling rate, the Nyquist-Shannon sampling theorem states that instead the {\it average} sampling rate can be used to limit the bandwidth \cite{Shannon}.  The trouble with this methodology is that we must apply the same, fixed frequency sampling to all communications, regardless of their average sampling rates (which surely vary tremendously across communication samples).  To illustrate what happens as a result, we show in Figure \ref{3} a simulated communication with a period of 3096 seconds. The average sampling rate is $\bar{f}_s = 0.0016$ Hz, and so the prominent harmonic at $f_* = 1/3096$ sec $= 0.0032$ Hz $> \bar{f}_s$ is aliased multiple times leading to the series of spikes in the spectrum at integer multiples of $f_*$.  Though troubling to an engineer seeking a unique harmonic, this does not seem to present a problem for us, who are merely interested in classification of a high-dimensional function.  As far as the classifier is concerned, aliasing just means more peaks in the feature vector, which might even aid in its ability to discriminate classes.  

\section{Classifier Ensemble}
We consider four different one-class algorithms in this study: kernel density estimation, one-class support vector machine, isolation forest, and a $k$-nearest neighbors-based method.  We first give a brief description of each of these techniques.
\subsection{Kernel Density Estimation}
Kernel methods (KDE) provide a way of non-parametrically approximating the probability distribution of multi-dimensional data.  The estimated density is essentially a smoothed binning of the data, where the width of each bin is controlled by the kernel function, which determines how strongly the full data set influences the density at the point ${\bf x}_i \in \mathbb{R}^\ell$.  A range of kernel functions are in use: we employ a Gaussian, $K({\bf x}_i,{\bf x}_j) \propto \exp(-||{\bf x}_i - {\bf x}_j||^2/2h^2)$, where the free parameter $h$ controls the {\it bandwidth} of the kernel.  When $h$ is large, many points contribute to the density at ${\bf x}_i$ and the resulting distribution is very smooth (potentially suffering from high bias), whereas when $h$ is small, only points in the immediate neighborhood of ${\bf x}_i$ contribute to the density there, resulting in a spiky estimation (potentially suffering from high variance). In practice $h$ is usually tuned by cross-validation. 
\subsection{One-class support vector machine}
We implement the one-class support vector machine (SVM) of Sch\"olkopf {\it et al.} \cite{Scholkopf}, which is a modification of the multiclass support vector machine useful in the one-class setting.  The technique maps all normal points ${\bf x}_i \in \mathbb{R}^\ell$ into a new feature space, $\Phi({\bf x}_i): \mathbb{R}^\ell \rightarrow F$, separating them from the origin with a maximum margin: new points that fall on the origin-side of the separating hyperplane are classified as novel.  The maximum margin is found subject to some tolerance for the misclassification of normal points, quantified in terms of a parameter, $\nu \in (0,1)$, which places an upper bound on the percentage of misclassified points; because of this parameter's key role, these machines are referred to as $\nu$-SVMs.  As with multi-class SVMs, data points that lie on the boundary are called {\it support vectors}. The mapping $\Phi$ can be a complicated, high-dimensional mapping that it furnishes a nonlinear boundary in the feature space, so that we have considerable freedom in separating the normal points.  The inner product is called the {\it kernel} of the SVM; we choose a Gaussian kernel,
\begin{equation}
K({\bf x}_i,{\bf x}_j) = \Phi^T({\bf x}_i)\cdot\Phi({\bf x}_j) = \exp(-\gamma||{\bf x}_i - {\bf x}_j||^2),
\end{equation}
where the parameter $\gamma$ controls the influence of each support vector on the placement of the boundary. 
\subsection{Isolation Forest}
The isolation forest  \cite{Liu} is based on the premise that anomalous points are both few and different.  These properties can be captured via tree architecture in the sense that fewer binary questions must be asked to isolate anomalies than the bulk of normal points. Starting with a subset of data, a tree is recursively generated by randomly selecting a feature and a split value; the tree completes when each datum either occupies its own node or a node with other data with the same value of the split feature.  A normalized outlier score is derived from the average path length, from root to terminal node, of the instance such that those with comparatively small path lengths are deemed more anomalous.  The anomaly threshold can be tuned. 

A tree can use only a single feature or a subset of features, and can test either a subset of data or the entire data set.  By forming trees from subsets of data, the isolation forest can leverage subsampling to deal better with outlier masking and swamping.  The number of trees to grow is also tunable.  
%We find that the performance of the isolation forest is rather insensitive to these parameters, and settle on 100 trees, each using a single feature and built from 256 instances.  We tune the performance by adjusting the anomaly threshold.    
\subsection{Nearest Neighbors}
It is possible to apply the $k$-nearest neighbors ($k$-NN) method of supervised learning to one-class problems.  A point is considered anomalous if its average distance to $k$ nearest neighbors surpasses some threshold; both $k$ and the threshold value can be tuned to adjust the accuracy.  The distance function can also be customized to the problem; for high-dimensional spaces, the {\it cosine similarity}, ${\bf x}_i^T{\bf x}_j/||{\bf x}_i||\,||{\bf x}_j||$, can be used to compare two points ${\bf x}_i$,${\bf x}_j$ in lieu of the Euclidean distance which loses discriminatory power in many dimensions.  For this application, though, we find that the Euclidean distance gives superior performance despite the considerable dimensionality of the feature spaces.  
\subsection{Ensemble} 
It is possible to combine multiple classifiers to yield a single outlier score, the hope being that the whole is greater than the sum of its parts.  Such a combination, called an {\it ensemble}, works best if the component classifiers have decorrelated errors, viz. they are wrong in different ways.  Ensemble classifiers have been studied most extensively in the supervised setting \cite{Kittler,Rokach}, but there has been some work applying them to both one-class \cite{Tax,Giacinto,Shoemaker,Aggarwal,Chiang} and unsupervised problems \cite{Portnoy,Eskin,Leung,Mazel}.  Unfortunately, some of the more powerful methods of ensemble learning in high-dimensional spaces, like feature bagging \cite{Lazarevic}, are not applicable to one-class learning.  But, at bottom, all methods seek to decorrelate the component classifiers in some way.  We can hope to decorrelate by combining different breeds of component classifiers: here we've selected one distribution-based (KDE), two boundary-based ($\nu$-SVM and $k$-NN) and one model-based (isolation forest) classifier.  Further help might be had by training different classifiers on different representations of the data, perhaps chosen so that each is learning a different aspect of the problem.  We employ both techniques in this work. 

Before a single, unified score can be generated by the ensemble, we must consider how the various component outlier scores should be combined.  The raw scores of each method considered here are quite different both in distribution and scale, making a straight combination (by, say, adding or averaging them) meaningless.   While there are ways of normalizing and regularizing scores for combination \cite{kriegel11,schubert12} (for example, by converting them to probabilities \cite{gao06,schubert12}), we opt instead to apply logic to the final binary (outlier/no outlier) scores of each classifier's decision function.  For example, applying logical OR to a two-classifier ensemble means that the ensemble fires on an outlier if either (or both) of the component classifiers do.   

The training data used to develop the detector is assumed normal (no exfiltration), as is typical in the one-class setting\footnote{Exfiltration is a rare event, and so if the presumed normal data happens to contain exfiltration flows, these will be very few in comparison with normal traffic and so will be very unlikely to skew the classifier, as all of our classifiers permit false positives during training.}. But, the lack of true exfiltration samples poses a challenge to performance testing: while we can build a detector with the desired low false positive rate (FPR), there is generically a trade-off with a decreased true positive rate (TPR).  To ensure that our classifiers are effectively catching exfiltration, we simulate a broad collection of different exfiltration communications and include them in the testing.  
\section{Simulating Exfiltration}
We consider four families of data exfiltration, differentiated by timing and data egress characteristics. We choose generous parameter ranges for each exfiltration scenario in an effort to cover the space of the possible, keeping in mind that some exfiltration samples won't be realistic from an attacker's perspective.   We stochastically generated 2000 samples from each family by randomly selecting values of these characteristics.  The ranges and sampling schemes used in the simulations were determined by analyzing the ingress and egress statistics of a set of real exfiltrations composed with common tools on the very systems explored in this paper. 
Table \ref{exfil} provides a summary of the different types.
\begin{table}
\begin{center}
\begin{tabular}{|c|l|l|}
\hline
Type&Timing&Egress Data \\ \hline
1&Periodic&Constant \\
2&Nonperiodic&Constant \\
3&Periodic&Variable \\
4&Nonperiodic&Variable \\
\hline
\end{tabular}
\end{center}
\caption{Exfiltration scenarios.}
\label{exfil}
\end{table}

The simplest scenario, called {\it type 1}, is characterized by perfectly periodic egress traffic drawn in equal-sized (constant) chunks.  This scenario would be implemented by an agent set to call back at fixed time intervals, and to transfer fixed amounts of data each time.  The simulated traffic amounts are actually not strictly constant from one moment to the next, but vary slightly to account for the idiosyncrasies of different network stacks, data buffering characteristics of different exfiltration tools and operating systems, and the time-variable latency of networks.   These effects are generically modeled by drawing each byte amount from a Gaussian distribution: for type 1 exfiltration, it is the mean and variance of the distribution that are held constant over time. Each type 1 sample has a mean egress and ingress amount separately drawn logarithmically from the range $[10^3,4\times 10^6]$ bytes\footnote{The maximum is bounded by the bandwidth of the connection, here chosen to be 4 MB/sec.}  and a standard deviation between 250 and 550 bytes depending on the size of the mean.  Each type 1 sample has a period, $T$, drawn logarithmically from the range $[1,8.64\times 10^{4}]$ seconds and a total duration of exfiltration activity sampled uniformly from between one hour and one month.  

{\it Type 2} scenarios have constant egress traffic as in type 1, but with nonperiodic connection times.  A ``nominal'' period, $T$, is drawn logarithmically from the range $[1,8.64\times 10^4]$ seconds and a {\it jitter factor}, $j$, is sampled uniformly from the range $(0,1)$.  Each connection occurs after the last with a time sampled from the uniform range $[T-jT,T]$.  This type of jitter is used, for example, by the Cobalt Strike Beacon.  All other aspects of type 2 communications are the same as type 1.

{\it Type 3} scenarios are periodic (same temporal parameters as type 1) but with varying amounts of data exfitrated in each connection.  For each communication, a maximum egress byte value, $b_{E,\text{max}}$, is logarithmically drawn from $[10^3, 4 \times 10^6]$ bytes, and a minimum egress byte value, $b_{E,\text{min}}$, drawn similarly from the range $[10^3, b_{E,\text{max}}]$ bytes. The egress data of each connection is then a random value drawn from $[b_{E,\text{min}},b_{E,\text{max}}]$.    

The type 3 scenario is rather unrealistic because it might be more difficult to implement than it is worth.  An easier way to vary egress data per connection is simply to buffer the data on the target and exfiltrate at random times, emptying the buffer when it is in varying states of fullness.  This scenario is generalized in {\it type 4} exfiltration, which is nonperiodic (same temporal parameters as type 2) with different amounts of data exfiltrated each connection (same traffic parameters as in type 3). 

\section{Testing the ensemble}
We configure the ensemble for a single host, since what is normal traffic for one host might not be normal for another of a different type.  We test three different kinds of systems: a workstation, email gateway, and web server. The workstation is a typical client, the email gateway exhibits both client and server behavior. It is certainly possible that even the normal traffic of a single host will be quite variegated, perhaps even ``multimodal'' if it performs several functions with distinct traffic characteristics.  At a broader level, transmission control protocol (TCP) and user datagram protocol (UDP) data might fall into distinct regions of traffic parameter space, making the learning of what normal looks like for a given host challenging.  We did not face these issues, since TCP traffic made up the vast majority of our traffic samples, and among the TCP traffic different protocols did not differ markedly in their characteristics.  However, it might be necessary in extreme cases to train multiple ensembles for a single host, each dealing with a subset of traffic.  Clustering and other unsupervised learning techniques will likely be helpful to identify natural traffic groups, but we do not carry out such a study here.  
%\begin{figure}[htp]
%\includegraphics[width=0.65\textwidth,clip]{powel10.pdf}
%\caption{ROC curves of each classifier trained on the transfer function representation of client workstation communications, tested against the workstation's normal data and simulated type 1 exfiltration. The nearest neighbors classifier has $k=40$ in this plot.}
%\label{5}
%\end{figure}

\subsection{Experimental set-up}
Training data included all external traffic involving the internal system over a period of two weeks. This time window is chosen to test the method's ability to identify long tracts of data exfiltration occurring over days, but is also sensitive to exfiltration over shorter time-scales.   NetFlow was collected into communications  $(b_I(t),b_E(t))$ for each triplet \texttt{(src\_ip,dst\_ip,dst\_port)}.  

A number of hyperparameters control the models and the feature generation, including: the dimensionality of the frequency binning of $|Y_{E,I}(f)|$; the size, $a$, of the STFT window; the dimension of the frequency binning of the STFT; bandwidth, $h$, of the KDE; $\nu$ and $\gamma$ of the $\nu$-SVM; contamination ratio and number of trees of the isolation forest; number of neighbors, $k$, and distance metric of the $k$-NN classifier.  An ``outer'' grid search over the feature space parameters was conducted, and for each combination, an ``inner'' grid search over model parameters was then performed.  For each full combination of parameters, each classifier was tested using 10-fold-cross validation on the normal data and against the full sample set of each exfiltration type. For $Y_{I,E}(f)$, the frequency binning was varied in the range dim {\bf f} $\in [250,500,1000,2000]$, and for $S(\tau,f)$, $\text{dim} \,{\bf \text{f}} \in [20,40,80,160]$ and $a \in [2,4,8,16]$ hours.  Parameter ranges for the different classifiers were taken as follows: 1) KDE: $\log h \in [-2,1]$ in steps of 0.01; 2) $\nu$-SVM: $\log \nu \in [-4,0]$ in steps of 1, $\log \gamma \in [-5,-1]$ in steps of 1; 3) isolation forest: logarithm of the contamination ratio $\in [-4,0]$ in steps of 1, number of trees to grow $\in [100,200,500]$; 4) $k$-NN: $k \in [1,2,5,10,25]$ and distance metric $\in$ [`cosin', `euclidean'].  The final ensemble includes the KDE trained on $(\log\,\overline{|Y_{I}(f)|},\log\,\overline{|Y_E(f)|})$, isolation forest on $|Y_{I,E}(f)|$, and the $\nu$-SVM on $S(\tau,f)$. Hereafter, for brevity, we refer to the $(\log\,\overline{|Y_{I}(f)|},\log\,\overline{|Y_E(f)|})$ component as avgDFT, $|Y_{I,E}(f)|$ as DFT$_{I,E}$, and $S(\tau,f)$ as STFT.
\subsection{Client Workstation}
\begin{figure*}[htp]
\begin{tabular}{cc}
\subfloat[]{\includegraphics[width = 3in]{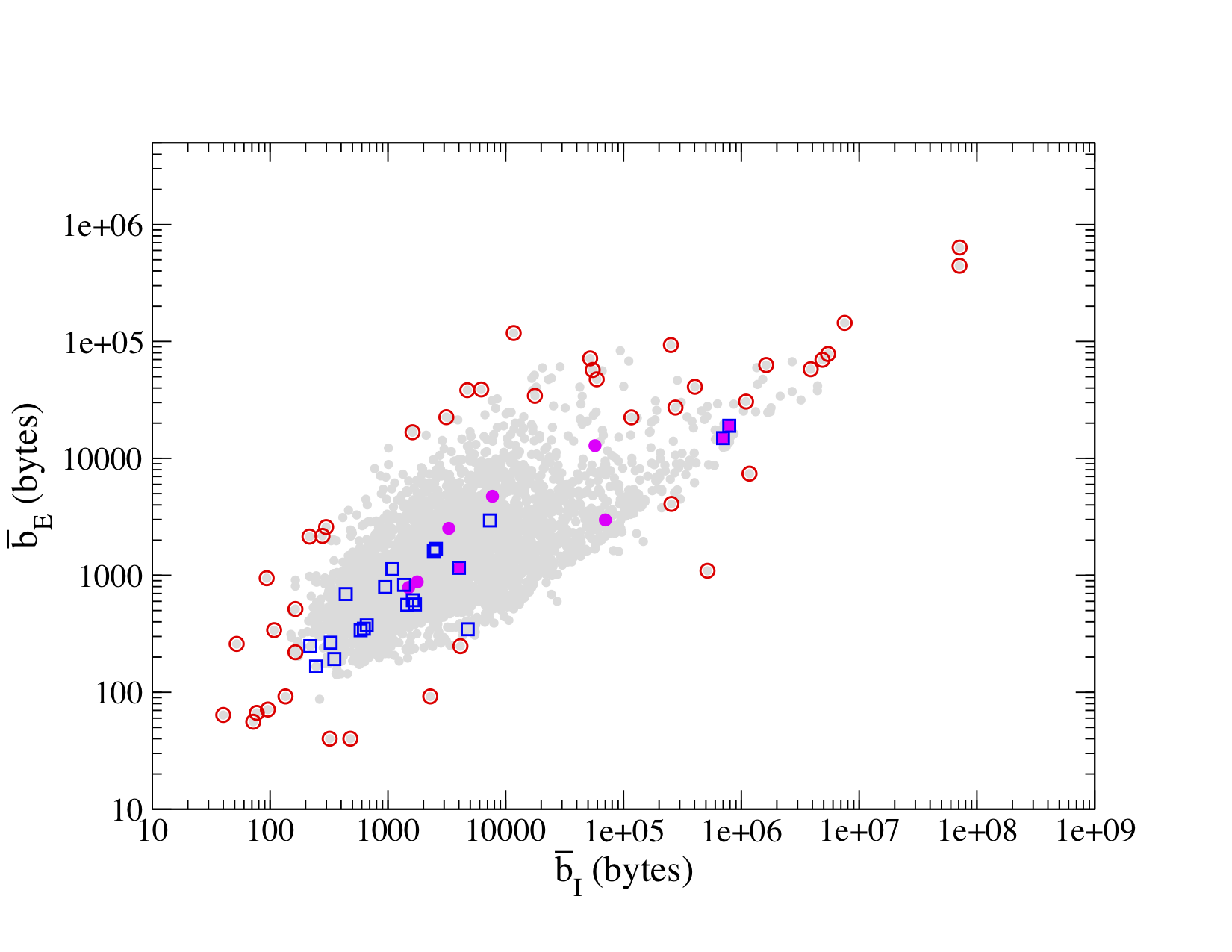}} &
\subfloat[]{\includegraphics[width = 3in]{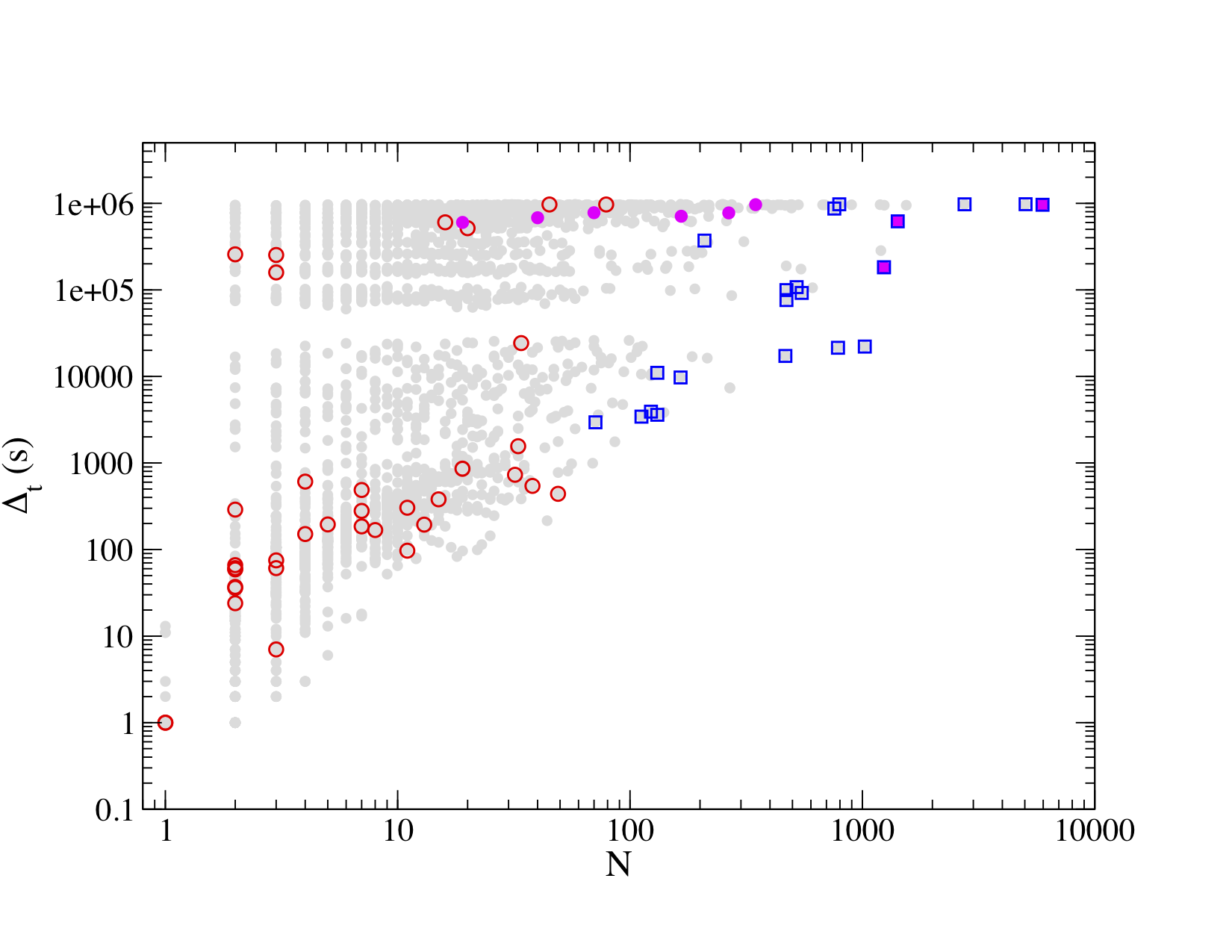}}\\
\subfloat[]{\includegraphics[width = 3in]{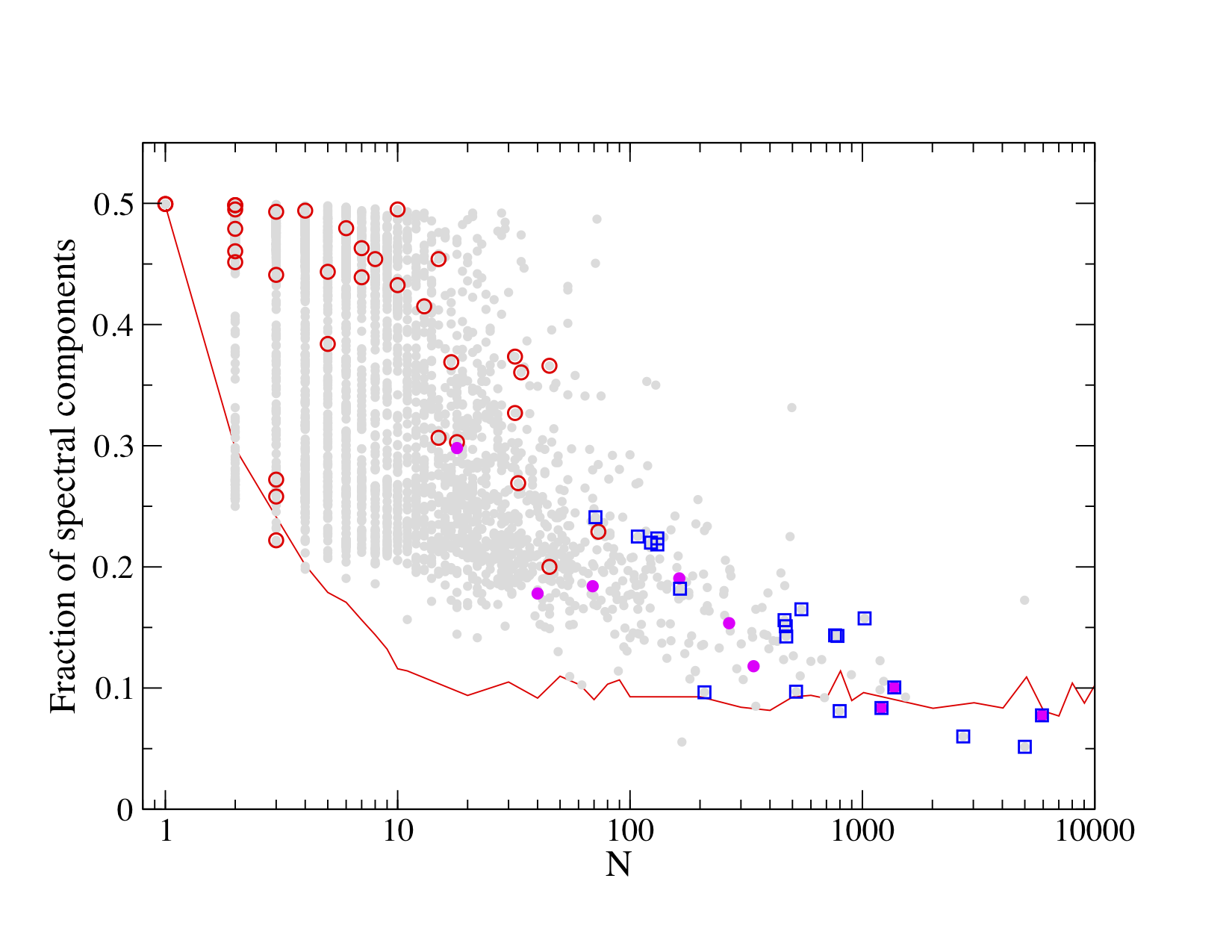}}&
\subfloat[]{\includegraphics[width = 3in]{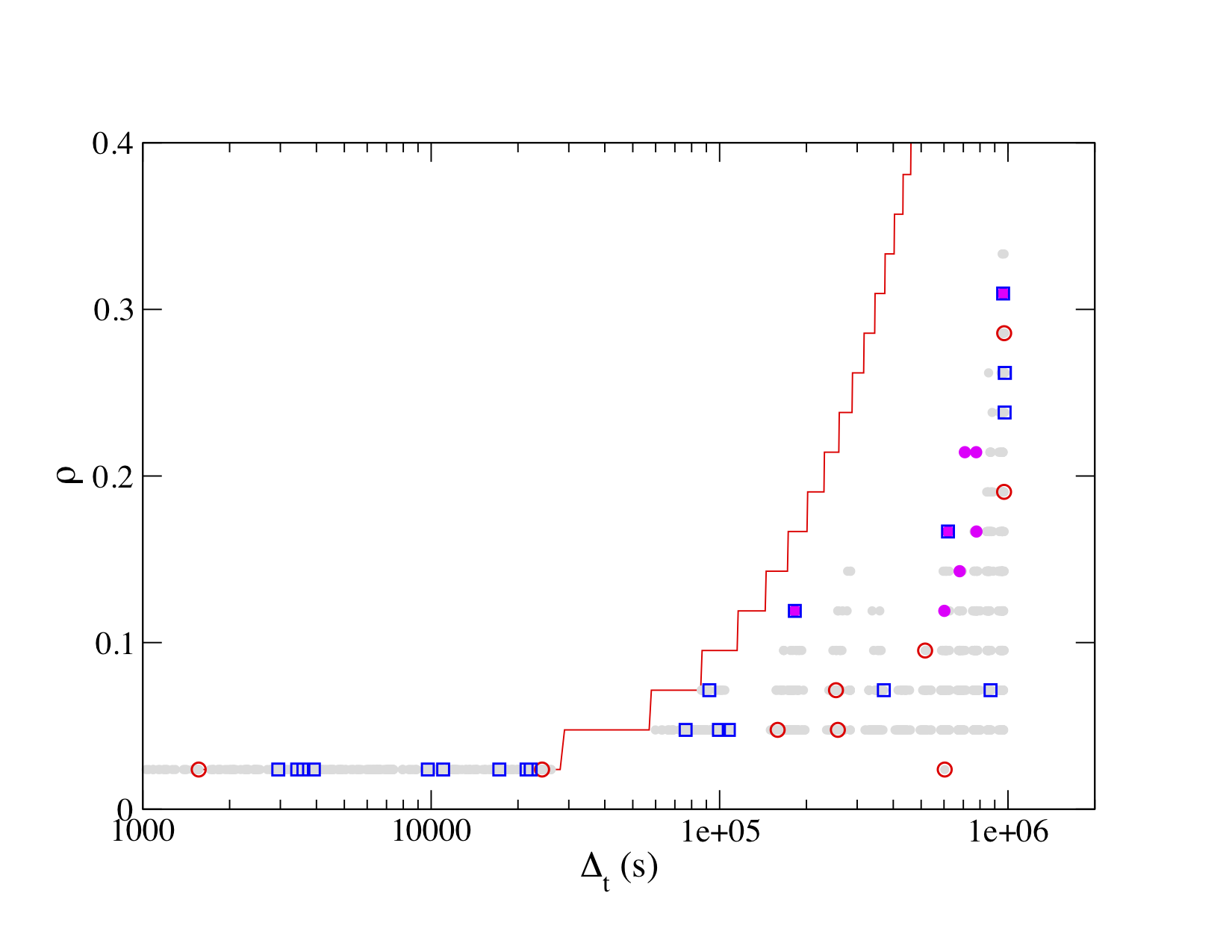}}
%\subfloat[ers.]{\includegraphics[width = 3in]{NORMmissedlvN.pdf}}\\
\end{tabular}
\caption{The 4143 normal workstation communications (gray points), with avgDFT false positives (1\%) shown as red circles, DFT$_E$ false positives (0.05\%) as blue squares, and  STFT false positives (0.02\%) as magenta points. (a) Average egress bytes vs average ingress bytes, (b) duration vs. number of connections, (c) fraction of Fourier components comprising 50\% of waveform energy vs. number of connections, with the average fraction computed from the periodic type 1 exfiltration samples shown for reference, (d) 8-hour coarse-grained density vs. duration with maximum density per duration shown by the red line.}
\label{5}
\end{figure*}
%\begin{figure}[htp]
%\includegraphics[width=1.0\textwidth,clip]{NORM_all.pdf}
%\caption{.}
%\label{5}
%\end{figure}
%\begin{figure*}[htp]
%\begin{tabular}{cc}
%\subfloat[Average egress bytes vs. average ingress bytes for the 4143 normal communications of the client workstation.]{\includegraphics[width = 3in]{powel8.pdf}} &
%\subfloat[Duration vs. number of connections per communication of the normal communications of the client workstation.]{\includegraphics[width = 3in]{powel9.pdf}}\\
%\end{tabular}
%\caption{Characteristics of client workstation traffic.}
%\label{4}
%\end{figure*}
We begin testing  with the client workstation, using it as the case study to introduce our testing approach and performance metrics.  The workstation is a typical user's desktop system on APL's enterprise network.  It is used primarily for data analysis, email, and web surfing.  Over the course of two weeks, there were 4143 unique communications, approximately 75\% of which were over secure web (port 443), with the remainder over nonsecure web (port 80).  The median duration was 4.7 minutes, and the median number of egress connections was 5, with 11\% of communications consisting of just a single egress connection.  The features $|Y_{I,E}(f)|$ have dimension dim {\bf f} = 500 and $S(\tau,f)$ has dim {\bf f}= 40 and $a=8$ hours.  In what follows, the ensemble has been tuned to a 2\% FPR.
%Since we are using high-dimensional and fairly abstract feature spaces, it's not possible to visualize the distribution of communications in these spaces.  We therefore propose a number of lower-dimensional parameterizations useful for visualizing the variability of normal communications on a given device.  Each parameterization corresponds to one of the characteristics being tested for by the detector using the more sophisticated, hig-dimensional features: ingress/egress bytes balance, periodicity, and local randomness.   

It is useful to characterize traffic according to a small set of properties: the average ingress and egress bytes per connection, $\bar{b}_{I,E}$; number of connections per communication, $N$; duration of the communication, $\Delta_t$, defined as the time from first to last byte exchanged; a coarse-grained communication density, $\rho$, which we'll define below; and a rough measure of signal periodicity, also defined below. These quantities are useful for understanding what the different classifiers are learning about the traffic, and are good proxies for the high-dimensional and rather abstract feature spaces being used\footnote{Consider the unintuitive and impractical characterization, ``the classifier often misclassifies spectra with a large harmonic at 0.003 Hz and a small one at 0.01 Hz''; this is unlikely to be useful to an operator or analyst.}.  In Figure \ref{5}, the 4143 communications of the client workstation are shown, with the false positives of the various classifiers indicated.  Figure \ref{5} (a) shows that the connections tend to be ingress-heavy, forming a dense, unimodal distribution in the $\bar{b}_I$-$\bar{b}_E$ plane (76\% of communications had $\bar{b}_I/\bar{b}_E > 1$ with an average of $\bar{b}_I = 5\bar{b}_E$).  The $\log \,\overline{|Y_{I}(f)|}$-$\log\,\overline{|Y_E(f)|}$ distribution is almost identical, and the KDE model trained on it finds false positives among the outliers (red circles in Figure \ref{5} (a).)  Meanwhile, the DFT$_E$ (blue squares) and STFT (magenta triangles) models, being insensitive to byte values, miss apparently arbitrary points in the $\bar{b}_I$-$\bar{b}_E$ plane.  Figure \ref{5} (b) shows the distribution of communications in the $\Delta_t$-$N$ plane, with most communications having fewer than one hundred connections, regardless of duration. DFT$_E$ appears to miss samples with large $N$ and short duration, $\Delta_t$, though it's not clear, yet, why.  The STFT component misses a few samples among the very longest-duration communications, but not sufficiently many to suggest a relationship.  

To understand the DFT$_E$ misclassifications better, we introduce a simple measure of periodicity defined to be the fraction of harmonics comprising 50\% of the spectrum energy, $E = \sum_j |Y_E(f_j)|^2$.  The idea is that periodic functions have most of their energy stored in relatively few dominant harmonics.  Figure \ref{5} (c) shows how this quantity is distributed vs. number of connections, $N$, for the workstation's traffic.  This quantity is also computed from the set of periodic type 1 exfiltration samples for reference (red line, averaged over logarithmic bins in $N$).  As naively expected, most of the normal traffic does not exhibit strong periodicity, though the communications tend to small fractions at large $N$.  It is in this region that DFT$_E$ erroneously catches some samples. 

To understand the STFT classifications, we argue that it should be good at discriminating communications by the number of $a$-hour lulls it has, $n_a$, where a lull is any contiguous period of time during which there are zero connections, and where $a$ is the size of the window function (cf. Eq. (\ref{window})). This is because over these time periods the Fourier transform is zero, and so the number and timing of lulls in the communication correspond to the number and placement of these vanishing Fourier transforms in the STFT.  The number of lulls can then be used to determine a coarse-grained communication {\it density}, $\rho = 1 - n_a/n_{a,{\rm max}}$, where ``max'' denotes the maximum number of lulls possible given the communication window, here taken to be two weeks.  The coarse-graining is controlled by the STFT window size, $a$. Figure \ref{5} (d) shows how the duration of communications varies with their density: relatively high-density communications are necessarily longer,
\begin{equation}
\Delta_t \geq \frac{a\rho}{n_a},
\label{density}
\end{equation} 
and so the STFT is not sensitive to long duration flows {\it per se}, but dense ones.  Eq. (\ref{density} reveals that there is a maximum density per duration, since the communication only has traffic within the duration time window, leaving the remainder of the two-week period devoid of connections.  The maximum density per duration is indicated by the red line, revealing that the workstations normal communications are of relatively low-density.  

To summarize: the avgDFT model learns the ingress/egress byte imbalance of normal traffic and so considers as anomalous those points on the outskirts of the distribution; the DFT$_E$ model seems to be sensitive to periodicity, which, for normal traffic emerges in communications with many connections (though not plotted, the DFT$_I$ component shows qualitatively similar behavior to DFT$_E$); and the STFT model appears to learn something about the coarse-grained density of the communications.   We will explore these relationships further in this section as we examine the role of each representation in detecting exfiltration of varying types.

%We now test the ability of the classifiers trained on the workstation's normal data to detect the different types of exfiltration.  We quote numerical results throughout this section, but all are summarized for easy reference in Table \ref{results}. 
\subsubsection{Type 1: periodic, constant egress data}
\begin{figure*}
\begin{tabular}{cc}
\subfloat[]{\includegraphics[width = 3in]{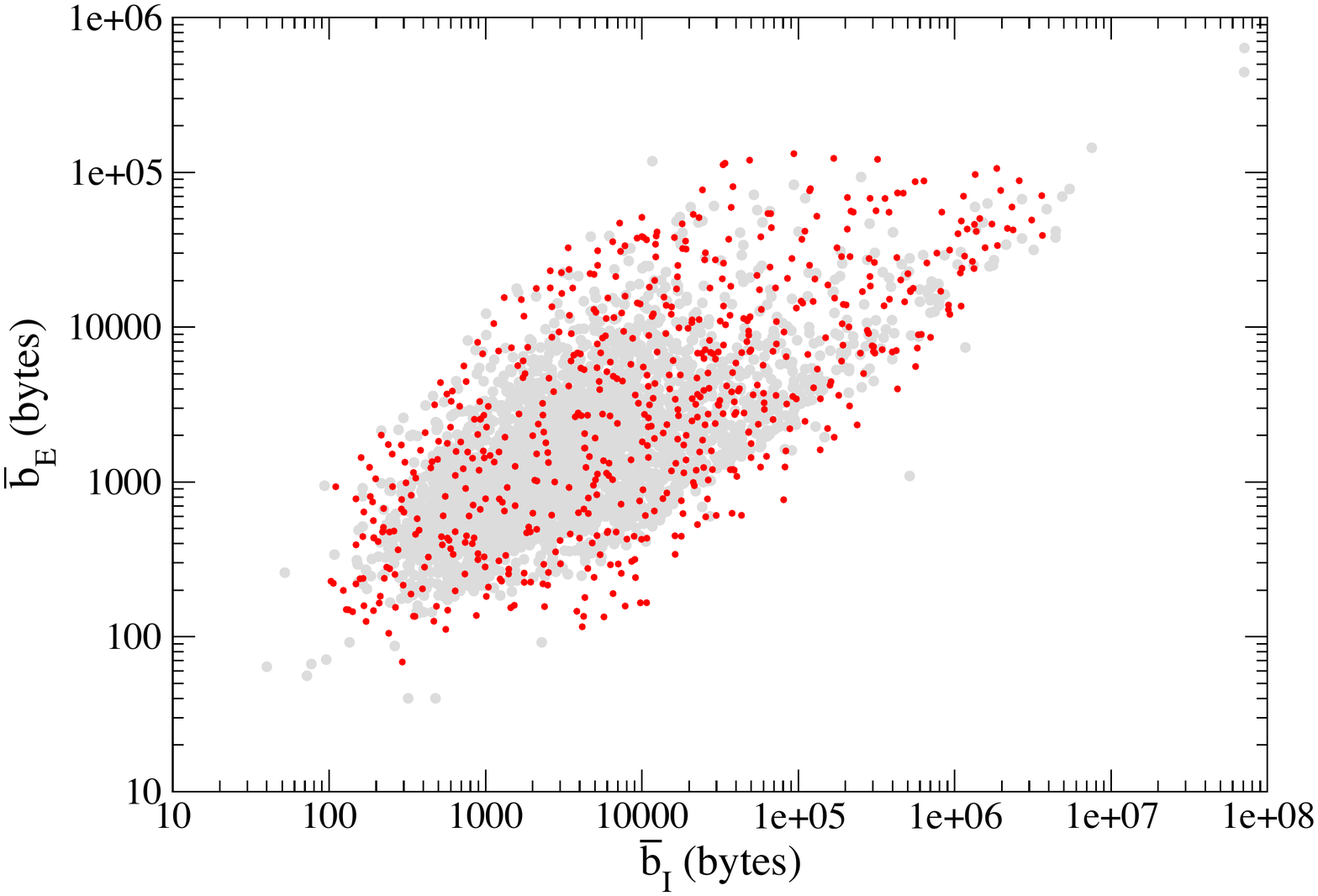}} &
\subfloat[]{\includegraphics[width = 3in]{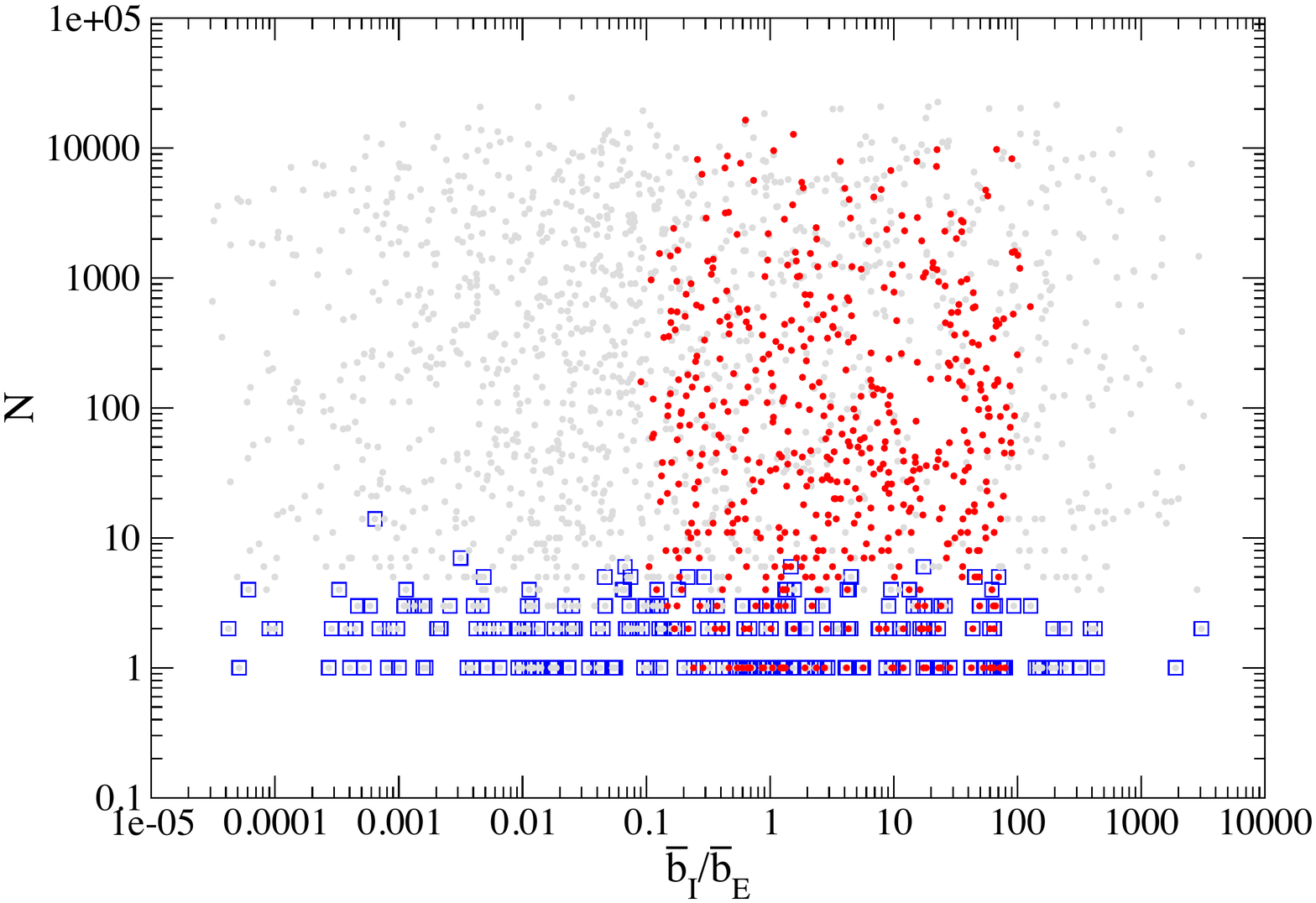}}\\
\end{tabular}
\caption{(a) Normal workstation communications (gray points) in the $\bar{b}_I$-$\bar{b}_E$ plane with type 1 exfiltration samples missed by avgDFT (red points). (b) Type 1 exfiltration samples (gray points) in the $\bar{b}_I/\bar{b}_E$-$N$ plane. Red points indicate samples missed by the avgDFT and blue squares those missed by DFT$_E$.}
\label{6}
\end{figure*}
We begin with the simplest exfiltration scenario with constant data egressed at periodic intervals.  We examine each ensemble component in turn to understand how it performs against each kind of exfiltration. The avgDFT misses those exfiltration communications with $(\bar{b}_I,\bar{b}_E)$ that overlap the distribution of normal traffic, Figure \ref{6} (a).  It catches everything else, including all egress-heavy exfiltration with $\bar{b}_E \gtrsim 10 \bar{b}_I$.  We think of this set of missed exfiltration samples as our starting point, as it marks our departure from conventional methods of exfiltration detection based on volume or ingress/egress byte balance \cite{Marchetti,Ramachandran}.  Of particular interest is how information on the temporal structure of the traffic, derived from our frequency domain representations, can help shrink this set of false negatives.  

The DFT$_E$ is applied first in the hopes that it detects the strong periodicity of type 1 communications.  In Figure \ref{6} (b), the missed exfiltration samples from Figure \ref{6} (a) are shown in the $\bar{b}_I/\bar{b}_E$-$N$ plane, along with those samples missed by DFT$_E$ (blue boxes).  It is immediately clear that the two representations have almost perfectly decorrelated errors, with avgDFT sensitive to the average byte values and DFT$_E$ focusing on harmonic structure that is evidently dependent on $N$, the number of connections.  Of the samples missed by avgDFT, DFT$_E$ catches 85\% of them. Though perfectly periodic, DFT$_E$ still misses almost all communications with fewer than ten connections.  Periodic signals are evidently most easily detected if there are many connections, since it is hard to tell if a signal with so few samples has any periodicity at all.  Indeed, for small numbers of connections ($N \lesssim 3$)  both type 1 and normal traffic have a similar fraction of Fourier components comprising 50\% of the signal energy, \ref{5} (c).

%\begin{figure}[htp]
%\label{WSEvI}
%\includegraphics[width=0.5\textwidth,clip]{type1count.pdf}
%\caption{Number of egress connections per communication in type 1 exfiltraiton samples.  The uncolored histogram is the full sample set, and blue histogram are those samples missed by the ensemble.  The ensemble tends to miss communications with small numbers of connections becasue these lack a strong periodic signal.}
%\end{figure}

Application of the DFT$_I$ component yields a tiny improvement, catching another 1\% of the samples missed by DFT$_E$.  The reason is that type 1 ingress connections are also periodic and in sync with the egress data, and once normalized, the egress and ingress Fourier spectra are highly-correlated.  The STFT component is also not sufficiently decorrelated with both avgDFT and DFT$_E$ to be of any help, and so we won't explore it further here.  In more complex exfiltration types, however, these two components will prove to be key contributors to the ensemble. 

In summary, on the client workstation which has primarily ingress-heavy communications, the complete ensemble detects periodic, constant-egress exfiltration that is not simultaneously ingress-heavy (within the approximate range $0.1 < \bar{b}_I/\bar{b}_E < 100$) and connection-sparse ($N \lesssim 10$). This is an interesting and rather constrained combination: exfiltration communications with so few connections must have large $\bar{b}_E$ in order to transfer a reasonable amount of data.  In general, an attacker must then {\it serve} an equal amount of data back to the victim system in order to keep $\bar{b}_I/\bar{b}_E$ sufficiently high to avoid detection, and, they must be careful not to egress more data than is typical of the system.   This is certainly possible, but perhaps unlikely for an adversary employing such simplistic (type 1: periodic, constant egress) exfiltration.  

%all exfiltration communications with average ingress and egress byte values that do not overlap the $(\bar{b}_I,\bar{b}_E)$ distribution of normal traffic, including all egress-heavy exfiltration.  Furthermore, 86\% of communications overlapping the distribution of normal traffic are detectable with the help of DFT$_E$ and DFT$_I$, which recognize the periodicity of type 1 exfiltration.  We now test the ensemble against more complex scenarios.

\begin{figure*}[htp]
\begin{tabular}{cc}
\subfloat[]{\includegraphics[width = 3in]{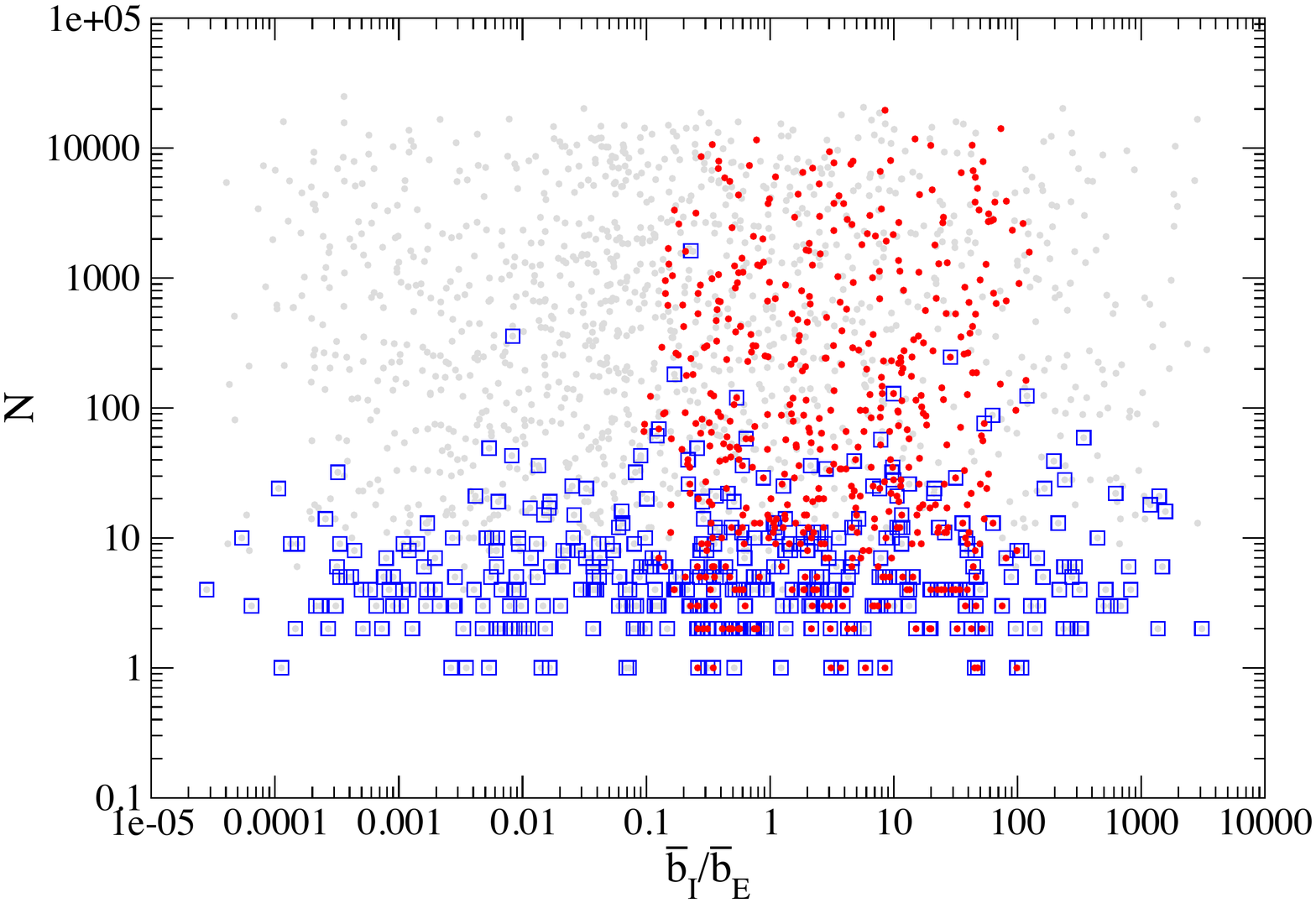}} &
\subfloat[]{\includegraphics[width = 3in]{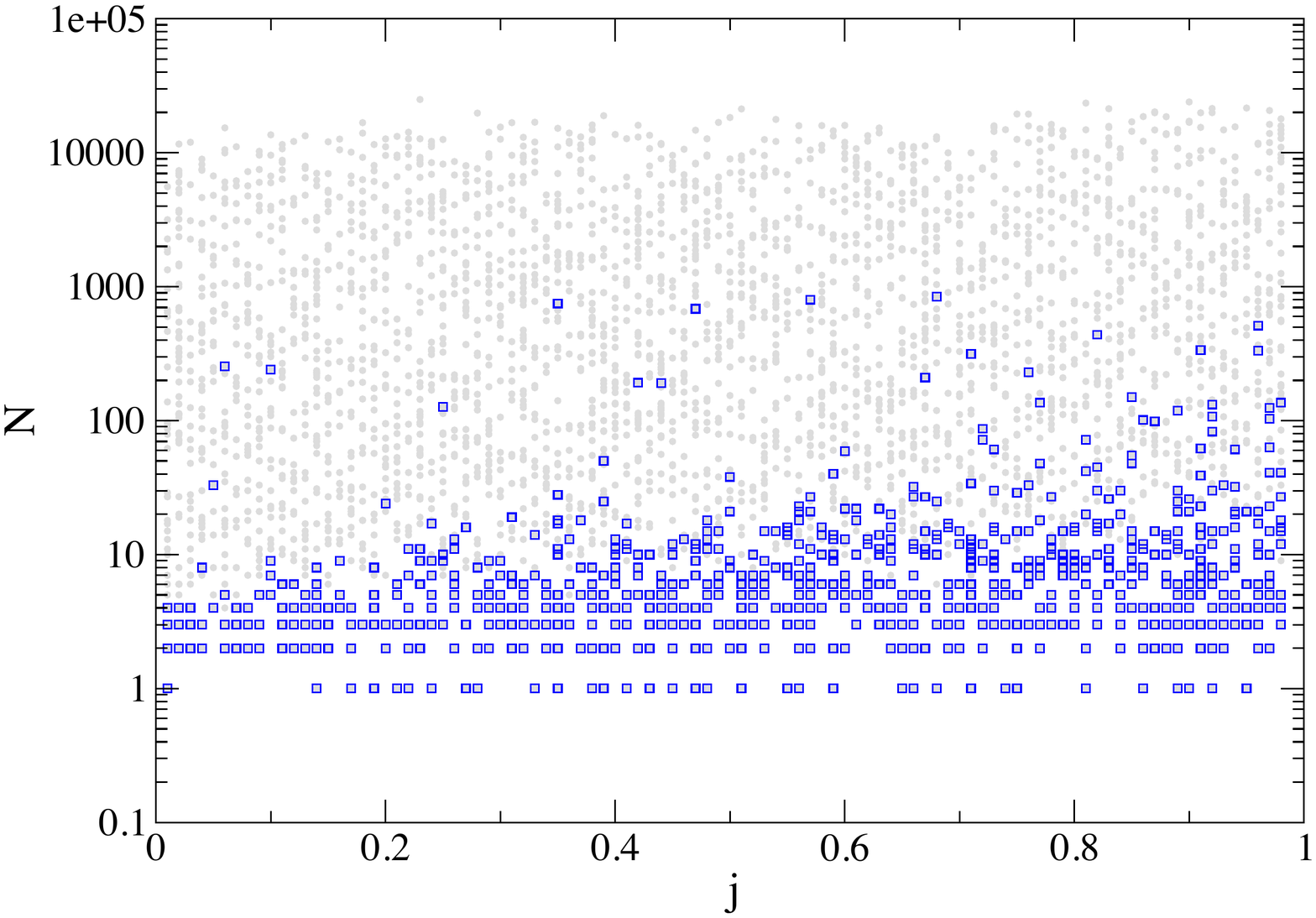}}
\end{tabular}
\caption{(a) Type 2 exfiltration samples (gray points) in the $\bar{b}_I/\bar{b}_E$-$N$ plane.  Red points indicate those samples misclassified as normal by avgDFT and blue squares indicate those points misclassified by DFT$_E$. (b) Type 2 exfiltration samples (gray points) in the $j$-$N$ plane.  Blue squares are those samples misclassified by DFT$_E$.  For a given $N$, a larger jitter factor $j$ results in a greater chance of misclassification.} 
\label{8}
\end{figure*}
\subsubsection{Type 2: random jitter, constant egress data}
As a generalization of type 1 exfiltration, we next consider periodic egress modulated by jitter: the egress timing is drawn randomly from the range $[T-jT,T]$ where $T$ is the nominal period and $j \in [0,1]$ is the {\it jitter factor}.  Since the egress data is constant as in type 1, the performance of avgDFT is the same as in type 1 and so we do not discuss it further.  Meanwhile, the DFT$_E$ misses communications with larger $N$ than it does for type 1 exfiltration, Figure \ref{8} (a), but still succeeds in catching 77\% of the points missed by avgDFT.  This is perhaps surprising, since the jitter factor might be expected to spoil any strong periodic signal in the communication.  But, if there are enough connections and the jitter factor isn't too large, the classifier is still able to identify dominant modes sufficiently to distinguish these samples from normal traffic.  Indeed, Figure \ref{8} (b) shows the mild decline in performance with increasing $j$.

\begin{figure*}[htp]
\begin{tabular}{cc}
\subfloat[]{\includegraphics[width = 3in]{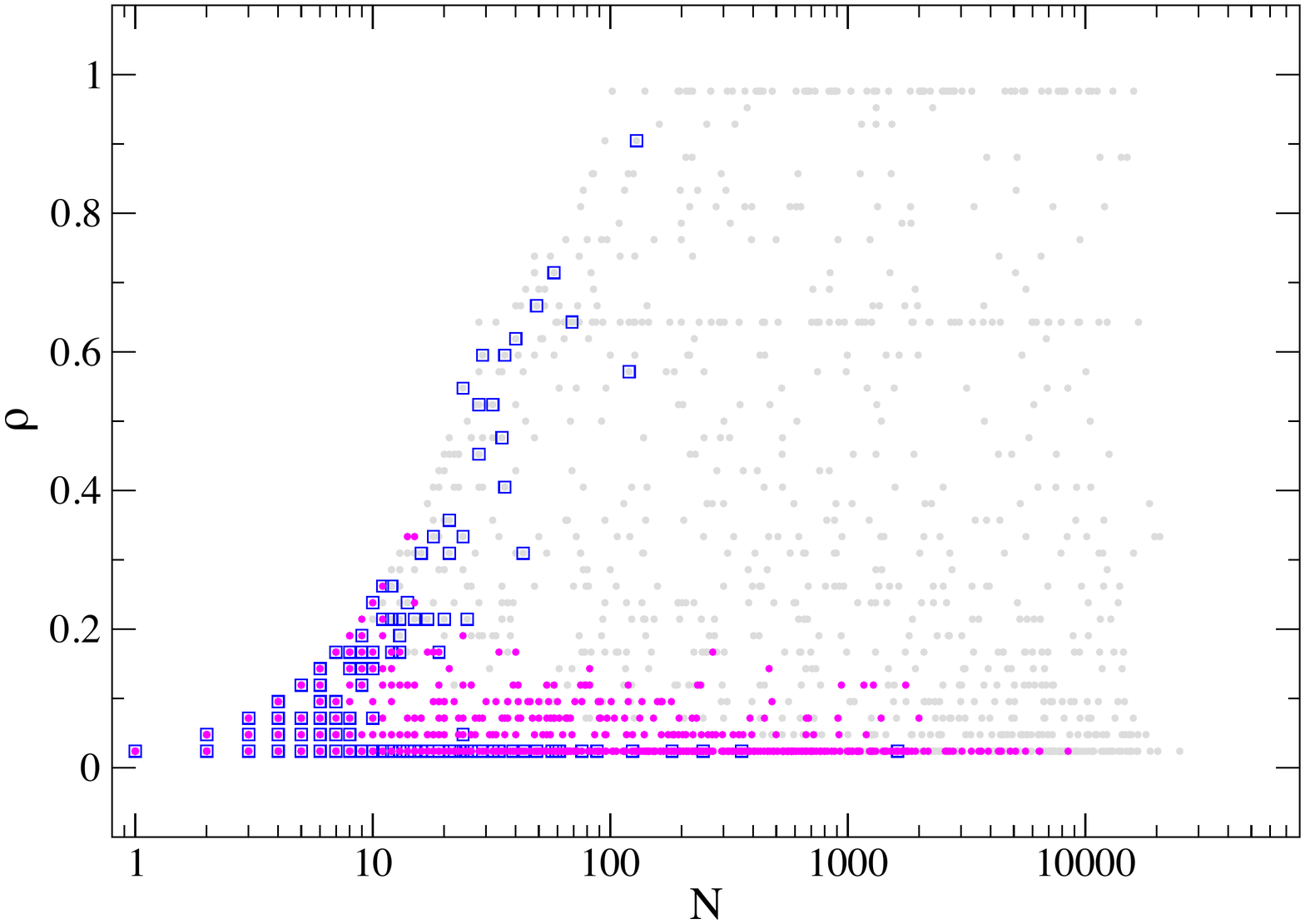}} 
\subfloat[]{\includegraphics[width = 3in]{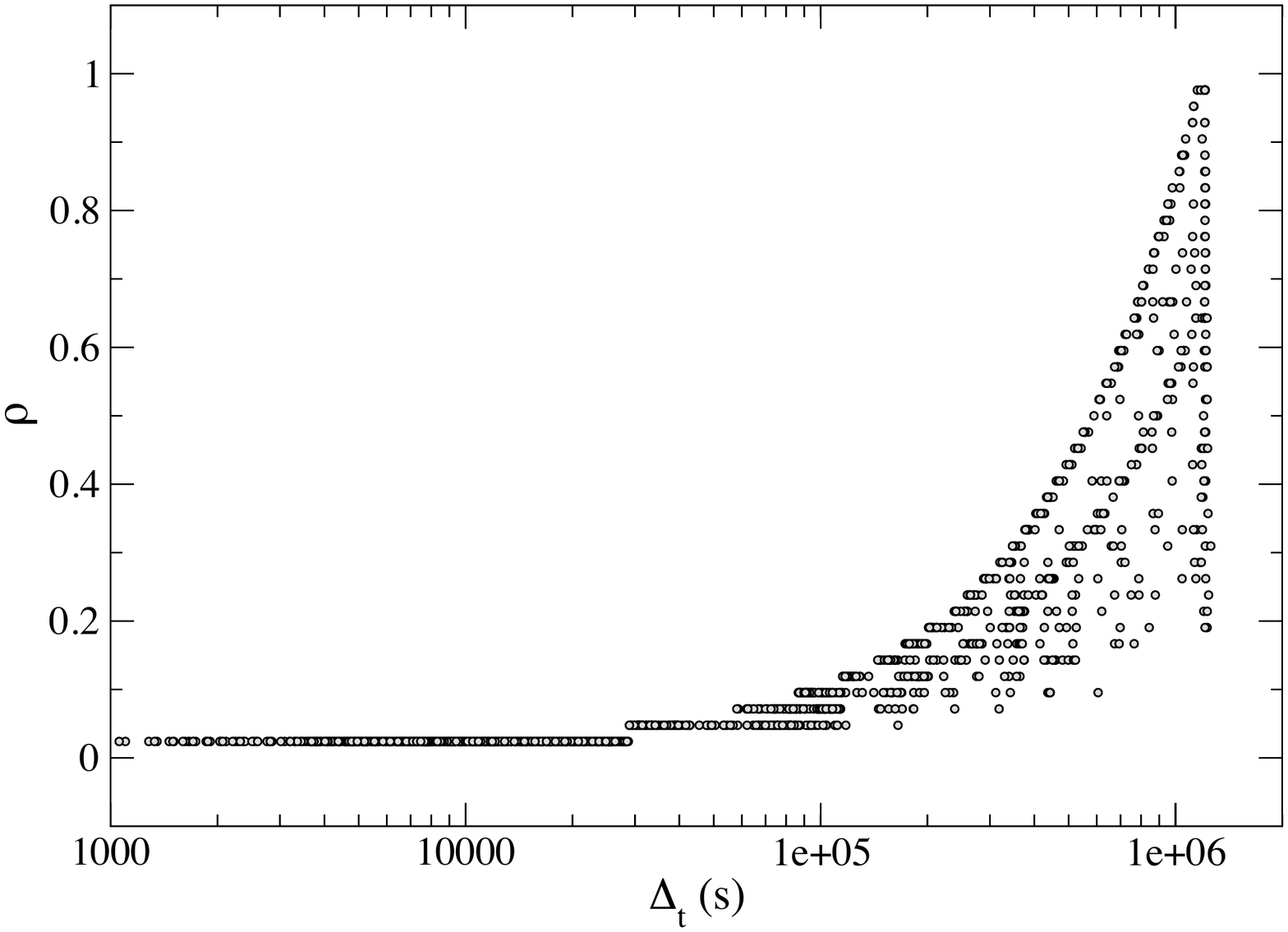}}\\
\end{tabular}
\caption{Type 2 exfiltration samples (gray points) in the $N$-$\rho$ plane.  Magenta points are those samples misclassified by STFT and blue squares by DFT$_E$. The two components miss two distinct branches of parameter space. (b) Relationship between communication duration and density: to be dense, a communication must have traffic ocurring in many $a$-hour windows across the full two week period, {\it i.e.} only long-duration communications can be dense.}
\label{9}
\end{figure*}
While dominant harmonics emerge if there are many connections, the jitter causes the communication to look generally quite irregular over short time scales.  We therefore introduce the STFT compoenent to see if these time irregularities can be leveraged to improve performance.  The STFT might be helpful for detecting traffic with jitter because it doesn't focus on dominant harmonics across the full span of the communication, but rather employs coarse-graining at $a$-hour intervals to smooth out the jitter.  But, for this to payoff, the communication must be sufficiently long and dense to contain enough of these intervals.  If we examine $\rho$ vs. the number connections per communication, $N$, we see in Figure \ref{9} (a) that the STFT component tends to miss low-denisty communications, along a distinct branch of parameter space from those samples missed by DFT$_E$.  Despite this, the STFT is not particularly decorrelated with the {\it intersection} of avgDFT and DFT$_E$ errors, contributing only a modest improvement of 3\%.  

Overall, the ensemble detects 80\% of those samples missed by the avgDFT.  It is successful at detecting low-jitter ($j \lesssim 25\%$) exfiltration that is not both ingress-heavy ($0.1 < \bar{b}_I/\bar{b}_E < 100$) and connection-sparse ($N \lesssim 10$).  With increased jitter ($j \gtrsim 25\%$), more connections ($10 \lesssim N \lesssim 100$) are needed for detection unless the communication is of low density, $\rho \lesssim 0.3$, in which case $N \gtrsim 10$ regardless of jitter.  From Figure \ref{9} (b), these low-density communications are of short duration ($\Delta_t \lesssim 3\times 10^5$ sec, or around 3.5 days).

%Notice that what might be considered a separate category of exfiltration---that of totally random, unconstrained, timing---is actually a subset of type 2 communications reached in the limit that $T \rightarrow T_\text{max}$ and $j \rightarrow 100\%$.  These highly irregular communications almost perfectly deceive the $Y_I(f)$-trained classifier, since they are essentially random signals without any strong harmonics.  However, in this limit $T-\langle Tj\rangle \rightarrow 0$ making the inequalities above hard to satisfy for any duration exfiltration, and so we expect the $S(\tau,f)$-trained classifier to do well here.  Indeed, we find that the TPR $< 90\%$ for this sub-type. While there is certainly room for successful exfiltration, the traffic patterns are unpredictable and cannot be controlled to avoid detection, and so it is unlikely that an attacker would devise an exfiltration scenario with totally random timing. 
\begin{figure*}[htp]
\begin{tabular}{cc}
\subfloat[]{\includegraphics[width = 3in]{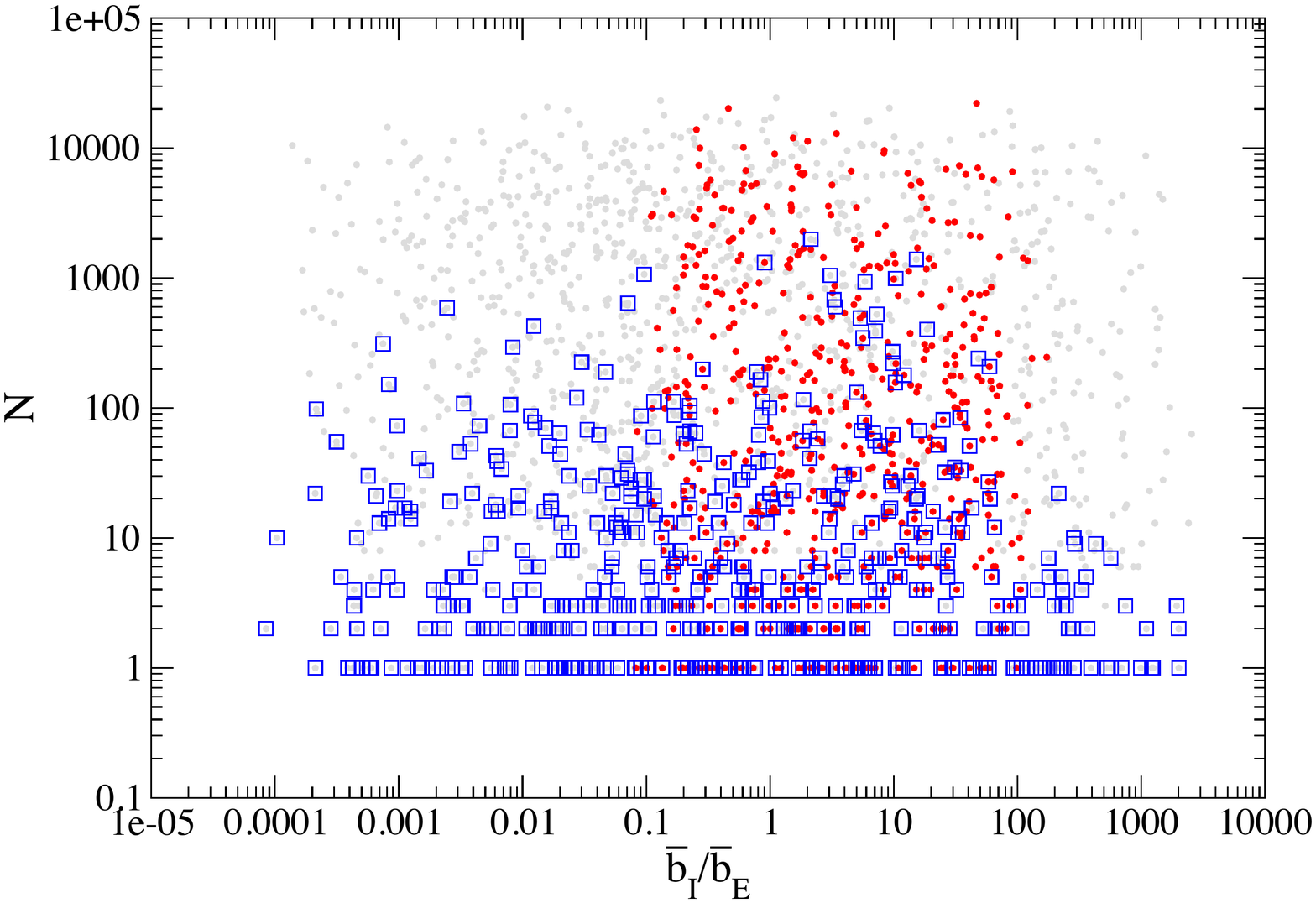}} &
\subfloat[]{\includegraphics[width = 3in]{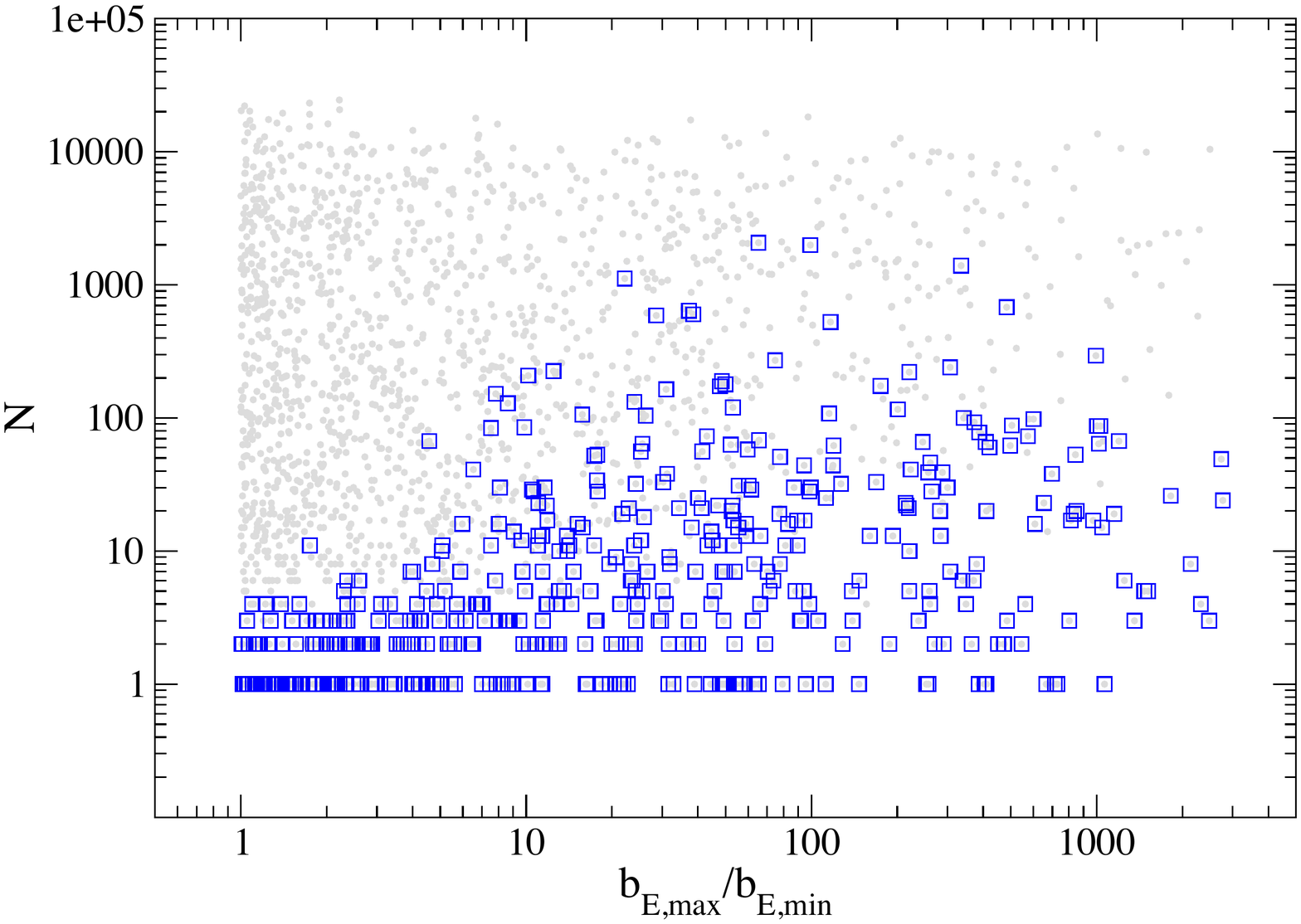}}\\
\end{tabular}
\caption{(a) Type 3 exfiltration samples (gray points) in the $\bar{b}_I/\bar{b}_E$-$N$ plane.  Red points indicate those samples misclassified as normal by avgDFT and blue squares indicate those points misclassified by DFT$_E$. (b) Type 3 exfiltration samples (gray points) in the $b_{E,\text{max}}/b_{E,\text{min}}$-$N$ plane.  Blue squares are those samples misclassified by DFT$_E$.  For a given $N$, a larger value of $b_{E,\text{max}}/b_{E,\text{min}}$ results in a greater chance of misclassification.}
\label{11}
\end{figure*}
\subsubsection{Type 3: periodic, variable egress data}
We next examine exfiltration scenarios in which the amount of data transferred varies over time but with periodic connections.  Variable egress could be readily implemented on the compromised host by configuring how the remote agent packages and extracts data.  For each communication, $b_{E,\text{max}}$ is drawn logarithmically from the range $[10^3, 4 \times 10^6]$ bytes, $b_{E,\text{min}}$ from the range  $[10^3, b_{E,\text{max}}]$, and at each time step a random value is drawn from $[b_{E,\text{min}},b_{E,\text{max}}]$.  The period of each communication is drawn logarithmically from the range $[1,8.64\times 10^4)]$ seconds, as in type 1. 

We first discover that avgDFT is moderately affected by the variable egress data, missing 10\% more exfiltration samples than it did for type 1 or type 2.  This is expected, since statistics like $\overline{|Y_E(f)|}$ won't reliably characterize the full communication if the data variance is large.  Further, DFT$_E$ does a little bit worse than it did for the {\it aperiodic} type 2 traffic, catching 72\% of the traffic missed by avgDFT, Figure \ref{11} (a).  Though periodic, the type 3 the signal varies in amplitude and so the dominant harmonics are weakened relative to the case of constant signal; as Figure \ref{11} (b) shows, the classifier performance worsens with increased $b_{E,\text{max}}/b_{E,\text{min}}$; in particular, for $b_{E,\text{max}}/b_{E,\text{min}} \gtrsim 10$, the classifier begins to occasionally miss communications with $N > 10$.  But, similar to type 2, we can catch some of the missed communications if they are of sufficiently dense (and so of long duration) by employing the STFT component.  For type 3, STFT catches an additional 5\% of the samples missed by both avgDFT and DFT$_E$, giving an overall ensemble performance of 77\%. 

It is perhaps worthwhile to now reconsider the component DFT$_I$ which has hitherto been highly correlated with DFT$_E$.  Whereas egress data might be deliberately staggered or varied by configuring the call-back agent, ingress data comprises the attacker's server's response.  When not containing instructions or other code, this traffic is generically TCP control data, with packet sizes set by the protocol and the system's network stack.  Of course, this is configurable, but perhaps not as easily as the egress traffic, and it is especially more difficult if the remote server is hosted by a cloud or file sharing service that does not permit such tuning of the response traffic.  In any case, if we assume that the ingress data is constant, then DFT$_I$ picks up the periodicity of the ingress communications that is effectively hidden in the egress data, decorrelating the representations. The ensemble performs essentially as it did against type 1 exfiltration, catching 85\% of the traffic missed by avgDFT. 

%\begin{figure}
%\includegraphics[width=0.65\textwidth,clip]{powel20.pdf}
%\caption{Type 3 exfiltration samples (gray points) in the $N$-$\langle C(S(\tau,f)=0)\rangle$ plane.  Orange points are those samples misclassified by the $S(\tau,f)$-based classifier and blue squares are those samples misclassified by the $|Y_I(f)|$-trained classifier.}
%\label{12}
%\end{figure}
\subsubsection{Type 4: random jitter, variable egress data}
In this last category of exfiltration, the timing characteristics of type 2 exfiltration are combined with the egress variability of type 3.  We therefore expect, and find, that the performance of DFT$_E$ is severely challenged by the combined aperiodicity and variability of the egress data, catching only 65\% of the communications missed by avgDFT.   The STFT catches an additional 9\% of the traffic missed by both avgDFT and DFT$_E$, for a total 74\% of all missed communications. Inclusion of DFT$_I$ is again beneficial under the assumption of constant ingress data, though here it must contend with the variable timing characteristics of the ingress data, leading to an overall ensemble performance of 80\%.  No new qualitative regimes emerge with type 4 exfiltration, however, and so the performance of the ensemble can be understood via the preceding analysis.  

%While we have introduced classifiers and built the ensemble in stages as we tested the different exfiltration scenarios, this was primarily to make the presentation pedagogical.  In practice, the full ensemble is implemented; its performance against each exfiltration scenario is summarized in Table 2. 
%The results of the ensemble for each exfiltration type for workstation A is summarized in Table 2.
\begin{figure*}[htp]
\begin{tabular}{cc}
\subfloat[]{\includegraphics[width = 3in]{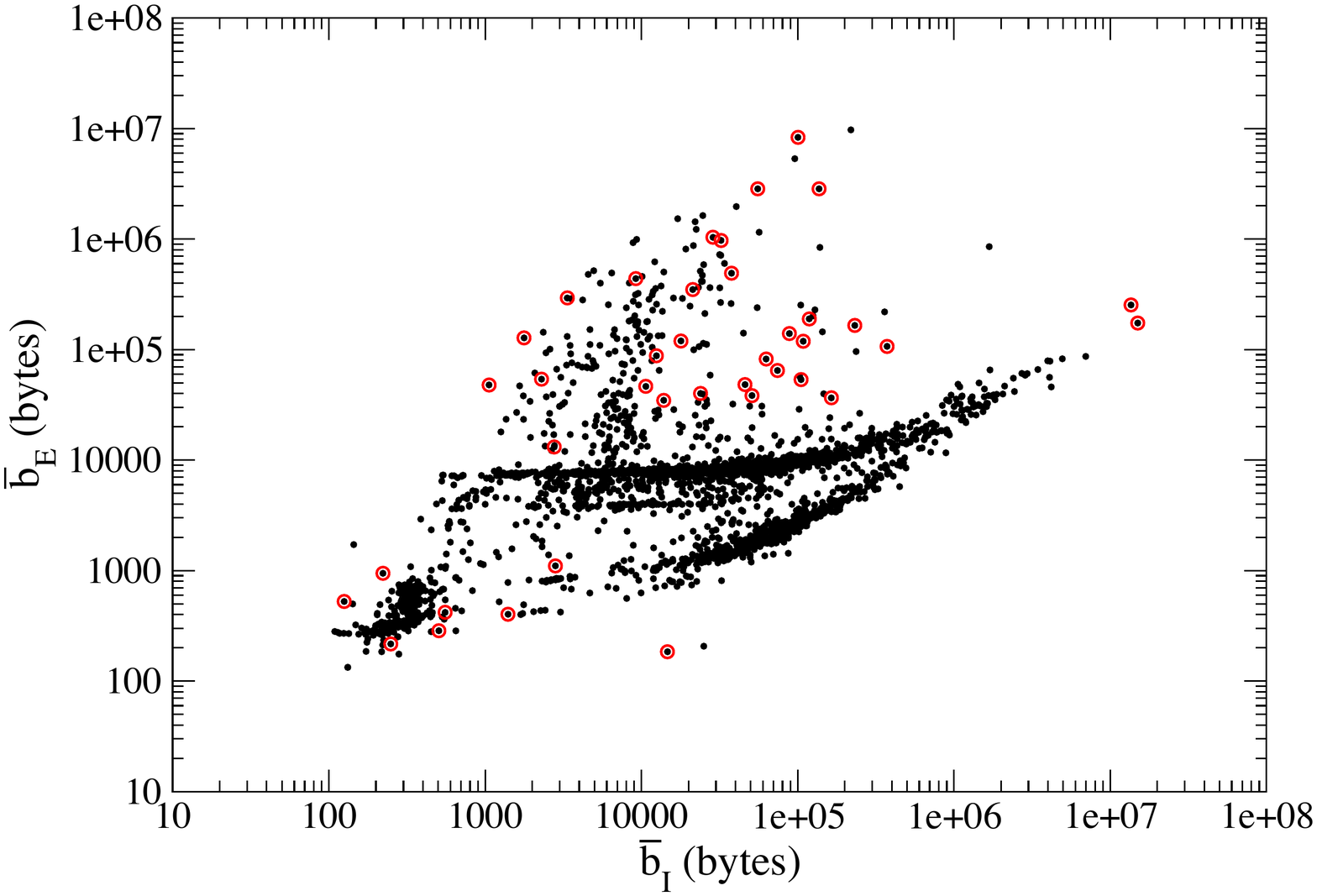}} &
\subfloat[]{\includegraphics[width = 3in]{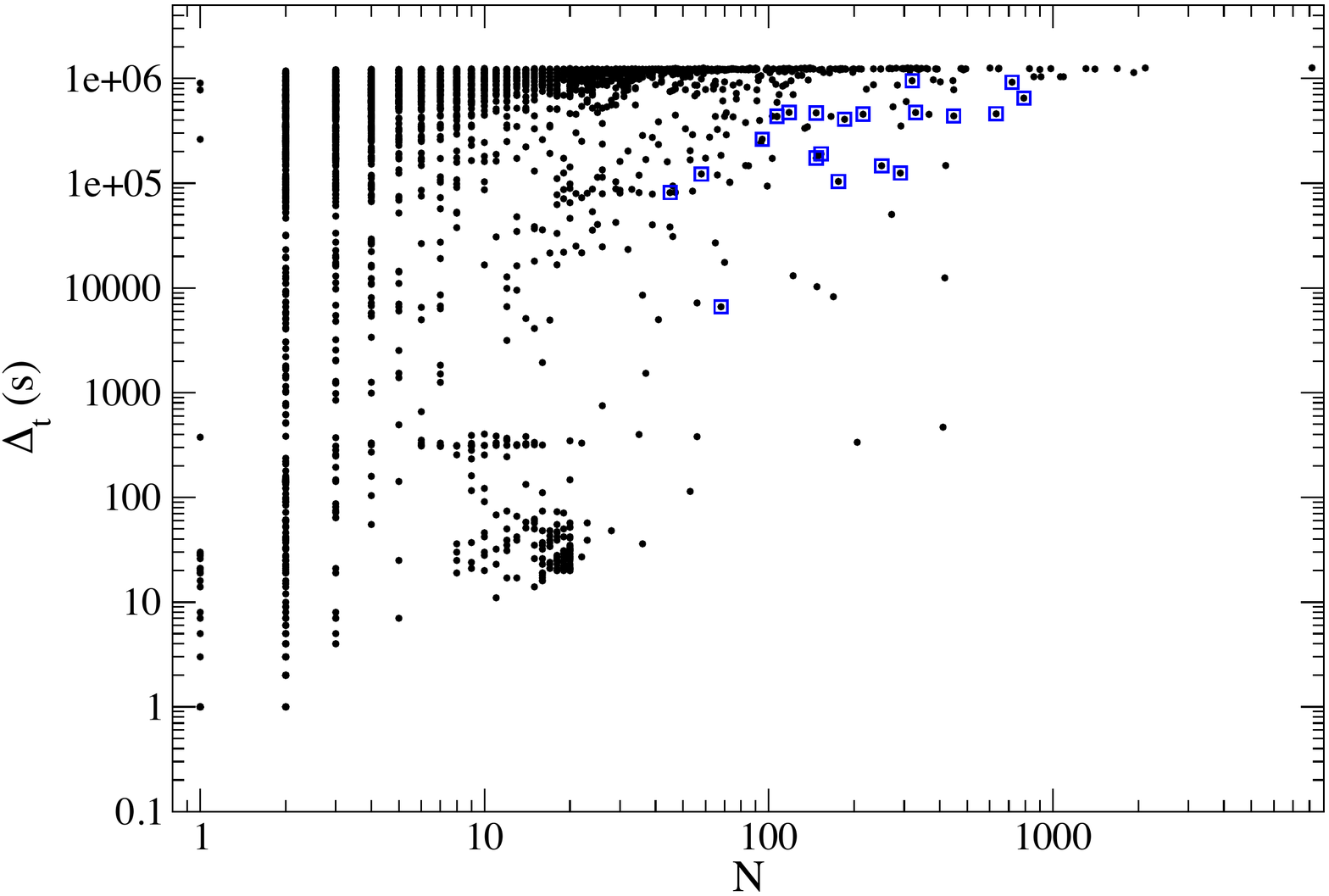}}\\
\subfloat[]{\includegraphics[width = 3in]{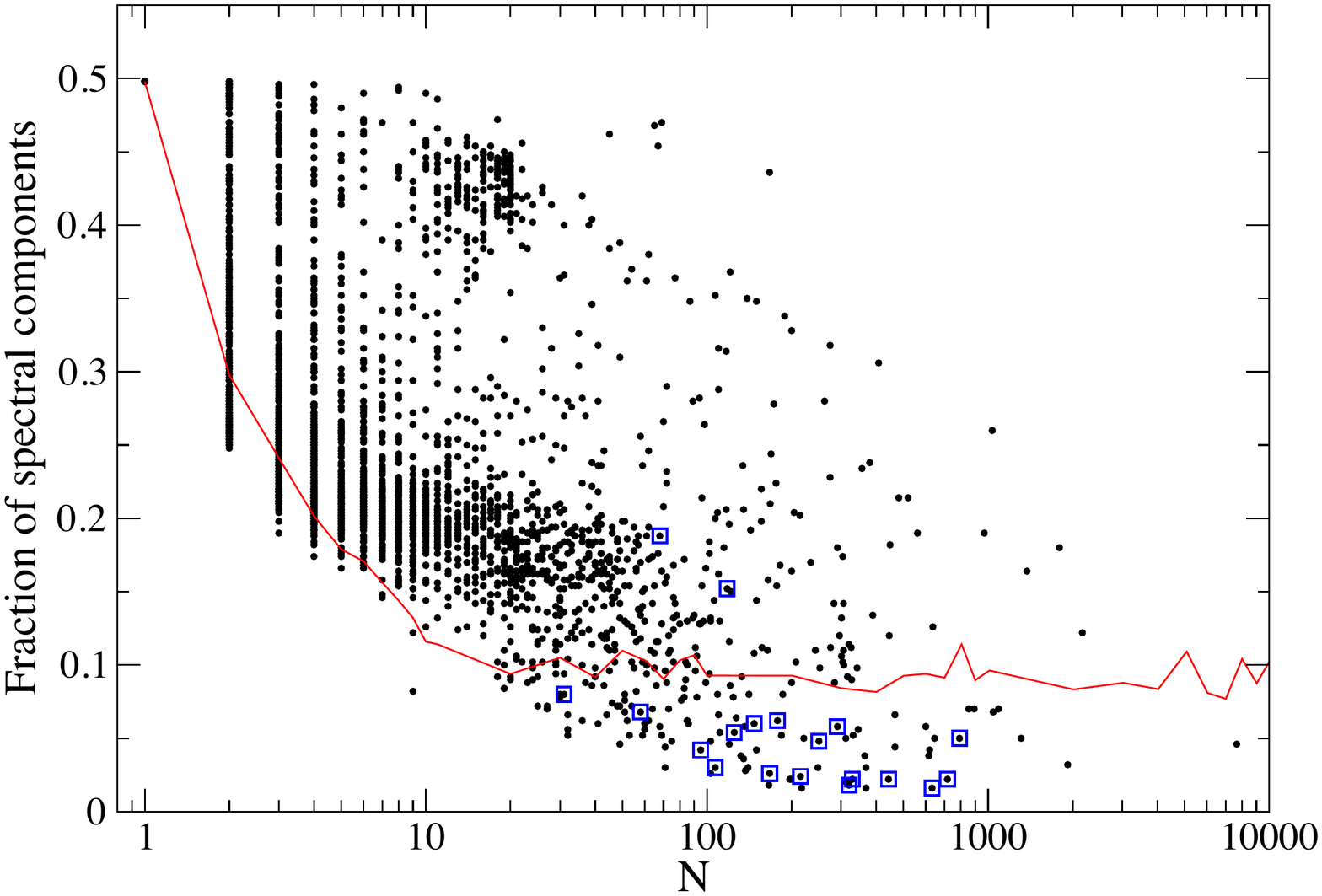}}& 
\subfloat[]{\includegraphics[width = 3in]{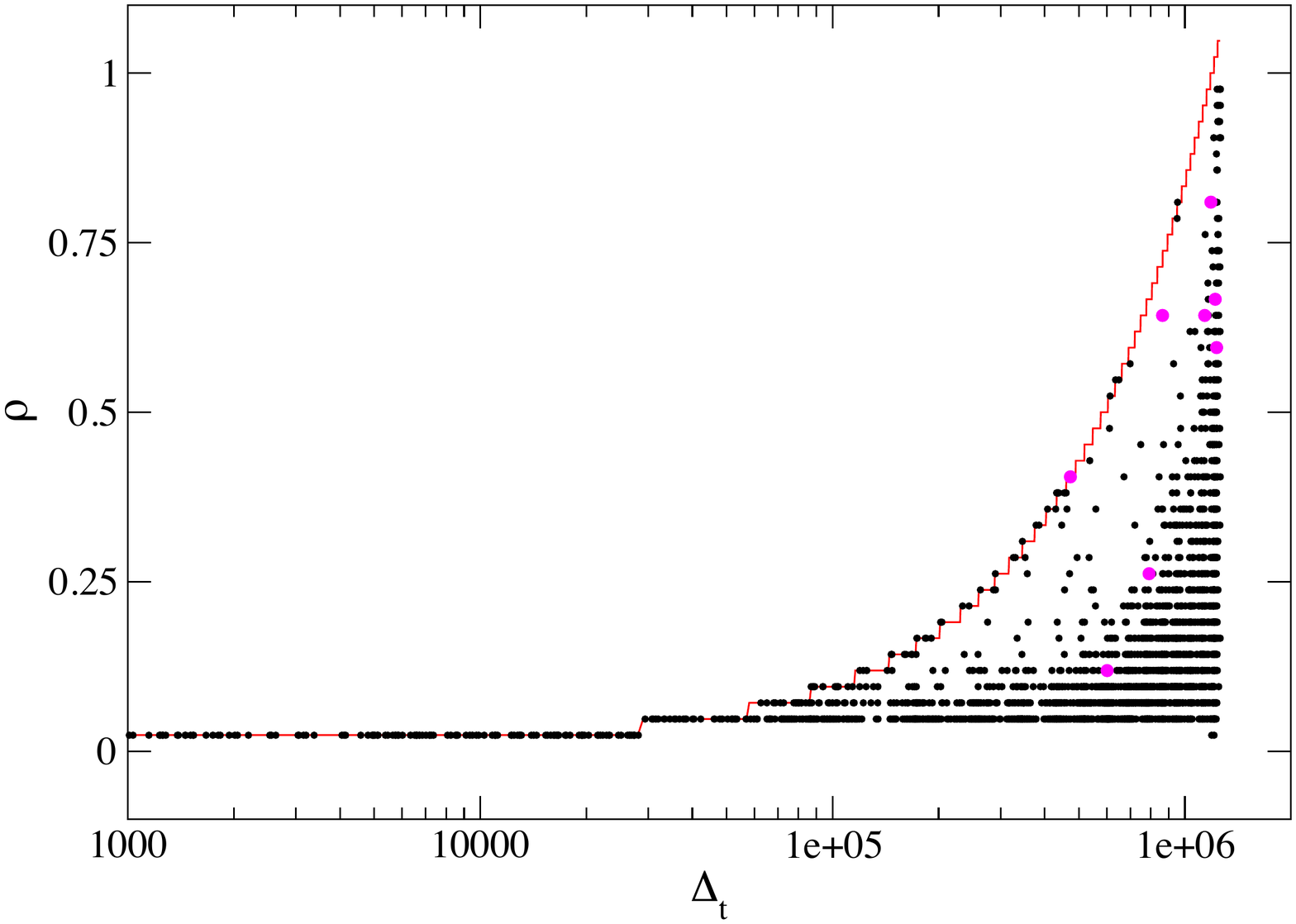}} 
\end{tabular}
\caption{4000 normal email gateway communications (black points), with avgDFT false positives (1\%) shown as red circles, DFT$_E$ false positives (0.05\%) as blue squares, and STFT false positives (0.02\%) as magenta points. (a) Average egress bytes vs average ingress bytes, (b) duration vs. number of connections, (c) fraction of Fourier components comprising 50\% of waveform energy vs. number of connections, with the average fraction computed from the periodic type 1 exfiltration samples shown for reference, (d) 4-hour coarse-grained density vs. duration with maximum density per duration shown by the red line.}
\label{13}
\end{figure*}
\subsection{Email Gateway}
We next test the detection ensemble on a rather different kind of system: one of APL's externally-reachable, data-serving email gateways.  NetFlow was collected for a two-week period and assembled into communications as in the preceding analysis.  The big difference in working with servers is that they are not exclusively source systems, as the workstations were, but receive incoming connections from external systems.  The gateway is a hybrid of sorts, acting as a client and sending emails off to other gateways and servers, and acting as a server by receiving email from the same. This particular email gateway receives external data over simple mail transfer protocol (SMTP, port 25), and originates connections to other email servers over SMTP and HTTPS.  There is also a significant difference in the number of communications as compared to the workstations we analyzed: over 23,000 communications, 95\% involved SMTP traffic. The communications tend to be long: the median duration was 147 hours, and the median number of connections per communication was 5.  

Figure \ref{13} shows the traffic statistics of the email gateway.  Besides the tantalizing shape of the distribution of communications in the $\bar{b}_I$-$\bar{b}_E$ plane, it is clear that many communications are egress-heavy.  In particular, the near-vertical swoosh of points at $\bar{b}_I \approx 10^4$ bytes is outbound SMTP traffic.  A look at Figure \ref{13} (b) reveals an almost bimodal distribution of communication durations: they are either short, on the order of minutes, or long, on the order of days.   Like the workstation, communications of a given duration tend to be connection-sparse, that is, $N \lesssim 100$ mostly independently of duration. One notable aspect of the gateway traffic is that it tends to exhibit periodicity, Figure \ref{13} (c), and (d) shows that there are maximally dense communications at almost any duration, foretelling a challenge to the DFT and STFT components.

\begin{figure*}[htp]
\begin{tabular}{cc}
\subfloat[]{\includegraphics[width = 3in]{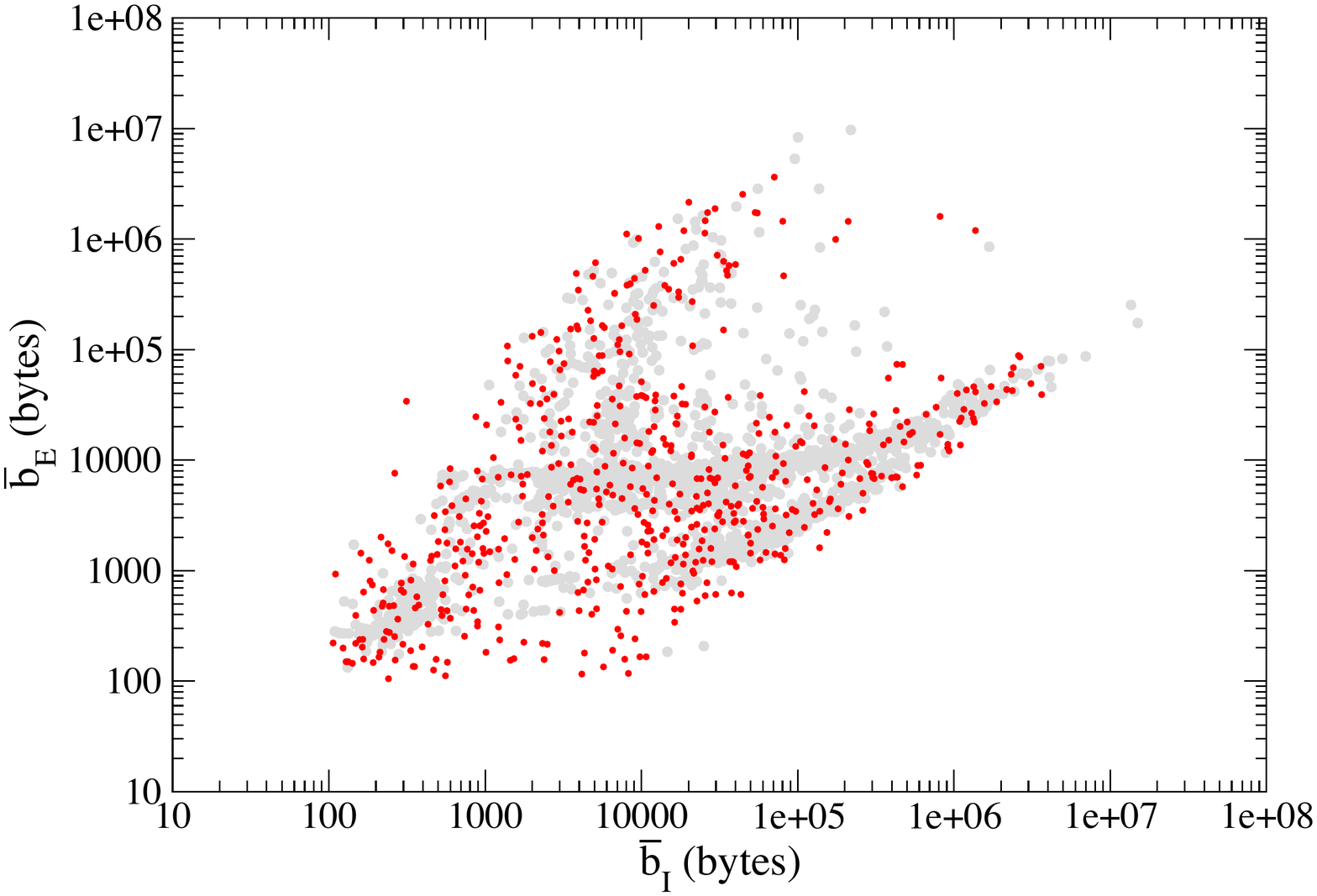}} &
\subfloat[]{\includegraphics[width = 3in]{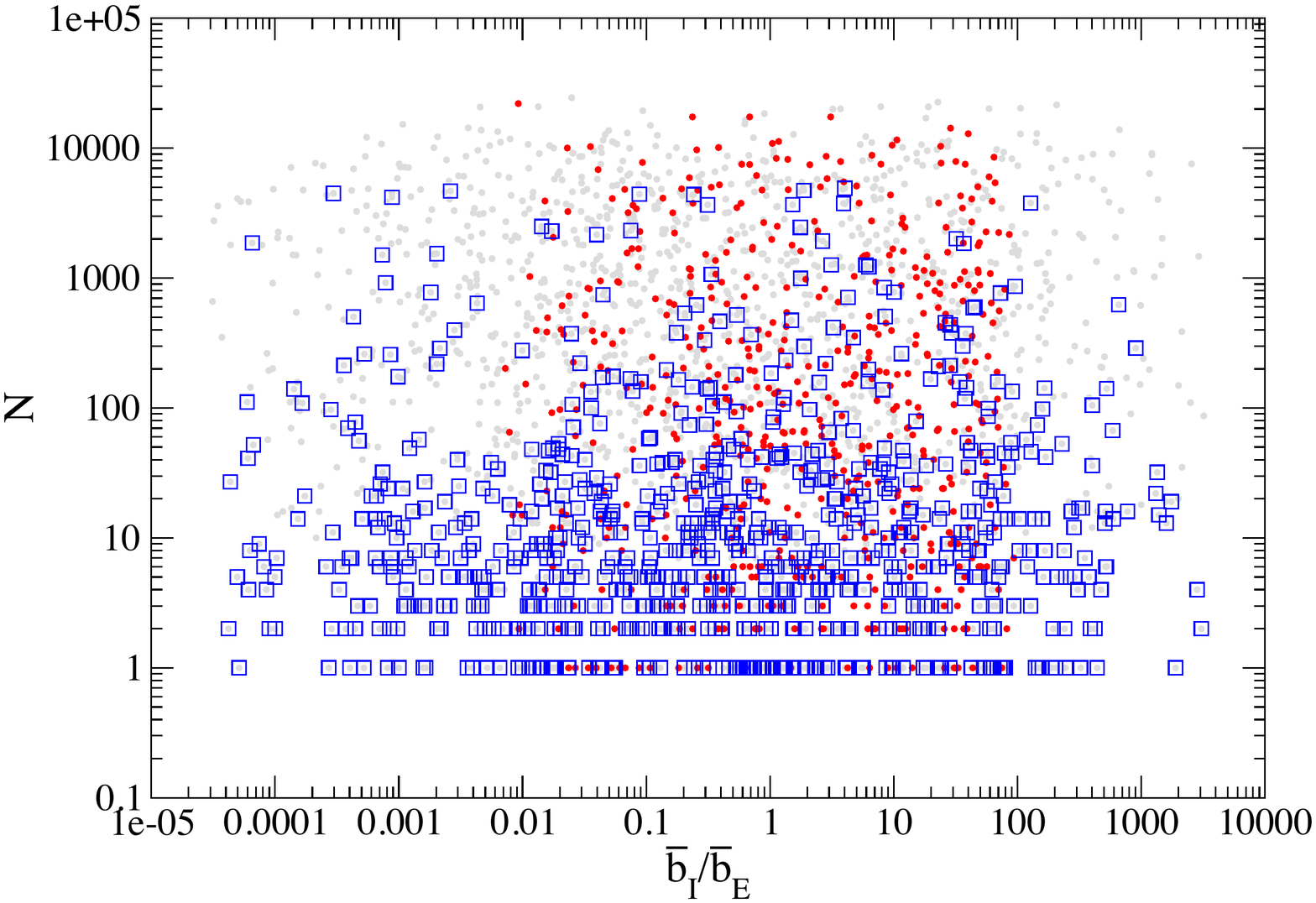}}\\
\end{tabular}
\caption{(a) Normal email gateway traffic from Figure \ref{13} (a) (gray points) with the type 1 exfiltration samples missed by avgDFT in red. (b) Type 1 exfiltration samples (gray points) in the $\bar{b}_I/\bar{b}_E$-$N$ plane.  Red points indicate those samples misclassified as normal by avgDFT and blue squares indicate those points misclassified by DFT$_E$.}
\label{14}
\end{figure*}

In what follows, we test a randomized 4000-sample subset of the full collection of 23,036 communications for the sake of efficiency; since subsampling might be necessary in practice for particularly traffic-heavy servers, this also serves to test the reliability of this practice.  We again present exfiltration detection performance at a 2\% FPR.  We find that an STFT window of $a=4$ hours performs marginally better than the 8-hour window used for the workstation.  The KDE does a good job of detecting all exfiltration communications (of any type) lying sufficiently far away from the distribution of normal samples, Figure \ref{14} (a), though there are plenty of egress-heavy normal communications to mask high-volume exfiltration.  We therefore pin our hopes on mining the timing characteristics to detect exfiltration, as we did successfully for the client workstation.

Concerns arising from Figure \ref{13} (c) and (d) are only partly confirmed with type 1 exfiltration, with DFT$_E$ catching 66\% of the communications missed by avgDFT.  Figure \ref{14} (b) shows that the DFT$_E$ misses communications across a wide range of $N$, though has a particularly difficult time against samples with $N \lesssim 100$. The STFT offers slight improvement, catching an additional 6\% for an ensemble total of 72\%.  Because the normal gateway traffic generally exhibits periodicity with many dense, long-duration flows, even type 1 exfiltration is difficult to detect.  The ensemble performs worse on the more sophisticated exfiltration types, achieving 46\%, 70\%, and 45\% against types 2, 3 and 4. 

In summary, given the egress-heavy branch of the average bytes distribution, Figure \ref{13} (a),  and the difficulty in detecting exfiltration with variable timing and data characteristics, servers like email gateways present a challenge to techniques seeking to detect advanced exfiltration scenarios. However, this method still succeeds in detecting approximately half of the wide range of exfiltration communications with the same byte distribution as normal traffic, which, together with the low false positive rate, makes it a potentially useful contribution to defense-in-depth.
%Indeed, by examining the normal samples in this region of parameter space, we find several strongly periodic communications.  Figure \ref{15} is an example of one such communication, which contains egress-heavy connections occurring at daily intervals, with other connections occasionally occurring at higher frequencies.   

%\begin{figure*}
%\begin{tabular}{cc}
%\subfloat[Email gateway communication exhibiting periodicity over a two-week period.]{\includegraphics[width = 3in]{powel25.pdf}} &
%\subfloat[Fourier amplitude of egress traffic in (a). The series of spikes at higher frequency are the result of aliasing of the $f = 1.15 \times 10^{-5}$ Hz and $f = 3.47\times 10^{-5}$ Hz signals.]{\includegraphics[width = 3in]{powel26.pdf}}\\
%\end{tabular}
%\caption{Example periodic email gateway communication and its egress Fourier amplitude.}
%\label{15}
%\end{figure*}

%\begin{table*}
%\begin{center}
%\begin{tabular}{|c|c|c|c|c|c|c|}
%\hline
%\multirow{2}{*}{System}&\multirow{2}{*}{Classifier}&\multirow{2}{*}{Normal Traffic}&\multicolumn{4}{|c|}{Exfiltration}\\
%&&&Type 1&Type 2&Type 3&Type 4\\
%\hline
%Workstation A&$H(f)$&0.1&40&38&40&35\\
%\cline{2-7}
%&$|Y_E(f)|$&0.53&86&79&75&72\\
%\cline{2-7}
%&$S(\tau,f)$&0.26&70&50&60&45\\
%\cline{2-7}
%&Ensemble ($H(f)\wedge |Y_E(f)| \wedge S(\tau,f)$)&0.8&91&88&88&86\\
%\cline{2-7}
%&Ensemble (all)&0.8&91&88&88&86\\
%\hline
%\end{tabular}
%\end{center}
%\caption{False positive rates (FPR) and true positive rates (TPR) of each classifier and ensemble on workstation A's normal traffic and each type of exfiltration.}
%\end{table*}
\subsection{Web Server}
\begin{figure*}[htp]
\begin{tabular}{cc}
\subfloat[]{\includegraphics[width = 3in]{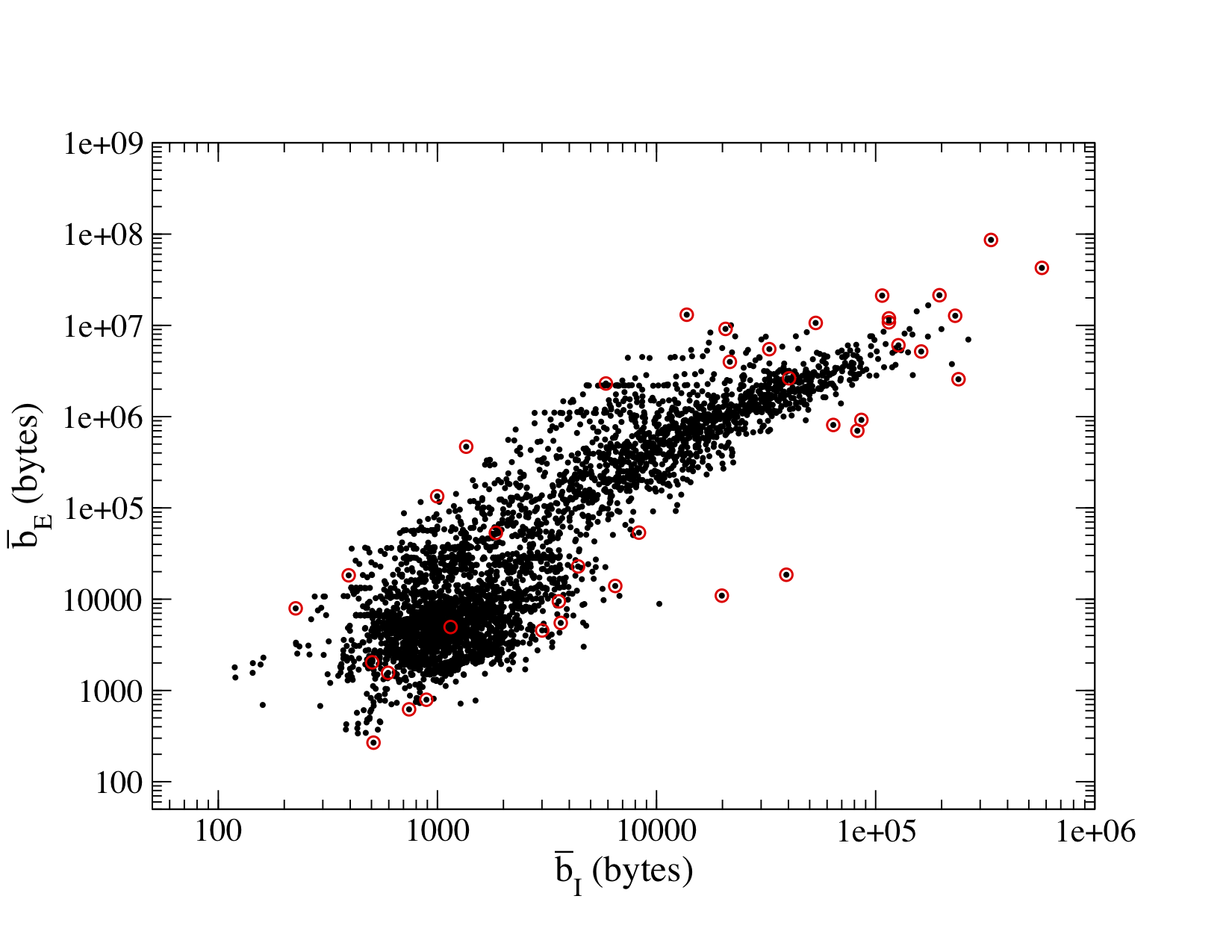}} &
\subfloat[]{\includegraphics[width = 3in]{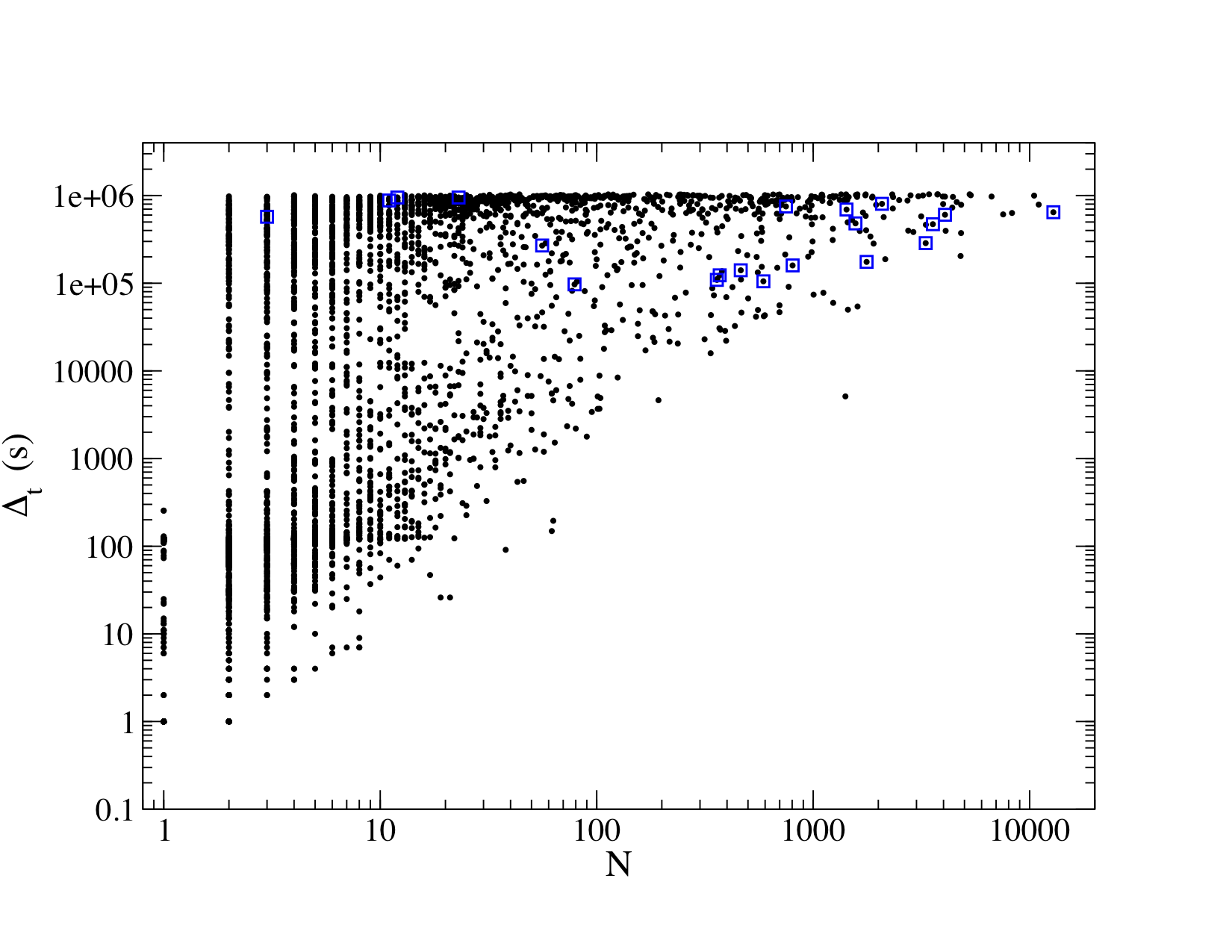}}\\
\subfloat[]{\includegraphics[width = 3in]{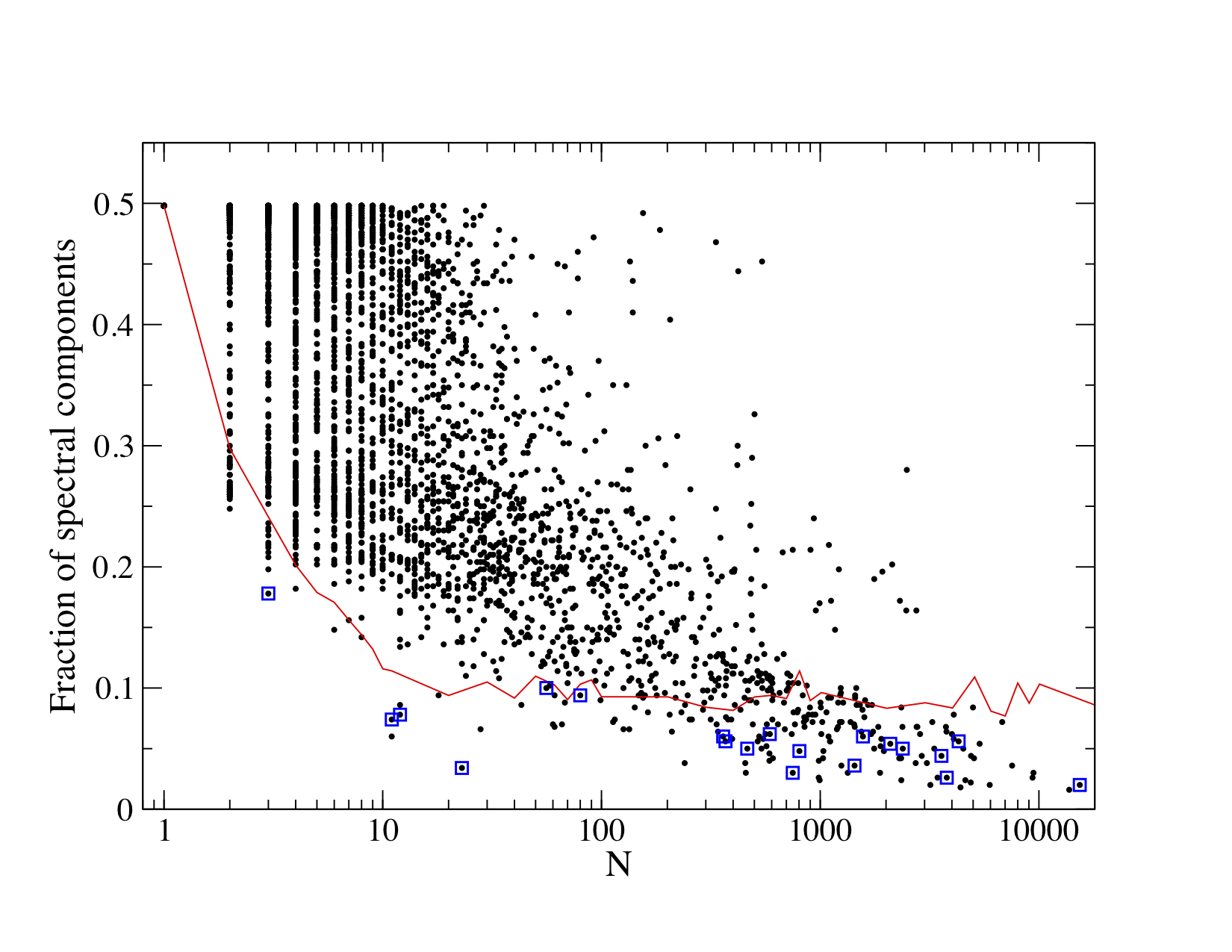}}& 
\subfloat[]{\includegraphics[width = 3in]{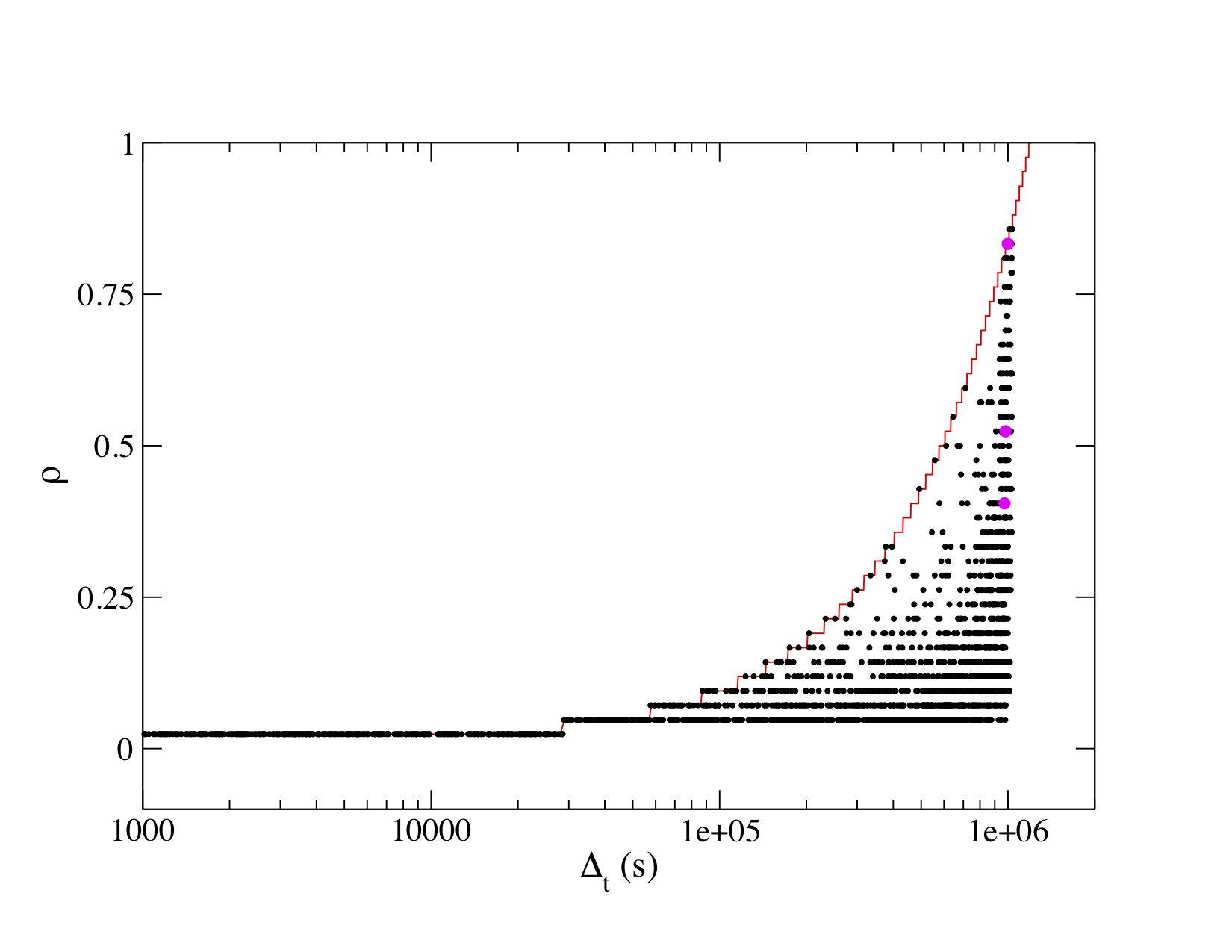}} 
\end{tabular}
\caption{4000 normal web server communications (black points), with avgDFT false positives (1\%) shown as red circles, DFT$_E$ false positives (0.05\%) as blue squares, and STFT false positives (0.01\%) as magenta points. (a) Average egress bytes vs average ingress bytes, (b) duration vs. number of connections, (c) fraction of Fourier components comprising 50\% of waveform energy vs. number of connections, with the average fraction computed from the periodic type 1 exfiltration samples shown for reference, (d) 4-hour coarse-grained density vs. duration with maximum density per duration shown by the red line.}
\label{15}
\end{figure*}

The last system tested is one of APL's heavily-trafficked, outward-facing web servers.  Over a two-week period we assemble 24,964 communications; the median duration was around 50 minutes and the median number of connections was five.  Given its primary role in serving web content, we expect and confirm that traffic is egress-heavy, Figure \ref{15} (a).    
Also of note is the periodic nature of normal communications as evident in Figure \ref{15} (c), and, like the email gateway, they are of relatively high density.  

As with the email gateway, we randomly select 4000 samples for training and testing.  We use an STFT window of $a=4$ hours, and report findings at 2\% FPR.  Against type I exfiltration, the ensemble detects around 50\% of samples, the primary difficulty stemming from the high-periodicity of normal samples, particularly those with many connections ($N \gtrsim 100$ in Figure \ref{15} (c)).  Performance against types 2 and 4, each with variable timing characteristics, is worse with around 20\% of samples missed by avgDFT caught.  The STFT component, which was key to detecting these kinds of exfiltration on the workstation, is blunted by the high-density of normal web traffic, Figure \ref{15} (d).  We conclude that web servers, which are egress-heavy and which tend to exhibit periodic, dense communications, are not optimal for detecting even basic data exfiltration.  For systems like this, which are excellent data egress candidates, the strategy should be to examine {\it internal} traffic flows {\it to} these systems {\it from} the primary data stores on the network, the idea being that it is too late to reliability detect data leaving the network once it gets to these staging points. 

\begin{figure}[htp]
\includegraphics[width=0.65\textwidth,clip]{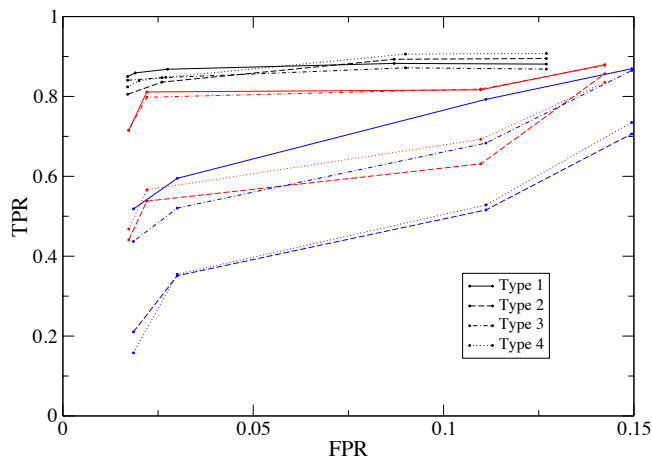}
\caption{ROC curves of each kind of exfiltration for the client workstation (black), email gateway (red), and web server (blue).}
\label{18}
\end{figure}

In summary, the receiver operating characteristic (ROC) curves of each kind of exfiltration for each system type studied in this section are shown in Figure \ref{18}.  The true positive rates (TPR) are the percentage of exfiltration communications missed by the avgDFT component but caught by the remaining components of the ensemble.  We do not consider as TPR the percentage of {\it all} exfiltration communications caught by the full ensemble because this percentage can be made arbitrarily small by simply widening the range of $(\bar{b}_I,\bar{b}_E)$ used to generate the samples.  Instead, the measure of TPR adopted here essentially assumes that all exfiltration sufficiently far away from the mass of normal communications of the system of interest will be caught by the KDE.   While this percentage is still sensitive to the parameter ranges and sampling schemes used to model the timing characteristics of the simulated exfiltration data, it lets the distribution of normal traffic set the bounds on the range of $(\bar{b}_I,\bar{b}_E)$, and it offers a means of comparing across models. 
\section{Detection of internal data movement}
An attacker would consider themselves lucky if the compromised system chosen for egress also happened to contain the targeted information.  More likely, the information of interest exists elsewhere on the network and so must be moved internally from its source to the egress system.  Internal data transfers are assuredly common in enterprise networks: email, data backups, file sharing, and even print jobs are everyday examples on most networks.  The hope is that systems that are good exfiltration candidates (lots of egress-heavy, dense outbound traffic with dominant harmonics) are not also good staging candidates (lots of incoming data from internal systems).  

Techniques for data staging are less sophisticated than those used for exfiltration.  There is no C2 to worry about, and so beaconing and other intermittent signaling is not needed.  To achieve stealth, an attacker will attempt to move data as it is typically moved across the network, for example, via file shares. SMB and other protocols used to access shared resources offer a natively available capability to attackers---no need to install customized agents with covert callbacks inside the network.   

The detection of data staging is therefore about finding unusual data ingress/egress characteristics rather than teasing out abnormal temporal structures, {\it i.e.} application of the KDE to avgDFT for internal traffic might prove useful for identifying large lateral flows of data.  As an example, we consider the internal traffic of the web server studied in the previous section: NetFlow was collected over a two-week period and communciations were formed as in the exfiltration study. There are roughly 12,000 communications predominantly over secure web. 
\begin{figure}[htp]
\includegraphics[width=0.65\textwidth,clip]{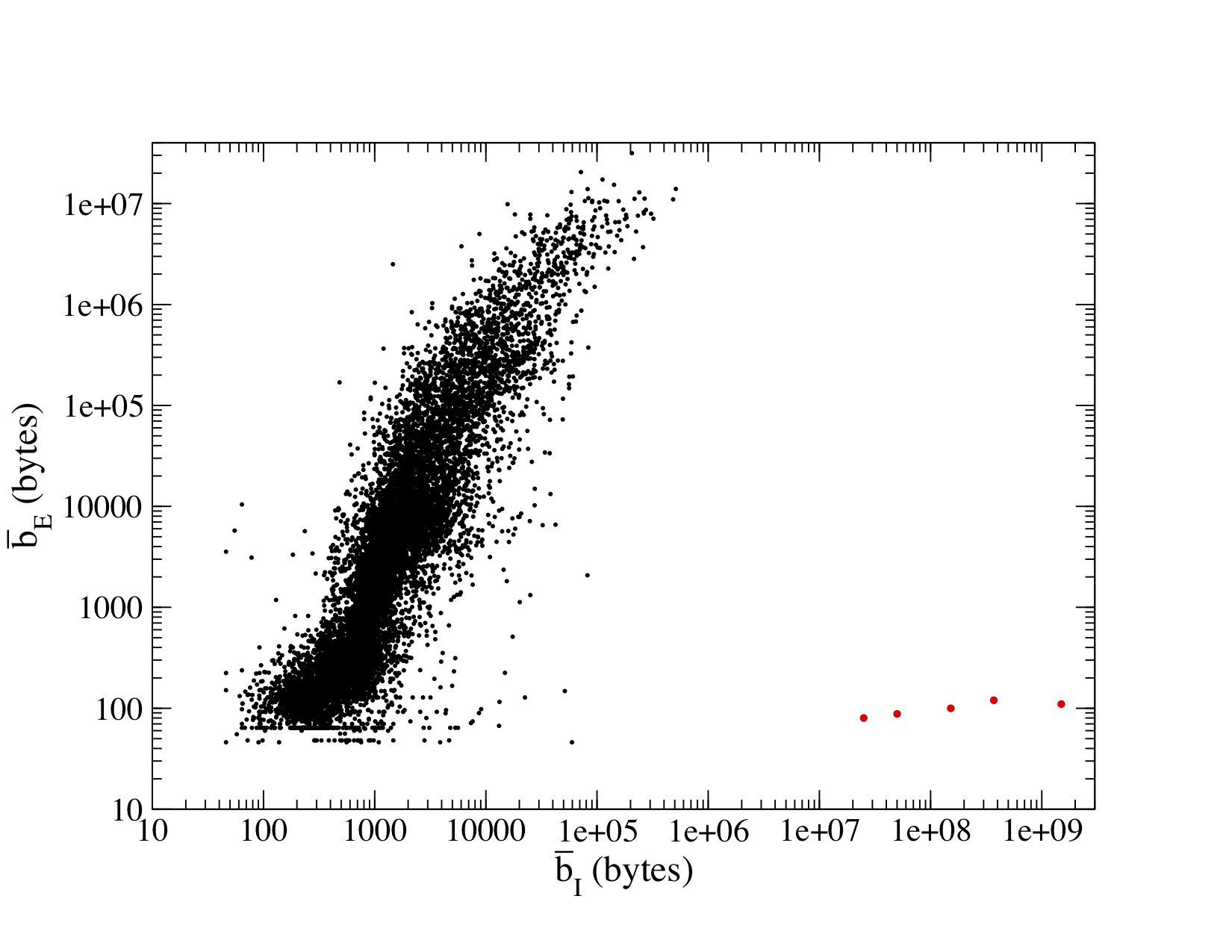}
\caption{12,000 normal internal communications of the web server (black) with five test file shares over SMB (red) of sizes 25, 50, 150, 375, and 1500 MB.}
\label{19}
\end{figure}

The normal internal traffic is generally egress-heavy, though a small, dense cluster of ingress-heavy traffic exists for relatively few bytes exchanged ($\bar{b}_I,\bar{b}_E \lesssim 1000$ bytes), Figure \ref{19}.  Next, files of different sizes were mounted to the server over SMB; the smallest test file was 25 MB, and the largest 1.5 GB.  These communications are clear outliers with respect to the distribution of normal traffic.  In fact, the KDE can be trained on the avgDFT of normal data with a very high tolerance such that none of the normal traffic is considered anomalous, yet still remain sensitive to file transfers of 25 MB or more at a confidence exceeding 99\%. 

Because this approach relies solely on avgDFT, its success hinges on the distribution of normal traffic.  In this case, while the web server is not a good candidate for exfiltration detection, it {\it is} a good candidate for detecting data staging.  This needn't be the case in general, however, and so critical systems should be analyzed for the optimal detection strategy.
\section{Comparison with other methods}
We now compare our approach with other relevant methods.  We first examine published methods focused on the detection of data exfiltration that are applicable to our dataset, namely, those that make use of data derived from flow records.  We then consider a one-class method applied directly to the time series representation of our data in order to compare performance in the temporal and transform domains.  Lastly, we develop a two-class supervised learning model trained on our data to compare performance against the one-class learning approach adopted in this work. Comparisons are done for the client workstation, which shows the most promise for application.

Of the known methods of exfiltration detection, the work of \cite{Marchetti} and \cite{Ramachandran} are most directly comparable to our proposal.  \cite{Marchetti} attempts to discover exfiltration by monitoring three flow-based quantities: the number of egress bytes, number of flows initiated by internal hosts, and the number of unique external address contacted by internal hosts within a given time period.  These metrics are assessed daily across the entire network and each host is given a suspiciousness score based on how the host's metrics compare with the rest of the network.  Since our approach applies solely to individual hosts, we cannot apply the method of \cite{Marchetti} directly to our case; but, the authors analyze a network similar to ours in size, bit rate, and flow rate, and find that exfiltration in excess of 500 MB per day is always detectable.  

In \cite{Ramachandran}, kernel density estimation is used to find outlier flows with excessively large byte or packet transfers.  This work also makes use of host-based parameters, including CPU and memory usage, which our method does not include.  The authors claim that large data transfers are always accompanied by spikes in these system parameters, making them redundant in the face of anomalously large data transfers, {\it i.e.} flow data alone is sufficient to detect data exfiltration.  We therefore apply kernel density estimation only to our normal traffic data, namely the distribution $\bar{b}_E$ and $\bar{b}_I$, and asses the simulated type 1 exfiltration for outliers with respect to the learned density.  We fit a Gaussian kernel to $(\log \bar{b}_E,\log \bar{b}_I)$ with a window size of 0.1 found via grid search with cross validation. This is the same approach taken for the avgDFT component in our ensemble, the primary difference being that avgDFT is trained on the average amplitudes of the egress and ingress Fourier spectra, rather than the average byte values.  

\begin{figure}[htp]
\includegraphics[width=0.65\textwidth,clip]{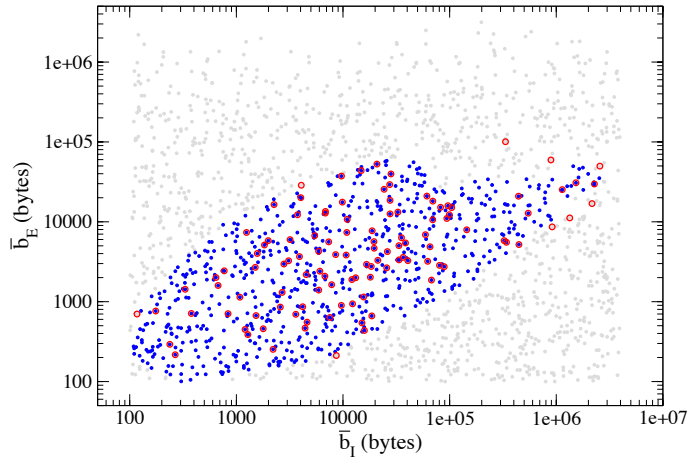}
\caption{Comparison of three detection methods against type 4 exfiltration: gray points are those missed by \cite{Marchetti}, blue by \cite{Ramachandran}, and red circles by our ensemble.  The methods miss 97\%, 37\%, and 6\% of samples, respectively.}
\label{16}
\end{figure}

In Figure \ref{16}, we compare the methods of \cite{Marchetti} and \cite{Ramachandran} with our ensemble, each applied to the set simulated type 4 exfiltration. This is the most general exfiltration scenario (variable data and timing characteristics) and so our method, which performs particularly well against periodic communications, will not gain an unfair advantage. Unsurprisingly, the method of \cite{Marchetti} identifies only 3\% of the samples (gray points),  because it is sensitive only to high volume, high data rate flows.  The method of \cite{Ramachandran} performs similarly to the first module of our ensemble, avgDFT, missing only those samples lying outside the bulk of learned density, or 37\% of the samples (blue points).  The avgDFT module of our ensemble performs slightly better, missing 33\%, and the additional modules reduce this to 6\% (red circles).   

This comparison demonstrates how our method is an extension of those based on byte characteristics, like volume and ingress/egress balance.  The components based on the DFT and STFT allows the ensemble a chance of detecting communications with normal byte characteristics by examining their temporal behavior for anomalies.

\begin{figure*}[htp]
\begin{tabular}{cc}
\subfloat[Type 1]{\includegraphics[width = 3in]{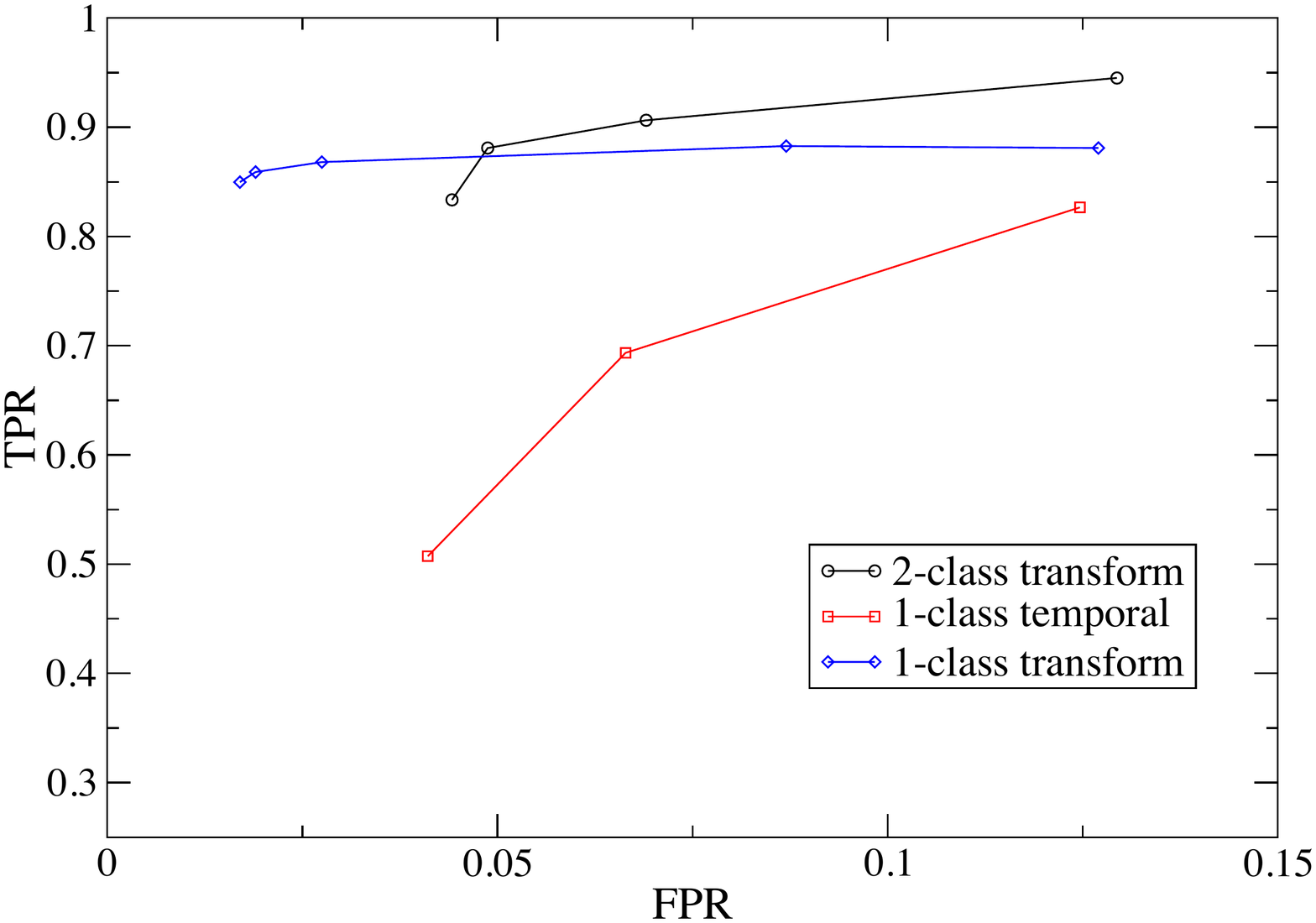}} &
\subfloat[Type 2]{\includegraphics[width = 3in]{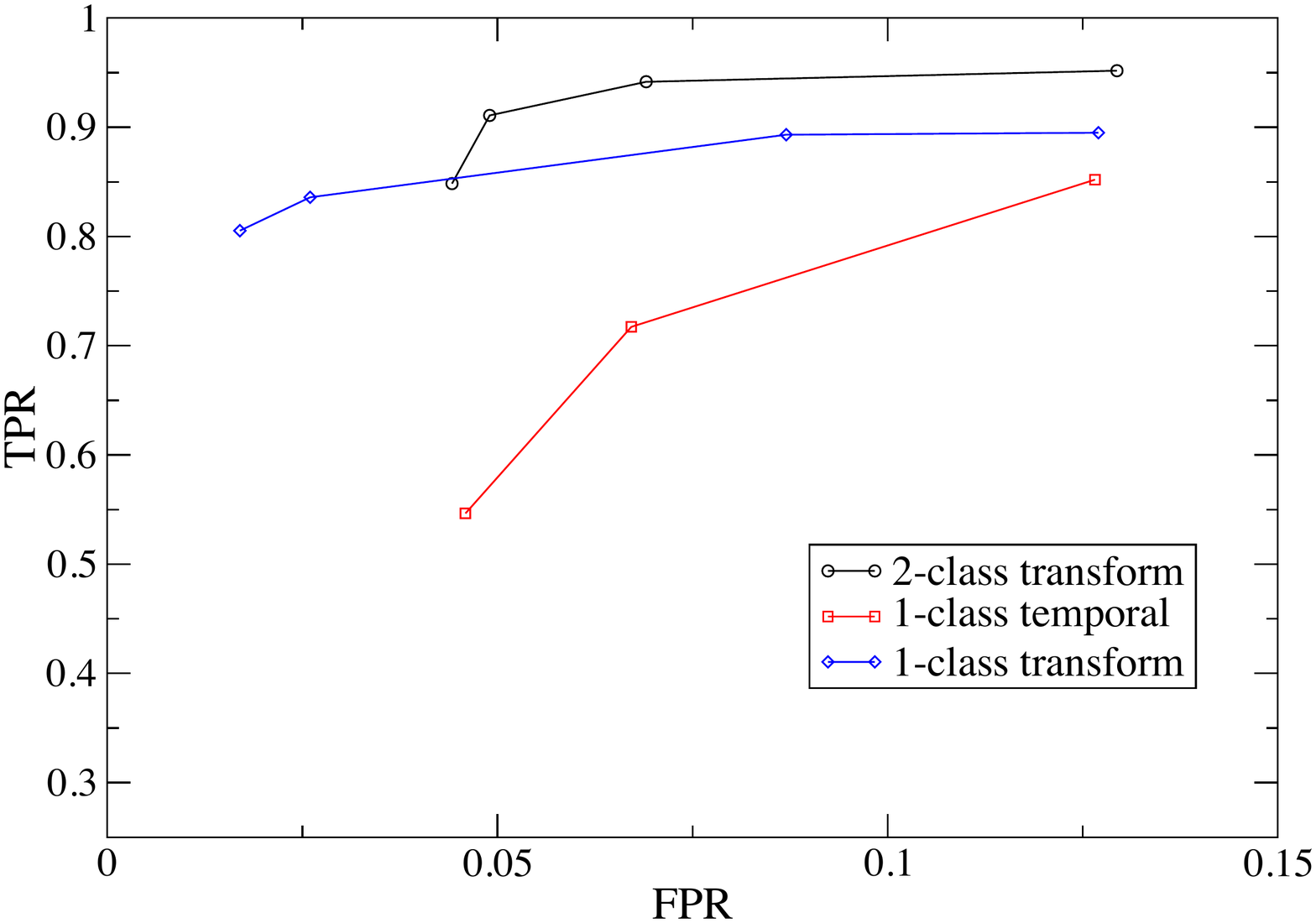}}\\
\subfloat[Type 3]{\includegraphics[width = 3in]{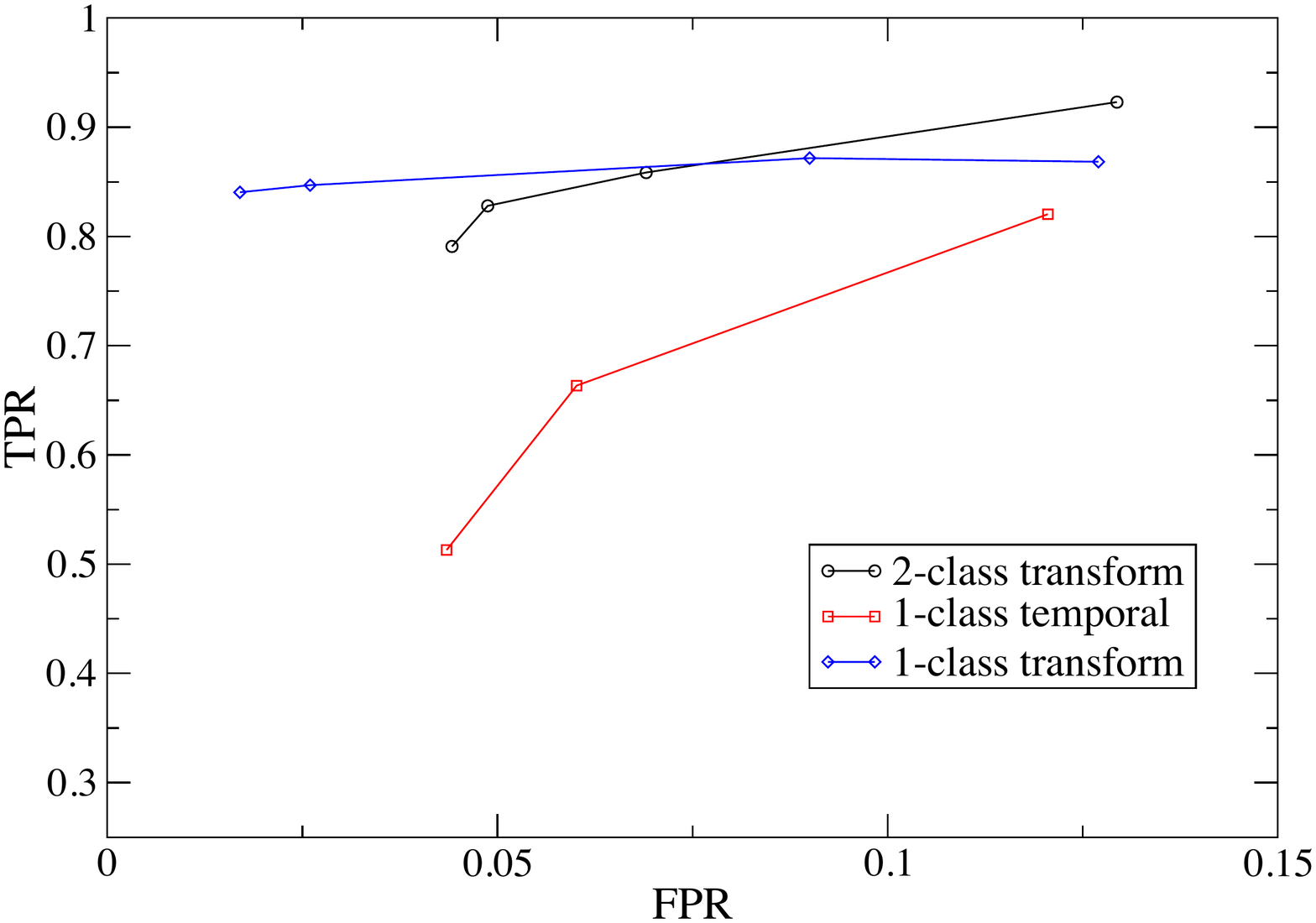}}& 
\subfloat[Type 4]{\includegraphics[width = 3in]{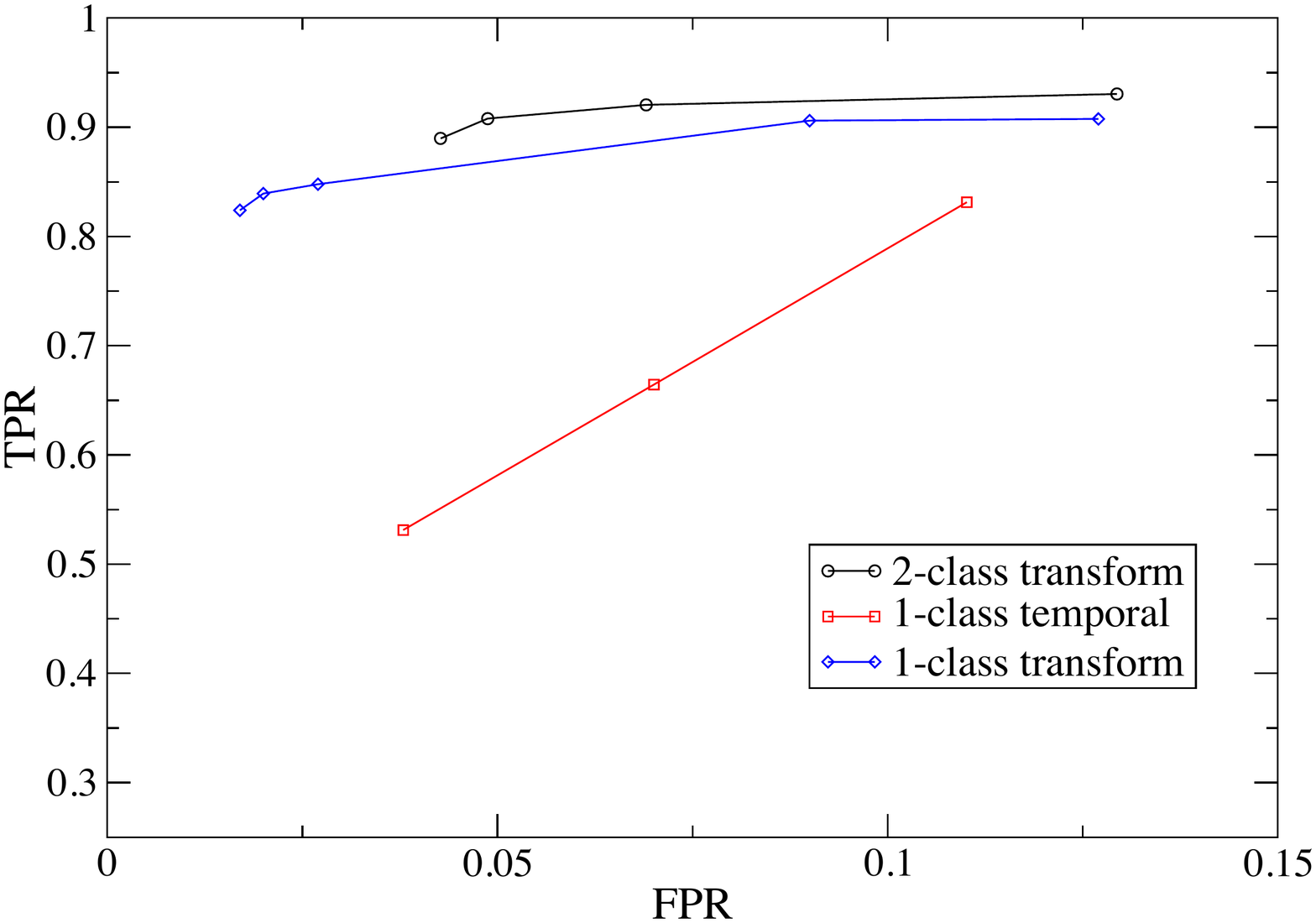}} 
\end{tabular}
\caption{ROC curves of three exfiltration detection methods: the one-class transform-domain ensemble described in this paper, a one-class time-domain ensemble, and a two-class transform domain model.  Models were tested separately against each kind of exfiltration, types 1 through 4. TPR is the rate of detecting exfiltration samples missed by the avgDFT component.}
\label{17}
\end{figure*}

We next compare our transform-domain based ensemble with an analogous one developed using only temporal domain features.  The immediate challenge of working in the time domain was cited earlier: long duration flows, at even seconds- or minutes-resolution, have lots of timesteps (over one million for the two week-long samples at 1 second-resolution used here).   In selecting the dimension of our feature vectors, there is a trade-off between the desire for high-resolution data and the performance of many classifiers in high-dimensional features spaces. For example, SVM's are liable to over-fit high-dimensional, low-sample size data through the phenomenon of {\it data piling}, in which many data points are projected to the same point in feature space \cite{Marron}.  To reduce the dimensionality of the time series, we downsampled by averaging byte values over non-overlapping windows of time: we tested window sizes of 500, 1000, and 2000 seconds.  The same one-class algorithms considered for use in the ensemble---$\nu$-SVM, isolation forest, and $k$-NN---were tested in the time domain.  The $\nu$-SVM with the 1000 second downsampling performed best.  The time domain ensemble was constructed of two $\nu$-SVMs, one trained on the ingress time series and one on the egress time series.  For the sake of fair comparison, we trained the time domain ensemble only on the samples missed by the avgDFT component of the transform domain ensemble, since the timing characteristics are principally used to detect traffic that cannot be identified as exfiltration by volume or byte imbalances alone. The ROC curves are shown in Figure \ref{17}, where it is shown that the temporal domain is outperformed by the transform domain.  TPR is the rate of detecting exfiltration missed by the avgDFT component. 

The sub-sampled discrete Fourier transform amounts to a compression of the time series, and the above comparison is useful because it sheds light on just how much time-domain information we throw away in adopting frequency space representations.  With respect to classifier performance, we can say that our transform-domain-based method retains more useful information than the original time series downsampled by a factor of 1000.

We next consider a two-class supervised model trained on the same transform domain features as the one-class ensemble.  In the two-class setting we must train the model on the ``anomalous'' class, which requires a collection of exfiltration samples that adequately cover the range of possible such communications.  Though it's possible to discover novelty with such a model ({\it e.g.} data exfiltration of a type not like any sample in the anomalous class), this requires the classifier to be equivocal about such samples (splitting probability roughly evenly across the normal and anomalous classes) and there is no natural decision threshold to determine just how equivocal the model should be.  In contrast, in the one-class setting we need only normal data---though we indeed used simulated exfiltration data in this study, we used it to {\it test} rather than {\it train} our model; of course, our confidence in the ensemble's performance is in proportion to the fidelity of the simulations, but it's ability to detect true novelty in a rigorous fashion is built into the method.  But, if we were to consider the exfiltration communications as comprising a fair and comprehensive sample of the kinds of malicious traffic we'd like to detect, could we train a two-class model competitive with our one-class ensemble?

We created nearly-balanced classes: the 4143 normal workstation samples comprise the ``normal'' class, and 1000 samples each of the four different types of exfiltration comprise the ``anomaly'' (or ``exfiltration'') class. We tested a small set of conventional supervised learning techniques (SVM, random forest, gradient boosted trees, $k$-NN) and found that $k$-NN performed best. A separate classifier was trained on the same features as the one-class ensemble, the egress and ingress Fourier amplitudes, $|Y_E(f)|$ and $|Y_I(f)|$, and the STFT, $S(\tau,f)$.  Like the time domain ensemble above, the model was trained only on samples missed by the avgDFT component of the transform domain ensemble.  ROC curves are shown in Figure \ref{17}.  Interestingly, the two-class model performs best at moderate accuracy (FNR $\approx$ FPR $\approx$ 10\%) but is unable to achieve the FPR of the one-class ensemble, and the ROC curves suggest that even if it did, TPR rates would not be competitive.  This difference in performance might have something to do with how the two models learn the classes: the two-class model learns the normal class {\it in relation} to the exfiltration class, while the one-class model attempts to learn the normal class without putting it in the context of other classes or data.  The decision boundaries are thus different, and we expect different performance particularly at extreme values for FPR and TPR. 
\section{Some thoughts on implementation}
The capability described in this paper has two potential implementations: as a near-real-time exfiltration detection suite, or as an analysis tool for targeted investigations or incident response.  For the latter case, implementation is straightforward: train the ensemble on a presumed-normal collection of flow data from the system(s) of interest, and test the suspect traffic.  The requirements for this application are a historic record of flow or packet captures, and some time is required to assemble the communications from the network data, compute the Fourier transforms, and train the classifiers.  A single thread running on a single system\footnote{2.9 GHz Intel Core i7, 16 GB 2133 MHz LPDDR3 RAM.} completes this task in around seven minutes per 100 communications on average with the parameterizations adopted in the preceding analysis, but it is easily parallelized.  If a system is tested more than once per time window (the two-week duration selected in the foregoing analysis), complete Fourier transforms need not be recomputed for each communication if the transforms are kept in storage.  Instead, those portions of the transform from times earlier than the start of the new analysis window can be dropped and only the transforms at new times need to be computed. Training the ensemble takes a matter of minutes, as does testing.  Training times are short enough to allow some kind of grid search to optimize parameters.

Next is the question of the number of ensembles that need to be trained: one for each system, or can systems be grouped?  We do not investigate this question here, but it is certainly worthy of future research.  It seems plausible that systems with similar roles could be grouped under a single classifier trained on data containing representative traffic samples from each system.  Such groupings could be obtained by applying clustering techniques to different representations of traffic data ({\it e.g.} \cite{Dewaele}).  On large and heterogeneous networks, there will likely remain several separate ensembles making training and grid searching more computationally intensive; furthermore, the training will likely need to be automated.

A key consideration of any machine learning application is the question of retraining and concept drift.  If a classifier is trained on a given workstation's data in early April, will it still be good in August?  One immediate consideration unique to the network environment is dynamic host configuration protocol (DHCP): if individual hosts are routinely assigned new IP addresses, this will need to be accounted for before any of the analysis described here is begun. This is more an issue of bookkeeping---of organizing the raw data---than retraining.  But, aside from configuration changes like DHCP, it is possible that the behavior of a particular system changes over time: perhaps the user gets re-tasked (changing the frequency or nature of internet usage) or does something simple like opens a streaming audio account.  These changes cannot in general be predicted across the network, but periodic monitoring of the classifier's false positive rates might reveal some kind of drift.  Even in the absence of an explicit change in system behavior, the statistics of a given system's traffic will fluctuate over time ({\it e.g.} Figure \ref{5} will change slightly from week to week).  We tested the prototype developed in the foregoing analysis on three other samples of traffic data from the same workstation, taken at one-month intervals from the initial training.  The ensemble had around a 2.5\% FPR on each data set, revealing no time dependence and an apparent slight overestimate of the FPR during cross-validation.

\section{Conclusions}
We describe and test the performance of an adaptive, behavior-based data exfiltration detection capability.  The method employs an ensemble of one-class learning algorithms trained on separate flow-oriented feature representations.  It extends known volume- and ingress/egress byte imbalance-based exfiltration detection schemes by testing traffic for unusual timing characteristics thought to be indicative of malicious, overt data transfers out of the network.  The ensemble is meant to be modular, so that classifiers trained to test different traffic characteristics can be mixed and matched to target exfiltration of varying sophistication.  The feature representations are in the frequency domain, capturing the harmonic structure of traffic over time; by working in the frequency domain, it is straightforward to detect data exfiltration occurring over days or weeks.  

The ensemble is trained on a single host's normal traffic collected over a chosen time interval.  Using NetFlow records, egress and ingress bytes over time between the host and each remote system to which it connects are extracted.  Each such communication is then transformed into the frequency domain to form three different feature representations:  the average amplitudes of the ingress and egress Fourier transforms, $(\log\,\overline{|Y_{I}(f)|},\log\,\overline{|Y_E(f)|})$, the full spectra $|Y_I(f)|$ and $|Y_E(f)|$, and the short-time Fourier transform, $S(\tau,f)$. Together, these features characterize the ingress/egress byte distribution of normal communications, their harmonic structure (including periodicity), local traffic irregularities, and their coarse-grained density.  We test out four different kinds of one-class learners: kernel density estimation, one-class support vector machine, isolation forest, and a $k$-nearest neighbors-based learner.  The best performing ensemble includes a mixture of kernel density estimation, isolation forest and one-class support vector machine classifiers: by training each on a different feature representation and selecting classifiers of fundamentally different type, errors are approximately decorrelated and we achieve a  low false positive rate of less than 2\%.  

To test the ensemble, we generate simulated exfiltration flows covering a wide range of different egress patterns, organized into four classes based on timing and data characteristics: periodic/constant egress data, aperiodic/constant egress data, periodic/variable egress data, and aperiodic/variable egress data.  We test the ensemble on three different kinds of systems: a client workstation, email gateway, and outward-facing web server.   On the client workstation, we find that the ensemble is successful at detecting exfiltration that is not simultaneously {\it ingress-heavy, connection-sparse, and of relatively short duration}---a combination that conspires to make it difficult for the adversary to get appreciable amounts of data out of the network.  In comparison with known exfiltration detection methods based on traffic volume and ingress/egress byte imbalance, the ensemble catches 85\% more of the simplest, periodic, constant-egress exfiltration samples, and 82\% more of the most general, aperiodic, variable-egress samples.  On systems with timing characteristics similar to exfiltration, like email gateways and web servers, we find that the ensemble does not perform as well: on the email gateway, 75\% of the simplest exfiltration samples are detected and 45\% of the most general; on the web server, 52\% of the simplest samples are detected and only 16\% of the most general.  Though the ensemble's effectiveness is challenged on these kinds of systems, given the low false positive rate it might still be an advantageous component of defense-in-depth targeting data theft.   

To improve prospects for detecting exfiltration from servers, the focus might be shifted to catching data staged to these servers from other internal systems before egress from the network.  The KDE can be applied to the avgDFT of internal traffic of systems to detect data flows that depart from the normal ingress/egress balance; since most internal data staging will occur with file shares over SMB, it will be generically ingress-heavy and should be detectable on systems that usually receive large data transfers from other internal systems. 

Other areas of potential improvement include the investigation of wavelet transforms as a means of obtaining high-resolution representations of traffic in the time-frequency domain.  While the STFT employed here had success in some areas, wavelets can more efficiently manage the time/frequency resolution trade-off.  

The traffic samples from the systems selected for testing in this work were sufficiently homogeneous that only a single ensemble was needed to accurately learn normal behavior, but this need not be the case in general.  In such circumstances, it might be necessary to perform initial clustering or other unsupervised learning on the traffic data to identify natural subsets on which to train separate ensembles.  A rigorous and systematic approach to this problem has been left to future research.

It is worth emphasizing that our approach is not a silver bullet.  Determined, well-resourced adversaries conducting targeted attacks will take the time to learn the normal communications patterns of a device before attempting any kind of data exfiltration.  They will adjust the timing and size of data pulls in order to mimic the behavior of common web activities or other network processes: it is entirely possible for an attacker to make malice look perfectly normal. This detection capability hopefully raises that bar, by challenging the adversary to be better than our learning algorithms at characterizing normal behavior.   

% use section* for acknowledgment
\begin{acks}
The author would like to acknowledge Charles Frick and Kristopher Reynolds who did early, inspirational work using supervised learning to detect data exfiltration in the frequency domain.  I would like to thank Christian Stoneburner for assistance generating exfiltration samples and providing expert input on data exfiltration tradecraft, Conrad Fernandez for making NetFlow data available in our big data environment, and Clark Updike for many helpful discussions and for support with database technologies.
\end{acks}

\bibliography{powell_exfil}
\bibliographystyle{ACM-Reference-Format}
\end{document}